\documentclass[12pt]{article}

\usepackage[T1]{fontenc}
\usepackage{lmodern}
\usepackage[utf8]{inputenc}
\usepackage[english]{babel}

\usepackage[maxbibnames=300, sorting=none, style=science, articletitle]{biblatex}
\renewbibmacro{in:}{}
\usepackage{newtxtext,newtxmath}
\usepackage{graphicx}
\usepackage{epstopdf}
\usepackage{CJKutf8}
\usepackage[letterpaper,margin=1in]{geometry}
\usepackage[T1]{fontenc}
\usepackage{nameref}

\renewenvironment{abstract}
	{\quotation}
	{\endquotation}

\date{}

\makeatletter
\renewcommand{\fnum@figure}{{Figure \thefigure}}
\renewcommand{\fnum@table}{{Table \thetable}}
\makeatother

\usepackage{url}
\usepackage{aas_macros}
\usepackage{longtable}
\usepackage{tablefootnote}
\def\scititle{\textcolor{black}{An axisymmetric shock breakout indicated by prompt polarized emission from the Type II supernova 2024ggi}}
\title{\bfseries \boldmath \scititle}

\author{
	Yi~Yang\begin{CJK*}{UTF8}{gbsn}
(杨轶)\end{CJK*}
$^{1\ast}$,
Xudong~Wen$^{1,2}$,
	Lifan~Wang$^{3,4\ast}$,
        Dietrich Baade$^{5}$,\and
        J. Craig Wheeler$^{6}$,
        Alexei V. Filippenko$^{7,8}$,
        Avishay Gal-Yam$^{9}$,
        Justyn Maund$^{10}$,\and
        Steve Schulze$^{11}$,
        Xiaofeng Wang$^{1\ast}$,
        Chris Ashall$^{12,13}$,
        Mattia Bulla$^{14,15,16}$,\and
        Aleksandar Cikota$^{17}$,
        He Gao$^{2,18}$,
        Peter Hoeflich$^{19}$,
        Gaici Li$^{1}$,
        Divya Mishra$^{3,4}$,\and
        Ferdinando Patat$^{5}$,
        Kishore C. Patra$^{7,20}$,
        Sergiy S. Vasylyev$^{7}$,
        Shengyu Yan$^{1}$\and
	\small$^{1}${Department of Physics, Tsinghua University, Qinghua Yuan, Beijing 100084, China}\and
	\small$^{2}${School of Physics and Astronomy, Beijing Normal University, Beijing 100875, China}\and
        \small$^{3}${Department of Physics and Astronomy, Texas A\&M University, 4242 TAMU, College Station, TX 77843, USA}\and
        \small$^{4}${George P. and Cynthia Woods Mitchell Institute for Fundamental Physics \& Astronomy,} \\
        {\small Texas A\&M University, 4242 TAMU, College Station, TX 77843, USA}\and
\small$^{5}${European Organisation for Astronomical Research in the Southern Hemisphere (ESO),} \\
{\small Karl-Schwarzschild-Str. 2, 85748 Garching b. M{\"u}nchen, Germany}\and
\small$^{6}${University of Texas at Austin, 1 University Station C1400, Austin, TX 78712-0259, USA}\and
\small$^{7}${Department of Astronomy, University of California, Berkeley, CA 94720-3411, USA}\and
\small$^{8}${Hagler Institute for Advanced Study, Texas A\&M University, 3572 TAMU, College Station, TX 77843, USA}\and
\small$^{9}${Department of Particle Physics and Astrophysics, Weizmann Institute of Science, Rehovot, Israel}\and
\small$^{10}${Department of Physics, Royal Holloway, University of London, Egham Hill, Egham, TW20 0EX, UK}\and
\small$^{11}${Center for Interdisciplinary Exploration and Research in Astrophysics (CIERA),} \\
{\small Northwestern University, 1800 Sherman Ave, Evanston, IL, 60201, USA}\and
\small$^{12}${Department of Physics, Virginia Tech, 850 West Campus Drive, Blacksburg, VA 24061, USA}\and
\small$^{13}${Institute for Astronomy, University of Hawai'i at Manoa, 2680 Woodlawn Dr., Hawai'i, HI 96822, USA}\and
\small$^{14}${Department of Physics and Earth Science, University of Ferrara, via Saragat 1, I-44122 Ferrara, Italy}\and
\small$^{15}${INFN, Sezione di Ferrara, via Saragat 1, I-44122 Ferrara, Italy}\and
\small$^{16}${INAF, Osservatorio Astronomico d'Abruzzo, via Mentore Maggini snc, 64100 Teramo, Italy}\and
\small$^{17}${Gemini Observatory/NSF's NOIRLab, Casilla 603, La Serena, Chile}\and
\small$^{18}${Institute for Frontier in Astronomy and Astrophysics, Beijing Normal University, Beijing 102206, China}\and
\small$^{19}${Department of Physics, Florida State University, Tallahassee, FL 32306, USA}\and
\small$^{20}${Department of Astronomy \& Astrophysics, University of California, Santa Cruz, CA 95064, USA}\and
	\small$^\ast$Email: yi\_yang@mail.tsinghua.edu.cn, lifan@tamu.edu, 	wang\_xf@mail.tsinghua.edu.cn
}

\begin{document} 

% Insert the title and author list
\maketitle

% Abstract, in bold
% There are strict length limits, and not all formats have abstracts.
% Consult the journal instructions to authors for details.
% Do not cite any references in the abstract.
\begin{abstract} \bfseries \boldmath
The death of massive stars is triggered by an infall-\textcolor{black}{induced bounce} shock that 
%severely 
disrupts the star. How such a shock is launched and \textcolor{black}{propagates through the star} is a decades-long puzzle. \textcolor{black}{Some models assume that the 
%re-bounce 
shock can be re-energized by absorbing neutrinos, leading to highly aspherical explosions.} \textcolor{black}{Other models involve jet-powered shocks that lead to bipolar explosions} reflected in the geometry of the shock-breakout emission. We report \textcolor{black}{measurement of the geometry of the shock breakout} through unprecedentedly early spectropolarimetry of the nearby Type II supernova 2024ggi starting $\sim$1.2 days after the explosion. \textcolor{black}{The measurement} indicates a well-defined symmetry axis 
\textcolor{black}{of the shock breakout}, which is also shared by the hydrogen-rich envelope that emerged after the circumstellar matter \textcolor{black}{was} engulfed by the ejecta, 
\textcolor{black}{revealing} a persisting and prominent symmetry axis throughout the explosion. These findings suggest \textcolor{black}{that} the physical mechanism driving the explosion of massive stars manifests a well-defined axial symmetry and acts on large scales.
\end{abstract}

% The first paragraph of any Science paper does NOT have a heading
% Nor is it indented
\section*{Introduction}
\textcolor{black}{``Since the beginning of physics, symmetry considerations have provided us with an extremely powerful and useful tool in our effort to understand nature \cite{1981PhT....34l..55L}.''} %%(T.D. Lee, Particle Physics and an Introduction to Field Theory, 1981, page 177).
\textcolor{black}{The geometry of a supernova (SN) explosion, which has been found aspherical, 
%asymmetric and diverse, 
provides fundamental information on stellar evolution and the physical processes leading to these cosmic fireworks \cite{2008ARA&A..46..433W}.}
Iron-core collapses of massive stars in \textcolor{black}{the} mass range of 8--20 M$_{\odot}$~\cite{2009MNRAS.395.1409S, 2015PASA...32...16S} are the dominant stellar explosions in the nearby universe~\cite{2011MNRAS.412.1441L}. 

%\textcolor{black}{Modern} 
\textcolor{black}{
Neutrino-driven models of core-collapse supernovae (CCSNe) have only become successful in recent years thanks to three-dimensional (3D) simulations.
In particular, the re-bounce shockwave may stall to accretion towards certain directions, while the accretion of in-falling matter onto the proto neutron star and neutrino energy deposition is continuous in other directions. Such a neutrino-driven explosion would result in a break of spherical symmetry~\cite{2013RvMP...85..245B, 2024ApJ...964L..16B, 2024ApJ...969...74W}. 
}
Nevertheless, explaining the details about the generation of the shock waves during the \textcolor{black}{collapse of the stellar core} and the \textcolor{black}{energy transportation via} a burst of neutrinos to produce \textcolor{black}{an explosion remains a challenge}. 
Alternative models include the deposition of energy in the stellar envelope through mechanisms such as magnetorotational processes during the formation of the protoneutron star. 
\textcolor{black}{This process, in which the progenitor iron core exhibits a short rotation period of $\lesssim10$ seconds~\cite{2006ApJS..164..130O, 2007ApJ...664..416B}, may launch} moderately relativistic jets into the outer core and the stellar envelope~\cite{1970ApJ...161..541L, 1999ApJ...524L.107K, 2003ApJ...598.1163M, 2014MNRAS.445.3169O, 2015Natur.528..376M, 2016ARNPS..66..341J}. %{2016NewAR..75....1S}. 
%In the latter case, the engine that drives the explosion is highly asymmetric, probably with a well-defined directionality as required by the observational evidence of pulsar kicks~\cite{1994Natur.369..127L, 1997MNRAS.291..569H, 2012ARNPS..62..407J, 2017ApJ...837...84J, 2017ApJ...846..170T}. 
Engines driving core-collapse explosions may follow well-defined global asphericities of the progenitor systems as hinted at by the observational evidence of SN remnants~\cite{1995Natur.373..587A, 2022RAA....22l2003S} and pulsar kicks~\cite{1994Natur.369..127L, 1997MNRAS.291..569H, 2012ARNPS..62..407J, 2017ApJ...837...84J, 2017ApJ...846..170T}. Bipolar/jet-driven models compatible with these observations have been proposed~\cite{1996ApJ...459..307H, 1999ApJ...524L.107K, 2001ApJ...550.1030W, 2003ApJ...598.1163M, 2009ApJ...705.1139M}. Explosion models adopting pure neutrino heating within the spherically symmetric scheme~\cite{2015ApJ...806..275P, 2018JPhG...45j4001O, 2020ApJ...890..127C} or driven by small-scale instabilities~\cite{2018MNRAS.479.3675M}, on the other hand, are expected to be amorphous or exhibit no symmetry axis. 
Recent 3D radiation-hydrodynamic simulations also illustrate that microscopic neutrino physics details in the early seconds can determine \textcolor{black}{the large-scale ejecta structure that is preserved 
\textcolor{black}{for} 
days~\cite{2025ApJ...982....9V}.} 
%, which indicates the morphology of the shockbreakout
\textcolor{black}{Modeling of the blueward color evolution of SN\,2023ixf, recorded within a few hours after the first light, infers an inhomogeneous emergence of the shock from the exploding star enshrouded by CSM that started from where the opacity yields the smallest~\cite{2024Natur.627..754L}. The critical link between the shock breakout and the explosion mechanism that drives the expansion of the ejecta may be facilitated by comparing their geometries. Whether the former}
%The geometry of the \textcolor{black}{shock breakout} of CCSNe, and whether it 
lines up with \textcolor{black}{that} of the SN ejecta and any explosion fingerprints left toward the core-collapse center, would thus provide a powerful probe of the explosion physics. 

Extremely early spectropolarimetry, taken within about one day after \textcolor{black}{shock breakout}, offers a unique \textcolor{black}{opportunity} to observe how the shock emerges on the surface of the exploding star and interacts with any surrounding circumstellar matter (CSM) as \textcolor{black}{evidenced} by short-lived photoionized (``flash ionization'') spectral features~\cite{2014Natur.509..471G, 2016ApJ...818....3K, 2017NatPh..13..510Y, 2017A&A...605A..83D, 2021ApJ...912...46B, 2023ApJ...952..119B, 2024ApJ...970..189J}. 
Because of the large distances of extragalactic supernovae (SNe), the regions concerned remain angularly unresolved, compressed to the radial velocity and time axes. Critical information about the 3D structure of the ejecta and their interaction with CSM is encoded in polarization spectra. Continuum polarization \textcolor{black}{measures} the deviations of the photosphere from spherical symmetry. 
\textcolor{black}{Line polarization traces the distribution of elements in the SN ejecta projected onto the plane of the sky~\cite{2008ARA&A..46..433W}. 
%In combination with radial velocities, 
Modulation of the polarization degree and position angle across a spectral feature probe the strength of departure from spherical symmetry and its orientation, respectively, delivering a low-resolution 3D map of the corresponding line-forming region~\cite{2003ApJ...591.1110W, 2003ApJ...593..788K, 2008ARA&A..46..433W}.}
The acquisition of such a dataset close in time to shock breakout only became feasible recently thanks to the transient-alert stream produced by 
sub-day-cadence wide-field sky surveys, 
%such as the `Asteroid Terrestrial-impact Last Alert System' (ATLAS~\cite{2018PASP..130f4505T}) and the Zwicky Transient Facility (ZTF~\cite{2019PASP..131a8002B, 2019PASP..131g8001G}), 
combined with rapid spectropolarimetric follow-up observations.
%\textcolor{black}{enabled with optical spectropolarimeters}~\cite{1998Msngr..94....1A}.
%, 1988igbo.conf..157M}.

\section*{Results}
\subsection*{Spectropolarimetry of Supernova 2024ggi}
SN\,2024ggi was discovered as a transient \textcolor{black}{with rapid intranight} rise~\cite{2024TNSAN.100....1S} in the spiral galaxy NGC\,3621 at a distance of $7.24 \pm 0.20$\,Mpc~\cite{2006ApJS..165..108S} and was quickly classified as a young Type II SN~\cite{2024TNSAN.104....1Z}. 
\textcolor{black}{The transient alert stream was produced by the `Asteroid Terrestrial-impact Last Alert System (ATLAS)~\cite{2018PASP..130f4505T}'.
}
The proximity of SN\,2024ggi \textcolor{black}{provides} a rare opportunity to investigate the pre-to-post-explosion properties of this CCSN in great detail. 
We initiated a spectropolarimetric time sequence of SN\,2024ggi (see Table~\ref{Table:log_specpol}), starting at UTC 05:57 on 2024-04-12 (MJD 60412.248) \textcolor{black}{following the immediate approval of the European Southern Observatory (ESO) Director's Discretionary Program (ID 113.27R1, PI Y.\,Yang).}
%\footnote{Spectropolarimetry of SN\,2024ggi has been conducted following the immediate approval of the European Southern Observatory (ESO) Director's Discretionary Program (ID 113.27R1, PI Y.\,Yang).}. 
\textcolor{black}{The first epoch was carried out at $\sim 1.1$ days after the discovery on MJD 60411.14~\cite{2024TNSAN.100....1S}, which is an objective observation, and 1.22$_{-0.05}^{+0.05}$ days after the estimated time of shock breakout on MJD 60411.03$_{-0.05}^{+0.05}$~\cite{2024ApJ...970L..18Z}, which is model dependent. Throughout this paper, all phases are given relative to the time of the SN discovery. The observing campaign on SN\,2024ggi harvested one of the two earliest spectropolarimetric datasets} 
%by far one of the two earliest spectropolarimetry 
of any transient, the other was 1.39$_{-0.02}^{+0.05}$ days after \textcolor{black}{shock breakout}~\cite{2024Natur.627..754L} of SN\,2023ixf~\cite{2023ApJ...955L..37V}. 
This rare early dataset enables \textcolor{black}{us to} measure the geometry of the shock breakout \textcolor{black}{(Methods:~\nameref{sec:vlt})}, which took place between days 0.7 and 1.2 as inferred from the early evolution of the ionization states of the CSM emission lines~\cite{2024ApJ...970L..18Z}. 

%A candidate red supergiant  progenitor has been identified from {\it Hubble Space Telescope} ({\it HST}) 1995--2003 archival images and the Dark Energy Spectroscopic Instrument (DESI) legacy imaging survey \cite{2024TNSAN.100....1S}.  The contradiction between an inferred mass-loss rate of $< 3\times10^{-6}$\,M$_{\odot}$\,yr$^{-1}$ from the {\it HST} and DESI archival spectral energy distribution  \cite[SED;][]{2024ApJ...969L..15X} and $\sim 10^{-2}$--$10^{-5}$\,M$_{\odot}$\,yr$^{-1}$ from the spectroscopic data and the X-ray evolution of the SN may imply significantly enhanced mass loss within a few years before the explosion \cite{2024ApJ...972L..15S}. 
%For example, both SNe\,1999em\cite{2008ARA&A..46..433W} and 2021yja\cite{2024MNRAS.527.3106V, 2024A&A...687L..17N} showed a well-defined symmetry axis over the observed wavelength range starting weeks after the explosion for different epochs and wavelengths. A sharp rise of polarization degree as a Type IIP SN evolves into its nebular phase also indicates an aspherical core \cite{2006Natur.440..505L}. 

Investigation of the geometry of the continuum and different spectral features can be facilitated by presenting spectropolarimetry on the normalized Stokes $Q–U$ plane~\cite{2001ApJ...550.1030W}. A prominent axial symmetry of an electron-scattering structure leads to a wavelength-independent polarization position angle (PA) of the continuum in the $Q-U$ plane. 
\textcolor{black}{
For data points with different wavelengths, their distance from the origin (polarization degree $p$) varies owing to different physical properties across the photosphere (e.g., temperature, density, and composition), resulting in a range of optical depths and scattering efficiencies.}
Together, they form a straight line known as the dominant axis~\cite{2003ApJ...591.1110W, 2010ApJ...722.1162M}. 

%Measurements over certain spectral ranges of interest can be projected onto the dominant axis, dubbed as $P_{d}$, and the axis orthogonal {($P_{o}$)} to it. 
\textcolor{black}{The polarization over certain spectral ranges can be decomposed into a component along the dominant axis ($P_{d}$) and another one along the orthogonal axis ($P_{o}$).}
The former captures the most dynamic range of the data~\cite{2003ApJ...591.1110W}. Its slope in the $Q-U$ plane \textcolor{black}{delivers} the spatial orientation of the axial symmetry. 
For ejecta with rotational symmetry, the dominant and orthogonal axes measure the axial asphericity of the ejecta and the deviations from such a geometry, respectively. Therefore, for any wavelength range or spectral line of interest, a clear dominant axis would indicate a prominent axial symmetry of the associated opacity distribution. 
\textcolor{black}{On the contrary, any clumpy, nonaxisymmetric structure will spread along the orthogonal axis, making the dominant axis less significant~\cite{2008ARA&A..46..433W}.}

After removal of the interstellar polarization (ISP) arising from the foreground interstellar dust \textcolor{black}{(Methods:~\nameref{sec:isp})}, in Fig.~\ref{fig:contpol} we present the temporal evolution of the intrinsic continuum polarization of SN\,2024ggi at eight epochs from days 1.1 to 80.8. In each panel, different symbols mark the \textcolor{black}{inverse 1$\sigma$ error weighted mean} polarization over the wavelength ranges identified in the color bar. In the top-left and bottom-right panels, the black dashed lines show the dominant axes of the first and last datasets. In these ISP-corrected data, $Q=0$, $U=0$ is between the red and blue wavelengths at day 1.1 and near the blue end of the dominant axis at day 80.8. The data at intermediate epochs do not show clear dominant axes. These data are substantially displaced from $Q=0$, $U=0$. A drastic change of the continuum polarization (from days 1.1 to 2.0) is followed by a gradual drift until a roughly stationary geometry is reached at day 10.9, indicating a large-scale transformation of the geometry as the CSM is swept up by the SN ejecta. Throughout all analyses and figures, the ISP has been subtracted \textcolor{black}{unless stated otherwise}.

\subsection*{Stage I -- The Shock-Breakout Phase} 
At very early epochs, the photosphere of SN\,2024ggi was most likely engulfed in the CSM, as evidenced by \textcolor{black}{several} highly \textcolor{black}{photoionized} narrow features superposed on a blue continuum (\textcolor{black}{Methods:~\nameref{sec:linepol}})~\cite{2024ApJ...972..177J, 2024ApJ...970L..18Z, 2024ApJ...969L..15X, 2024A&A...688L..28P, 2024ApJ...972L..15S}. 
The dynamical timescale is short on day 1, when the photospheric radius yields $\lesssim1.5\times10^{14}$\,cm~\cite{2024ApJ...970L..18Z} and the ejecta expand rapidly. 
%Later after day 10.9, the photosphere recedes into the H-rich envelope of the ejecta and stabilizes owing to the longer dynamical timescales. This is indicated by the development of P~Cygni features in the spectra (bottom-right panel of Figure~\ref{fig:contpol_qu}). 
At day 1.1, the $Q-U$ diagram shows a well-defined dominant axis \textcolor{black}{with} 
2\,PA$_{\rm day\,1.1}=132^{\circ}.7_{-3^{\circ}.7}^{+4^{\circ}.3}$ (Fig.~\ref{fig:contpol}), where PA $=$ 0.5\,tan$^{-1}(U/Q)$.
%, 2PA is given to represent the slope of the dominant axis of the continuum polarization displayed on the Stokes $Q-U$ plane. 
\textcolor{black}{The distribution of the day 1.1 polarization can also be described by an ellipse, whose semimajor and semiminor axes are defined by the scatter about the dominant and orthogonal axes, namely 
%$a=0.122$\% and $b=0.088$\%, 
$a \approx$0.12\,\% and $b \approx$0.09\,\%,
respectively (Fig.~\ref{fig:contpol_qu}).}
\textcolor{black}{As supported by the blueward $g-r$ color evolution and the continuous rise of the until about day 1.6~\cite{2024ApJ...970L..18Z, 2024ApJ...972L..15S, 2024ApJ...972..177J}, spectropolarimetry at day 1.1 measures the emission of the shock breakout, when photons promptly diffuse out of the optically thick CSM in certain directions. 
We note that such a geometry measurement is only feasible immediately after the onset of the shock breakout, during a brief moment when the shock has promptly emerged from the surface of the progenitor in some directions, while the remaining part of the shock is still embedded in the optically thick atmosphere or CSM. Therefore, the first epochs of spectropolarimetry of SN\,2023ixf did not infer the shock breakout geometry~\cite{2023ApJ...955L..37V, 2024ApJ...975..132S, 2025ApJ...982L..32S, 2025arXiv250503975V}, as the spread of the shock front to cover the entire surface of the SN\,2023ixf progenitor persisted only for the first few hours~\cite{2024Natur.627..754L}.}
%The photosphere shaped by the shock breakout displays a prominent symmetry axis. 

\textcolor{black}{We remark that the wavelength-dependent polarization on day 1.1 closely resembles the ISP as described by the empirical Serkowski law~\cite{1975ApJ...196..261S}. Our attempts to characterize such a time-invariant redistribution of the data points on the $Q-U$ plane imply that the day 1.1 polarization wavelength dependence is intrinsic to the SN (Methods:~\nameref{sec:isp}). 
Instead, a wavelength-dependent photosphere would be expected for a spherically asymmetric shock breakout. The total observed intensity is a summation of various emitting components, each having an intensity of $I_{j}(\lambda)$ at a given wavelength $\lambda$. The net polarization is thus the total polarized flux normalized by the total flux, i.e., 
\begin{align}
    p(\lambda) = \sum_{{j}} I_{j}(\lambda) p_{j}(\lambda) \bigg{/} \sum_{{j}} I_{j}(\lambda);
\end{align}
%(i): P(λ) = Σi (Fi(λ) * Pi (λ)) / Σi (Fi(λ)). 
Therefore, even if the polarization of each emission component with a characteristic blackbody temperature is wavelength independent, the net polarization can still be wavelength dependent.
}

Additional information on the geometry can be deduced from the polarization across spectral lines, which is especially sensitive to the geometric distribution of chemical species involved rather than the global shape represented by the photosphere and the continuum polarization. For a geometric structure with rotational symmetry, the $Q-U$ diagram representing the wavelength bins within a spectral line reflects the geometry of the atomic species producing the line. The emitting regions at the earliest epoch most closely trace the ionization front of the shock propagating in the CSM, as indicated by the common dominant axis determined from the continuum and from the spectral features with the highest ionization potentials. 

\textcolor{black}{
%At day 1.1 SN\,2024ggi as the central regions of the shock-photon-ionized spectral features are less affected by the electron scattering emission from the wings, which are sensitive to small-scale structures such as the lumpiness of the scattering regions, their polarization displayed on the Stokes Q$-$U plane would closely trace the geometry of the shock breakout ionization front with the least influence from other effects. 
Because electron-scattering emission wings are sensitive to small-scale structure such as lumpiness in the scattering CSM, we focus on the emission cores of lines on day 1.1 of SN\,2024ggi. The line cores are less affected by electron scattering than the line wings. The polarization of the line cores as displayed in the Stokes $Q-U$ plane should more closely trace the geometry of the shock-breakout ionization front with the least influence from other effects. 
As illustrated in the upper-right panel of Fig.~\ref{fig:linepol_early}, all photoionized spectral features line up with the dominant axis on day 1.1. The only exception is H$\alpha$; the excitation energy of H$\alpha$ ($\chi=13.6$\,eV) is the lowest among all \textcolor{black}{lines} identified in the earliest flux spectrum and can thus be emitted over a wide range of angles with respect to the direction of the shock breakout so that any geometrical information is strongly diluted. By contrast, the highest excitation state, O\,V ($\chi=113.9$\,eV), exhibits a clear dominant axis across the O\,V\,$\lambda$5597 feature similarly to the continuum (fig.~\ref{fig:linepol_early_method}).
}

This \textcolor{black}{observational signature}
can be understood as the associated high ionization potential required by the highly ionized species to be realized close to the shock front, where the highest temperature produces the highest excitation states. The fact that the high-ionization lines (e.g., C\,IV\,$\lambda\lambda$5801, 5812, N\,IV\,$\lambda\lambda$7109, 7123, and O\,V\,$\lambda$5597) in the spectra of SN\,2024ggi emerged after day 1.1, rather than \textcolor{black}{before} day 0.7~\cite{2024ApJ...970L..18Z}, \textcolor{black}{is compatible with} an early increase in photospheric temperature~\cite{2024ApJ...972..177J, 2024ApJ...972L..15S, 2025ApJ...983...86C}. This suggests a shock breakout within the CSM where a progressively hotter and stronger radiation field is emitted, in contrast to a shock-cooling 
\textcolor{black}{process}~\cite{2017hsn..book..967W, 2023MNRAS.522.2764M, 2024ApJ...972L..15S, 2025ApJ...983...86C}. 
Accordingly, the polarization of the continuum on day 1.1 traces the geometry of the emitting zone, where the shock breakout promptly \textcolor{black}{leaks into} the CSM. The line photons with the highest excitation potential on 
%days 1.1 
\textcolor{black}{day 1.1}
are formed close to the ionization front produced by the shock; thus, their polarization traces the pre-shocked CSM over the line-forming regions. 

\subsection*{Stage II -- The Ejecta-CSM Interaction} 
From days 1.1 to 2.0, \textcolor{black}{a clockwise rotation by 2PA $\approx 59^{\circ}$} is seen among the \textcolor{black}{data clouds in the Stokes $Q-U$ plane}, which represents the continuum polarization over 3800--7800\,\AA. The rotation continued at a slower rate after day 2.0 until the degree of continuum polarization settled at a roughly stationary level between days 10.9 and 33.0 (Fig.~\ref{fig:contpol}). \textcolor{black}{Such temporal evolution} does not necessarily imply a rotation of the symmetry axis in space, but it can be due to a change in the relative contributions of different structures to the total signal. Fig.~\ref{fig:contpol_qu} summarizes the temporal evolution of the continuum polarization by resampling 
\textcolor{black}{the observations at each epoch} into very broad \textcolor{black}{800\,\AA} wavelength bins. 
The green dashed line in the left panel of Fig.~\ref{fig:contpol_qu} shows the dominant axis as defined by the data on day 1.1. It represents the geometric axis of the photosphere at the earliest epoch, which is mostly within the CSM layer ionized by the shock-breakout flash. From days 2.0 to 6.9, the photosphere recedes into a deeper layer of the CSM where the emission produced by the expanding ejecta interacting with the CSM becomes progressively dominant. 
\textcolor{black}{The time-evolving continuum polarization during Stages I--II as displayed} in Fig.~\ref{fig:contpol_qu} clearly reveals a misalignment between the shock breakout and the later ejecta-CSM interaction processes \textcolor{black}{(Methods:~\nameref{sec:pcygni} and ~\nameref{sec:rotate}).} 

The dominant axis can no longer be identified during \textcolor{black}{Stage II} as seen in individual epochs compared to that on day 1.1 (Fig.~\ref{fig:contpol}). 
The temporal evolution of the continuum polarization measurements from days 2.0 to 6.9 follows a different path compared to the axial symmetry on day 1.1 (the blue dashed line in Fig.~\ref{fig:contpol_qu}), \textcolor{black}{demonstrating} that the ejecta-CSM interaction process manifests a geometry different from that inferred during the shock-breakout phase: 2\,PA$_{\rm CSM} = 109^{\circ}.8_{-5^{\circ}.2}^{+10^{\circ}.7}$ compared to 2\,PA$_{\rm day\,1.1}=132^{\circ}.7_{-3^{\circ}.7}^{+4^{\circ}.3}$, respectively. 
From days 2.0 to 6.9, \textcolor{black}{lines from ions such as O\,V, N\,IV, C\,IV, and H$\beta$} are much weaker, and their dominant axes \textcolor{black}{become} significantly less prominent. 

We sketch out four possible geometric configurations of the ejecta within the CSM in Fig.~\ref{fig:schem_early}. 
\textcolor{black}{The schematic drawing of the CSM exhibits a density variation that manifests as a moderate density enhancement toward a specific orientation, namely the CSM plane (i.e., by a factor of $\lesssim$2; \textcolor{black}{Methods:~\nameref{sec:diffusion}).}}
\textcolor{black}{The schematics represent only 
the transition of the emission from Stages I to II.}
%the relatively early phase of the shock-breakout emission. 
On day 1.1, the photosphere displays an axially symmetric structure with a dominant axis that \textcolor{black}{agrees with} the shape of the shock breakout from the CSM, eliminating the \textcolor{black}{doubly} spherically symmetric case illustrated by Fig.~\ref{fig:schem_early}A. 
The configuration evolves rapidly toward a geometry dominated by that of the CSM. However, we find that a spherical shock breakout sculpted by an aspherical CSM (Fig.~\ref{fig:schem_early}B) and an aspherical shock breaking out of a spherical CSM (Fig.~\ref{fig:schem_early}C) would both imprint a single symmetry axis at all times (\textcolor{black}{Methods:~\nameref{sec:contpolmodel}}, fig.~\ref{fig:schem}). 
Both configurations would manifest as a progressive shrinkage of the distance between the data cloud and the zero point in the $Q-U$ plane until \textcolor{black}{the data sequence} %it 
flips to the opposite direction (see the bottom row of fig.~\ref{fig:schem}) \textcolor{black}{instead of displaying the observed} gradual rotation (Fig.~\ref{fig:contpol}) that draws a loop-like trajectory (Fig.~\ref{fig:contpol_qu}). 
Therefore, we conclude that the symmetry axes of the shock breakout (day 1, green dashed line in Fig.~\ref{fig:contpol_qu}) and the ejecta-CSM interaction (days 2--7, blue dashed line in Fig.~\ref{fig:contpol_qu}) are misaligned, 
\textcolor{black}{requiring}
an aspherical shock breakout from the progenitor surface \textcolor{black}{as the explanation}. 
\textcolor{black}{We conclude that} Fig.~\ref{fig:schem_early}D, \textcolor{black}{where} the shock breakout and the CSM are both ellipsoidal but misaligned, is a more realistic representation of SN\,2024ggi.

\subsection*{Stage III -- Dominance of the Hydrogen-Rich Envelope}
%The Hydrogen-Envelope-Dominated Phase}
\textcolor{black}{
At later epochs (day 10.9 and thereafter), the characteristic P~Cygni profiles of the Balmer lines are fully developed (fig.~\ref{fig:linepol_late_iqu}), implying that the receding photosphere has passed the inner boundary of the CSM and resides in the hydrogen-rich envelope \textcolor{black}{of the exploding progenitor} (see Methods:~\nameref{sec:linepol} for the temporal evolution of the spectral features). 
Polarimetry on and after day 10.9 thus probes the geometry of the H-rich envelope of the outermost SN ejecta. The roughly circular, not elongated distribution of the data points in the Stokes $Q-U$ plane hinders the identification of a dominant axis of SN\,2024ggi at individual epochs.} 
The PA of the H-rich envelope \textcolor{black}{in} Stage III 
\textcolor{black}{estimated from} the error-weighted mean of the polarization on days 10.9, 19.9, and 33.0 yields 2\,PA$_{\rm ej} = -20^{\circ}.4_{-25^{\circ}.3}^{+32^{\circ}.4}$, which \textcolor{black}{differs} by $\sim153^{\circ}$ from the symmetry axis inferred for Stage I. 
%Such a change of 
\textcolor{black}{This change in PA} close to a flip \textcolor{black}{in} the direction in the $Q-U$ plane discloses a similar axial symmetry in Stages I and III, with a \textcolor{black}{geometric prolate-to-oblate transformation in between.}
%prolate-to-oblate transformation in geometry. 
As an example shown in Methods:~\nameref{sec:rotate} and the top row of fig.~\ref{fig:schem}, a small change of axial symmetry during Stage II would manifest itself as a gradually rotating data cloud in the $Q-U$ plane, which qualitatively accounts for the observed evolving continuum polarization of SN\,2024ggi. 
In contrast, a flip of the dominant axis would imply a geometric transformation with the same symmetry axis (Fig.~\ref{fig:contpol_qu}, the bottom row of fig.~\ref{fig:schem} in Methods:~\nameref{sec:rotate}, and Ref~\cite{2017ApJ...837..105T}).

By approximating the electron-scattering atmosphere with an ellipsoid and a $\rho(r)\propto r^{-12}$ density distribution~\cite{2003MNRAS.345..111C}, the temporal evolution of the continuum polarization suggests moderate asphericity if \textcolor{black}{viewed within} $\sim 30^{\circ}-60^{\circ}$ from the aspect angle of the observer, i.e., $\sim 0.8\lesssim A \lesssim 0.95$ and $\sim 1.2\lesssim A \lesssim 1.4$ for the prolate (before day 2.0, fig.~\ref{fig:model_prolate}) and the oblate (days 5.8 to 10.9, fig.~\ref{fig:model_oblate}) configurations, respectively \textcolor{black}{(Methods:~\nameref{sec:contpolmodel}).}  
From days 10.9 to 33.0 \textcolor{black}{(Stage III)}, the H$\alpha$ and H$\beta$ lines exhibit PAs of the dominant axes that are roughly consistent with the orientation of the data cloud and that of the shock breakout (Fig.~\ref{fig:linepol_late}. 
\textcolor{black}{The only apparent exception is the H$\beta$ line on day 33.0; however, it is caused by a blend}
%The only exception is the H$\beta$ line on day 33.0, which strongly deviates from a P~Cygni profile owing to the blending 
with the emerging blueshifted Fe\,II\,$\lambda$5018 line (figs.~\ref{fig:linepol_late_iqu},~\ref{fig:iqu_ep8}). This tends to confirm that except for \textcolor{black}{Stage II when the ejecta-CSM interaction is prominent,}
%the ejecta-CSM interaction phase, 
the axial symmetry derived from the continuum persists throughout the explosion of SN\,2024ggi.

The detection of SN\,2024ggi also in X-rays during the first few days~\cite{2024ATel16586....1L, 2024ATel16587....1M, 2024ATel16588....1Z} supports the notion that the early shock-breakout process is modified by a dense and confined CSM. 
The direct measurement of the shock-breakout geometry, which exhibits a \textcolor{black}{spatially elongated,} axially symmetric configuration (figs.~\ref{fig:schem} and~\ref{fig:model_prolate}), is also compatible with the blueward color evolution within the first day~\cite{2024ApJ...972L..15S, 2024ApJ...972..177J, 2025ApJ...983...86C}. The early polarization evolution of SN\,2024ggi is highly complementary to the existence of the CSM and the way \textcolor{black}{the CSM} modifies the shock breakout. More importantly, the symmetry axis defined by the shock breakout, which is aligned with that inferred for Stage III, suggests that the core collapse could be driven by
\textcolor{black}{a mechanism that shapes}
the explosion on large scales. 
Moreover, the continuum polarization of SN\,2024ggi shows a conspicuous time evolution but never exceeded $\lesssim$0.4\% ($A\lesssim1.4$, \textcolor{black}{Methods:~\nameref{sec:contpolmodel}),} which is lower \textcolor{black}{than}
the $\lesssim2$\% and $\sim$1\% observed in the early phases of the Type IIn SN\,1998S~\cite{2000ApJ...536..239L} and Type \textcolor{black}{IIL/IIP} SN\,2023ixf~\cite{2023ApJ...955L..37V, 2024ApJ...975..132S, 2025ApJ...982L..32S, 2025arXiv250503975V}. SN\,1998S can be adequately modeled with a pole-to-equator density ratio of $\sim$5~\cite{2025A&A...696L..12D}.
\textcolor{black}{In summary, the shock-breakout phase of SN\,2024ggi shows a well-defined symmetry axis. The moderate global asymmetry is overall consistent with an asymmetry induced by an emitting zone extended 
in a particular direction.}
%The moderate asymmetry with a well-defined symmetry axis in the shock-breakout phase of SN\,2024ggi is overall consistent with an asymmetry induced by an emitting zone extended towards a certain direction.

%, which could be shaped by neutrino-driven turbulence developed at the initial core-collapse and preserved for the first days~\cite{2025ApJ...982....9V}.
%For example, a bipolar explosion yields a natural production of least resistance through angularly-asymmetric accretion~\cite{2021Natur.589...29B}. Recent simulations based on the neutrino-driven mechanism, in which a star explodes owing to neutrino heating from a proto-neutron star~\cite{2012ARNPS..62..407J}, failed to reproduce sufficiently energetic explosions (e.g.,~\cite{2015ApJ...801L..24M, 2017hsn..book.1095J, 2023PhRvL.131f1401E}). 
%r1The regularized explosion structure of SN\,2024ggi may not favour the enhancement of the neutrino heating efficiency based solely on instabilities such as convective motion, 
%The regularised explosion structure of SN\,2024ggi, with an unambiguously identified symmetry axis, may not favour multidimensional explosion scenarios that account for the required enhancement of the neutrino heating efficiency based solely on instabilities such as convective motion. 
%1the standing-accretion-shock-instability (SASI~\cite{2003ApJ...584..971B}), and Rayleigh-Taylor instabilities~\cite{2018MNRAS.479.3675M}. 

\section*{Discussion}
%Implications for the Death of Massive Stars}
SN\,2024ggi enables measurement of the shock-breakout geometry soon after the explosion. During this brief earliest moment, the geometry reflects the asymmetry of the explosion itself, as the photons toward the preferred directions of the explosion diffuse out promptly (Fig.~\ref{fig:schem_early}D). 
SN\,2024ggi is also the second of two H-rich CCSNe after SN\,2023ixf~\cite{2024Natur.627..759Z, 2024Natur.627..754L} with spectrophotometric observations carried out days after \textcolor{black}{shock breakout}~\cite{2023ApJ...953L..16H, 2023ApJ...954L..42J, 2024Natur.627..754L, 2024Natur.627..759Z}, 
%in which extensive asphericity during the shock breakout phase and ejecta engulfing a certain amount of CSM with large-scale asymmetry have been suggested
for which significant asphericity during the shock breakout as well as ejecta engulfing CSM with large-scale asymmetry have been diagnosed~\cite{2023ApJ...955L..37V, 2024ApJ...975..132S}. 
This may suggest a general pattern for the shock breakout from 
\textcolor{black}{dying}
%the death of 
massive stars.

\textcolor{black}{
3D full-sphere SN simulations also suggest the development of large-scale asymmetries that manifest themselves as giant plumes of radioactive matter penetrating deeply into the helium and hydrogen envelopes~\cite{2022MNRAS.510.4689V, 2025ApJ...982....9V}. In contrast, \textcolor{black}{the standing accretion shock instability (SASI~\cite{2003ApJ...584..971B, 2013ApJ...770...66H})} and a rather steep density gradient near the degenerate core will result in small-scale 
%homogeneous explosions
\textcolor{black}{asymmetries in the ejecta 
~\cite{2020MNRAS.496.2039S}.}
%2013ApJ...770...66H
%The former 
\textcolor{black}{The shock breakout that evinces large-scale directional dependencies}
also indicates that the time \textcolor{black}{at which} the shock emerges on the progenitor surface along the plume-mixing or other directions could differ by $\sim+$0.7 days. 
Such a significantly aspherical explosion is also \textcolor{black}{supported} by very recent 3D hydrodynamic calculations, suggesting that the shock-breakout geometry could be shaped by neutrino-driven turbulence developed at the initial core collapse and preserved \textcolor{black}{during the following few} days~\cite{2025ApJ...982....9V}.
Although such a bubble-driven explosion is compatible with the observed large-scale asymmetry shared by the shock breakout and the SN ejecta, additional mechanisms that regulate the explosion to maintain a well-defined axial symmetry may still \textcolor{black}{be} needed.
}

The early axisymmetric configuration of SN\,2024ggi may also be \textcolor{black}{compatible with} a prompt outflow enhanced moderately toward the polar regions. Core collapses producing a neutron star and involving an amplified magnetic field through magnetorotational instability may lift matter \textcolor{black}{along} the rotational axis of the collapsing core~\cite{2012ApJ...750L..22W, 2014ApJ...785L..29M}. 
This process does not necessarily involve the formation of powerful jets that penetrate the helium and hydrogen envelopes, as \textcolor{black}{implied} by the moderate level of asphericity observed throughout the shock breakout and the ejecta expansion phases of SN\,2024ggi. Details on how such a Lorentz-force-driven mechanism would account for the prompt axial symmetric emissions of SN\,2024ggi require future quantitative \textcolor{black}{model calculations}.
%r1By contrast, core collapse producing a neutron star and involving magnetorotational mechanisms that are highly asymmetric and launch jets~\cite{1999ApJ...524L.107K, 2012ApJ...750L..22W, 2014ApJ...785L..29M} is compatible with the observed polarization properties of SN\,2024ggi. 
%r1Additional geometric clues include the spatially resolved jet structures in the Crab Nebula~\cite{2000ApJ...536L..81W} and Cassiopeia A~\cite{2006ApJ...644..260L, 2024ApJ...965L..27M} that can be traced into the explosion zone. 

\textcolor{black}{Additional geometric clues include the spatially resolved axisymmetric structures consistent with a bipolar outflow in the Crab Nebula~\cite{2000ApJ...536L..81W} and Cassiopeia A~\cite{2006ApJ...644..260L, 2024ApJ...965L..27M} that can be traced into the explosion zone.}
The explosion mechanism may be related to collapsar models for long-duration gamma-ray bursts~\cite{1999ApJ...524..262M} and even magnetar models of some superluminous SNe~\cite{2010ApJ...717..245K, 2010ApJ...719L.204W}. 

\textcolor{black}{The misalignment of the axes} of the CSM and the ejecta (Fig.~\ref{fig:contpol_qu}) deserves further attention. The mass loss from the progenitor star may be governed by processes related to the angular momentum of the progenitor system, either as a single star or \textcolor{black}{a binary companion, which may naturally produce the misalignment of the explosion and the CSM symmetry axes.}
\textcolor{black}{Binary mass transfer during the common-envelope phase tends to enrich the CSM towards the orbital plane~\cite{2024A&A...685A..58E}. Such a disk-concentrated CSM content, which is compatible with the polarization time series of SN\,2024ggi, could be ubiquitous considering $\gtrsim$80\% of massive stars are in multiple systems~\cite{2024NatAs...8..472L}.
}
A magnetic field, which becomes more toroidal with distance from the progenitor, can also play a key role in shaping the CSM as inferred from well-structured planetary nebulae~\cite{1994ApJ...421..225C, 2001Natur.409..485B}.
%%}
%Magnetic fields may also play a key role in shaping the CSM as inferred from highly-structured planetary nebulae~\cite{1994ApJ...421..225C, 1995MNRAS.272..800H, 2001Natur.409..485B, 2006Natur.440...58V}.
Unlike the symmetry axis defined by the angular momentum of the system, the origins of the magnetic fields may be more complex and exhibit axes that are significantly different from the stellar rotation axis. For instance, the ejecta symmetry axis of SN\,1987A is $\sim 28^{\circ}$ away from the CSM symmetry axis~\cite{2002ApJ...579..671W}, and the Type IIn SN\,1998S displayed conspicuously different PAs of the continuum polarization and the polarization across the Balmer lines~\cite{2000ApJ...536..239L, 2001ApJ...550.1030W}. 
During the core collapse and the formation of the protoneutron star, the neutrino-driven instabilities or the initiation of jets through magnetorotational instabilities would also follow the structures of the progenitor stars~\cite{2000ApJ...537..810W, 2003ApJ...584..954A, 2008ApJ...677.1091W}, not the CSM. The combination of rotation encapsulated in the explosion geometry and magnetic fields encapsulated in the CSM geometry may naturally account for the misaligned axial symmetry between the ejecta and CSM. 

\textcolor{black}{
Spectropolarimetry of SN\,2024ggi reveals a moderately aspherical explosion that shows a well-defined symmetry axis shared by the prompt shock-breakout emission and \textcolor{black}{the} SN ejecta. 
%Such a manner 
\textcolor{black}{This variability illustrates} that instead of an amorphous/spherical setup resulting from small-scale instabilities, the core-collapse explosion of SN\,2024ggi can be driven by a mechanism that shapes the explosion from the earliest shock breakout \textcolor{black}{throughout the entire} ejecta expansion.
}
\\

%%%%%%%%%%%%%%%% MAIN TEXT FIGURES %%%%%%%%%%%%%%%

%\begin{figure} % Do NOT use \begin{figure*}
%	\centering
%	\includegraphics[width=0.6\textwidth]{example_figure} % for an image file named example_figure.*
%	% Pick an appropriate width - in print, figures are usually one or two columns wide, which can
%	% be approximated by 0.3\textwidth or 0.6\textwidth respectively. Use appropriate label sizes.
%
%	% Captions go below figures
%	\caption{{All captions must start with a short bold sentence, acting as a title.}
%		Then explain what is being shown, the meanings of any line styles, plotting symbols etc. Multi-panel figures must label the panels A, B, C, etc. and refer to them in the caption like this: ({A}) Description of panel A. ({B}) Description of panel B. Captions are placed below figures.}
%	\label{fig:example} % give each figure a logical label name
%\end{figure}

\begin{figure}
    \centering
    \includegraphics[trim={0.0cm 0.0cm 0.0cm 0.0cm},clip,width=1.0\textwidth]{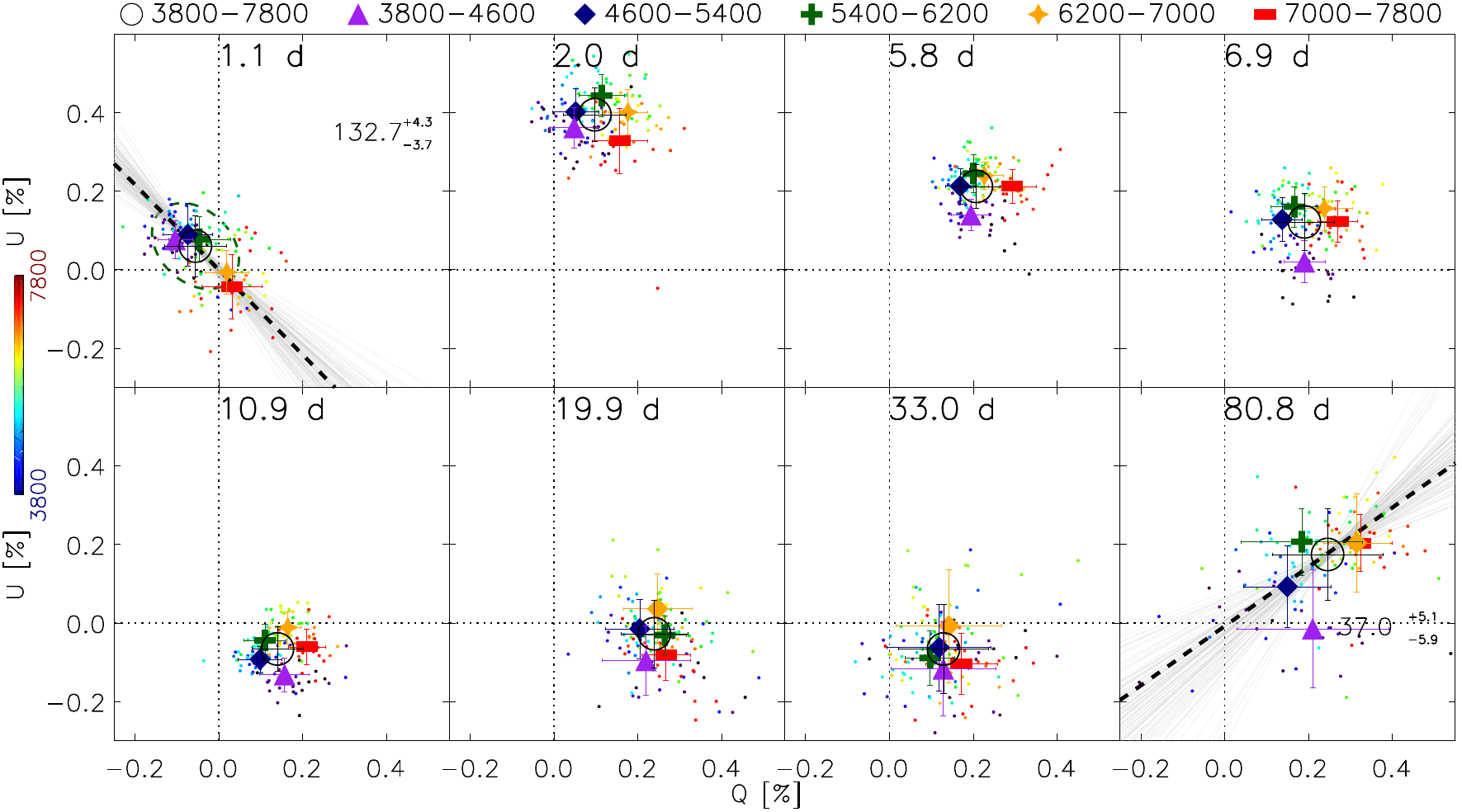}
    %{figures/sn2024ggi_contpol_regions_7800_30A.eps}
    \caption{\textbf{Temporal evolution of the polarization of SN\,2024ggi after subtracting the ISP.} In the top-left and the bottom-right panels, 
    \textcolor{black}{the black dashed line shows the dominant axis determined from linear fits to the small data points (the position angles and uncertainties are labeled), which cover the wavelength range 3800--7800\,\AA.} 
    \textcolor{black}{The orientation of the dominant axis in degrees with uncertainties is indicated in the subpanels for days 1.1 and 80.8.  
    A dashed ellipsoidal contour, whose major and minor axes respectively represent the 1$\sigma$ dispersion about the dominant and orthogonal axis, is also presented.
    }
    In each panel, different symbols mark the \textcolor{black}{error-weighted mean} polarization calculated over the wavelength ranges identified in the color bar. A drastic change of the continuum polarization (from days 1.1 to 2.0) is followed by a gradual drift until a roughly stationary geometry is reached at day 10.9. This behavior is accompanied by a clockwise rotation of the distribution of the data points, \textcolor{black}{revealing} a large-scale transformation of the geometry as the CSM is swept up by the SN ejecta. 
    \textcolor{black}{Light gray lines in the upper-left and lower-right panels present the dominant axes fitted to the data through a Monte Carlo re-sampling approach using the errors in $Q$ and $U$ measured at each wavelength bin.}
}
    \label{fig:contpol}
\end{figure}

\begin{figure}
    \centering
    \includegraphics[trim={0.0cm 0.0cm 0.0cm 0.0cm},clip,width=1.0\textwidth]{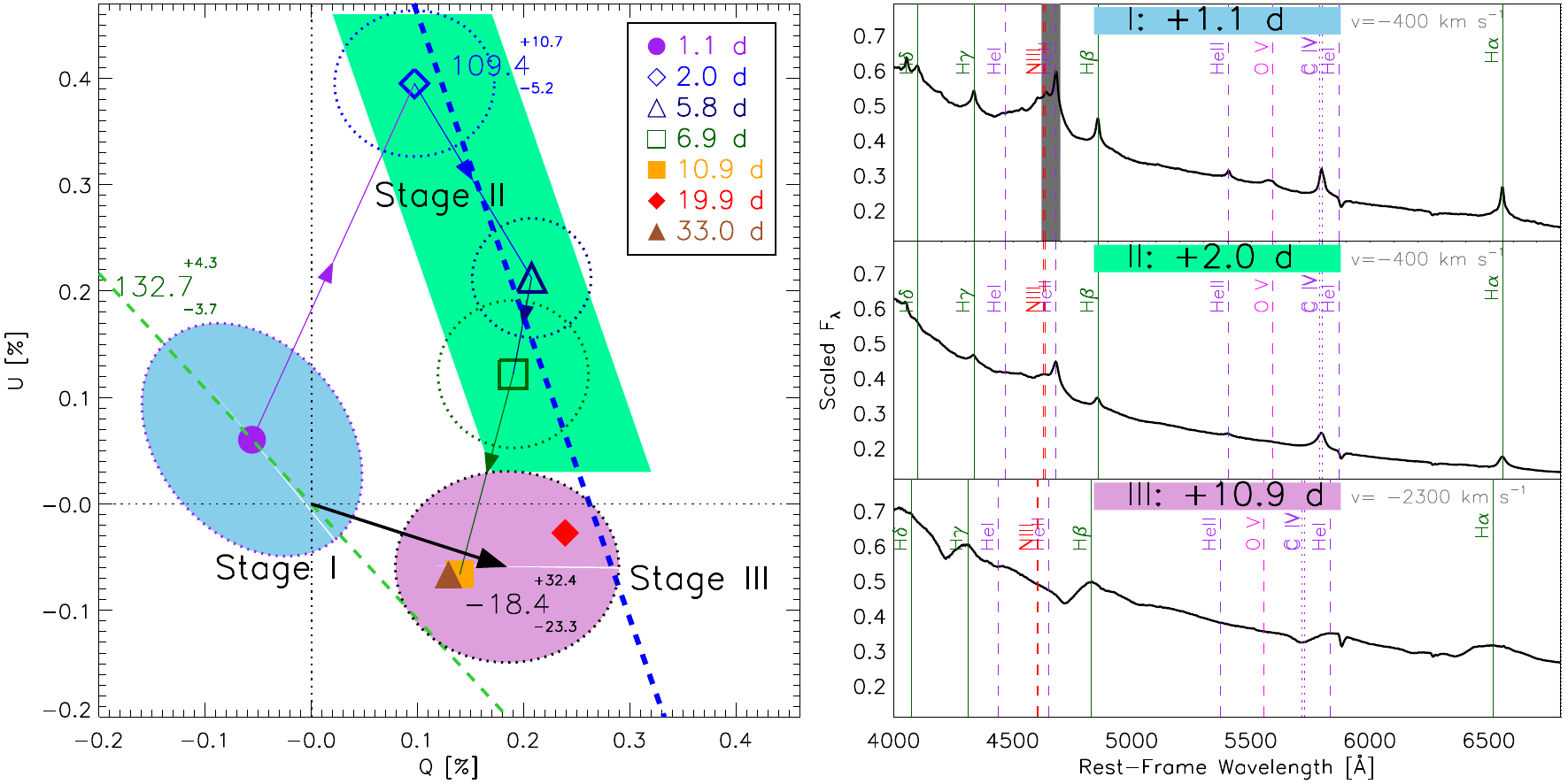}
    %{figures/sn2024ggi_poltime_30A_7800A_main.eps}
    \caption{\textbf{Temporal evolution of the continuum polarization of SN\,2024ggi displayed in the $Q-U$ plane.} 
    {\it Left:} The blue, green, and pink-shaded areas \textcolor{black}{mark} the three stages of the SN\,2024ggi polarimetry. Different symbols represent the continuum polarization of SN\,2024ggi 
    \textcolor{black}{at different epochs.}
    The thin green dashed line shows the dominant axis at day 1.1 for comparison. The blue dashed line approximately follows the Stage II locus (days 2.0--6.9), when the interaction between the ejecta and CSM led to a change in overall geometry. The black arrow represents the PA of the continuum polarization of Stage III, which was estimated by the error-weighted mean of days 10.9, 19.9, and 33.0. The size of each contour is determined by the standard deviation of the polarization measured at the encircled epoch(s). {\it Right:} The upper, middle, and lower-right panels show the scaled flux-density spectra ($F_{\lambda}$) at days 1.1 (Stage I), 2.0 (Stage II), and 10.9 (Stage III), respectively, with major photoionized lines from several species labeled at velocity $v$ 
    \textcolor{black}{in the rest frame}. The region of the dark-gray-shaded band at day 1.1 suffers from detector saturation. 
    \textcolor{black}{Observations at day 80.8 are not presented as the polarization is affected by strong outward mixing of the inner He-rich layer and nickel clumps.}
    }
    \label{fig:contpol_qu}
\end{figure}

\begin{figure}
    \centering
    \includegraphics[trim={0.0cm 0.0cm 0.0cm 0.0cm},clip,width=1.0\textwidth]{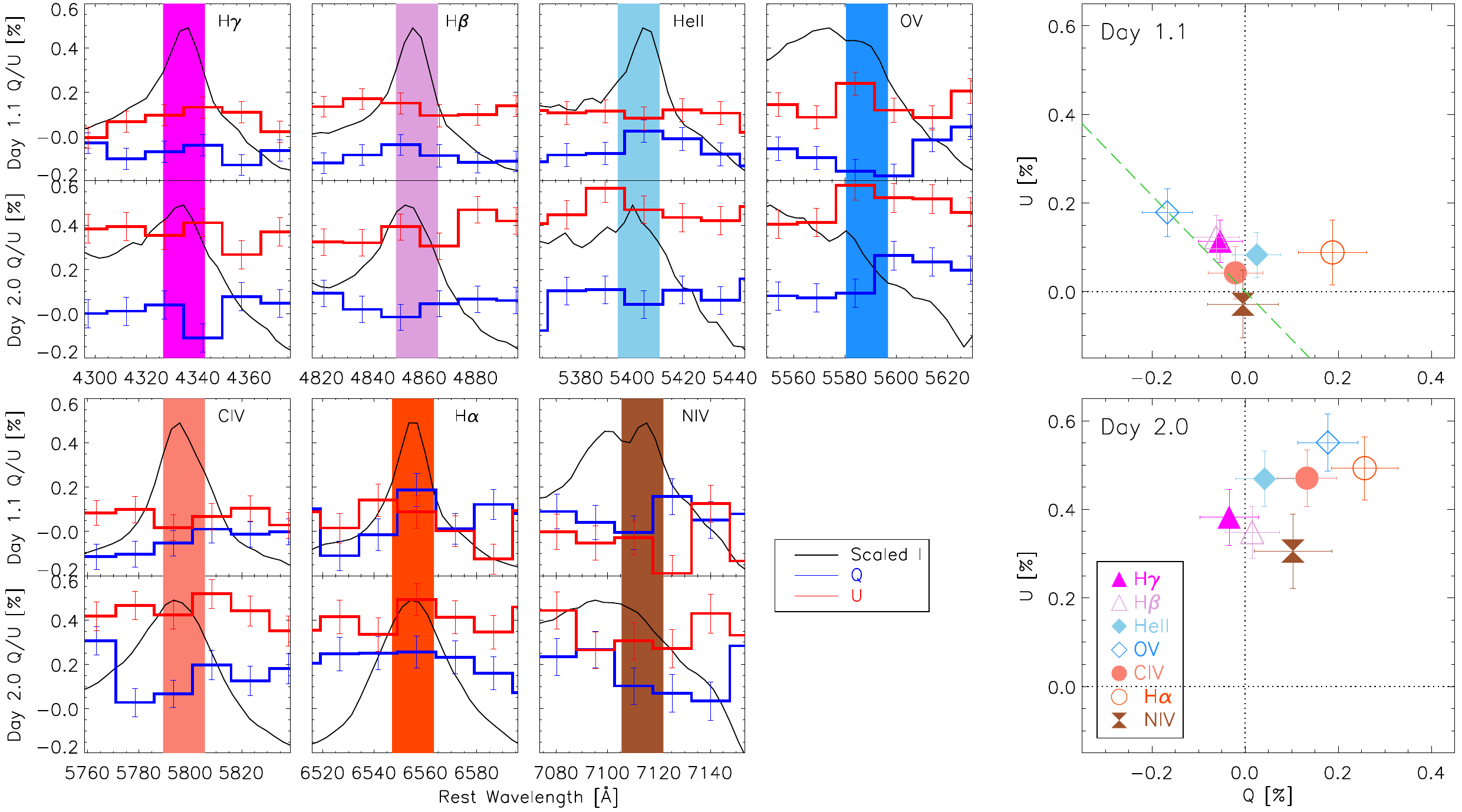}
%    {figures/sn2024ggi_line_core_15A.eps}
    \caption{
    \textcolor{black}{\textbf{Polarizations measured in the central $\pm10$\,\AA\ of various emission peaks at days 1.1 and 2.0.}} \textcolor{black}{The emission cores as highlighted by the color-shaded spectral regions in the left subpanels are less affected by the electron-scattering emission from the wings (which are sensitive to smaller-scale structures such as lumpiness of the scattering region).  Their distribution in the Stokes $Q-U$ plane as shown in the upper-right and lower-right panels for days 1.1 and 2.0 (respectively), traces the geometry of the shock-breakout ionization front with the least influence from other effects. In the upper-right panel, the green dashed line presents the dominant axes determined over the wavelength range 3800--7800\,\AA\ on day 1.1. 
}
%    \textcolor{black}{Polarizations measured in the central $\pm10$\,\AA\ of various emission peaks at days 1.1 (upper) and 2.0 (bottom rows), respectively. The emission cores are less affected by the electron scattering emission from the wings (e.g., lines are sensitive to smaller scale structures such as lumpiness of the scattering region), and their distribution on the Stokes Q$-$U plane would trace the geometry of the shock breakout ionization front with the least influence from other effects. In the upper-right panel, the green dashed line presents the dominant axes determined over a wavelength range 3800--7800\,\AA\ on days 1.1.}
}
    \label{fig:linepol_early}
\end{figure}

\begin{figure}
    \centering
    \includegraphics[trim={0.0cm 0.0cm 0.0cm 0.0cm},clip,width=1.0\textwidth]{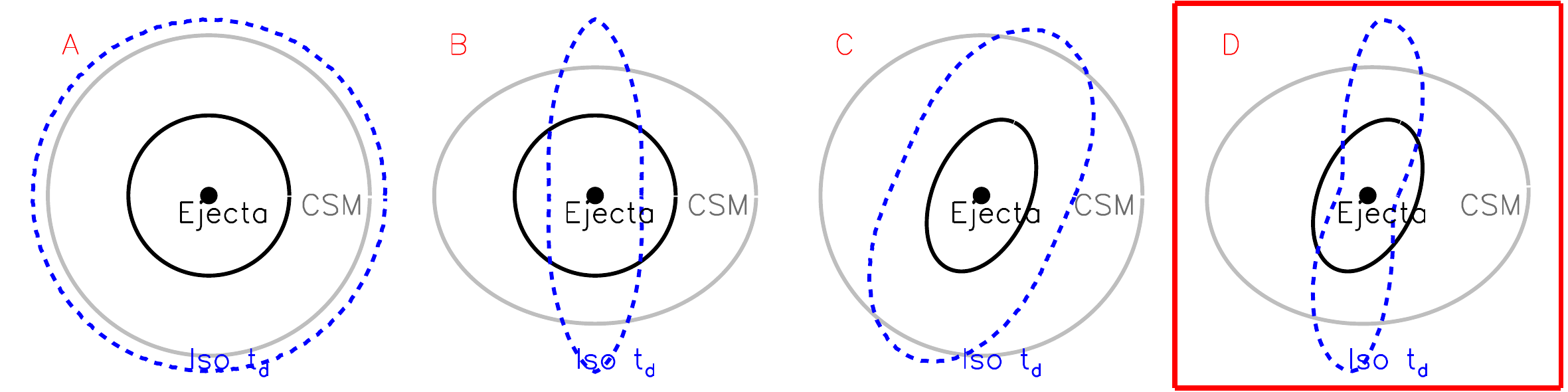}
%{figures/schem_sn2024ggi_early.eps}    
    \caption{\textbf{Illustration of the expanding ejecta and the invariant CSM for different explosion schematics.}  
    In each panel, the blue dashed contour \textcolor{black}{displays} the location of the ionization front estimated from the isodiffusion-time surface (Methods:~\nameref{sec:diffusion}) and the solid gray circle/ellipse represents the outer boundary of the CSM, and the solid black circle/ellipse \textcolor{black}{shows} the outer boundary of the SN ejecta embedded in the CSM. The different schematics are ({\bf A}) spherical ejecta + spherical CSM, ({\bf B}) spherical ejecta + disk-concentrated CSM, ({\bf C}) aspherical ejecta + spherical CSM, and ({\bf D}) aspherical ejecta + \textcolor{black}{disk-concentrated}
    %aspherical 
    CSM. \textcolor{black}{The axisymmetric prompt shock-breakout emission during Stage I and the time-dependent symmetry axis during the transition to Stage II suggest}
    %The time-dependent axial symmetry of SN\,2024ggi during stage II indicates 
    ({\bf D}) as the most plausible scenario. 
}
    \label{fig:schem_early}
\end{figure}

%\begin{figure}
%    \centering
%    \includegraphics[trim={0.0cm 0.0cm 0.0cm 0.0cm},clip,width=1.0\textwidth]{figures/sn2024ggi_linepol_regions_15A_main.eps}
%    \vspace{-0.4 cm}
%    \caption{Line polarization of SN\,2024ggi at day 1.1. $Q-U$ plots are shown for four spectral lines as labeled. In each panel, the magenta dash-dotted line fits the polarization measured at different velocity intervals identified by colour; note that the colour bars have different velocity ranges. The green dashed lines are the dominant axes of the continuum polarization (copied from Figure~\ref{fig:contpol}). The dominant axes of the spectral features with high excitation potentials (e.g., O\,V) or lower transition probability (e.g., H$\beta$) closely follow that of the continuum, while other lines, which are formed farther out in the ionisation front, have different orientations. 
%}
%    \label{fig:linepol_early}
%\end{figure}

\begin{figure}
    \centering
    \includegraphics[trim={0.0cm 0.0cm 0.0cm 0.0cm},clip,width=1.0\textwidth]{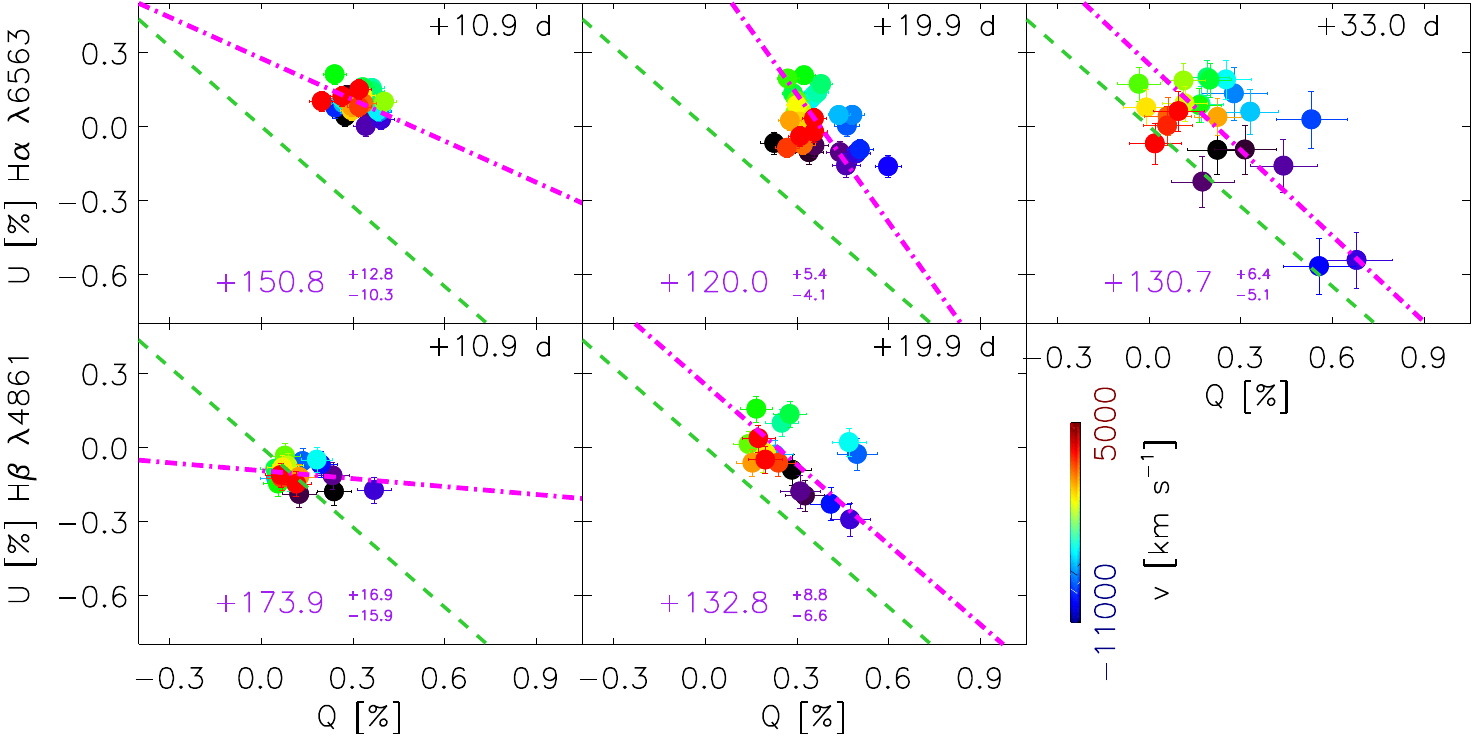}
%{figures/sn2024ggi_linepol_late_regions_hahb_15A_main.eps}
    \caption{\textbf{\textcolor{black}{Evolution in the $Q-U$ plane of H$\alpha$ (top row) and H$\beta$ (bottom row) of SN\,2024ggi from days 10.9 to 19.9.}} The colors encode \textcolor{black}{rest-frame} velocities according to the color bars. In each panel, 
    %the thick black line represents the shared symmetry axis between stages I and III 
    %shared axial symmetry between the prolate and the oblate configurations 
    %as fitted in Figure~\ref{fig:contpol_qu}. 
    the magenta dot-dashed line fits the polarization distribution measured at different velocities that cover the corresponding spectral feature. The green dashed lines overplot the dominant axis at day 1.1, which appears to be aligned with that of the H envelope that has progressively emerged after day 6.9.}~\label{fig:linepol_late}
\end{figure}
%%%%%%%%%%%%%%%% MAIN TEXT TABLES %%%%%%%%%%%%%%%

\begin{table}
\small
\begin{tabular}{c|cc|ccc|ccc}
\hline
Epoch &  MJD        & Phase [day]$^a$ & Grism & Exp Time [s]$^b$  & Airmass   & Grism & Exp Time [s]$^b$ & Airmass   \\
\hline
1     &  60412.246  &  1.1      & 300V  & 180$\times$4$\times$2  & 1.39 & --    & --  & --   \\
2     &  60413.144  &  2.0      & 300V  &  45$\times$4$\times$2  & 1.03 & 1200B &  80$\times$4$\times$2 & 1.04 \\
NA$^{c}$ &  60416.078  &  4.9   & 300V  &  90$\times$4$\times$2  & 1.02 & 1200B & 240$\times$4$\times$2 & 1.01 \\
3     &  60416.988  &  5.8      & 300V  &  90$\times$4$\times$2  & 1.22 & 1200B & 240$\times$4$\times$2 & 1.16 \\
4     &  60418.008  &  6.9      & 300V  &  50$\times$4$\times$2  & 1.13 & --    & --  & --   \\
5     &  60422.023  & 10.9      & 300V  &  70$\times$4$\times$2  & 1.07 & 1200R & 130$\times$4$\times$2 & 1.03 \\
6     &  60430.996  & 19.9      & 300V  &  75$\times$4$\times$2  & 1.08 & 1200R & 140$\times$4$\times$2 & 1.05 \\
7     &  60444.164  & 33.0      & 300V  &  40$\times$4$\times$2  & 1.41 & --    & --  & --   \\
8     &  60491.979  & 80.8      & 300V  &  65$\times$4$\times$2  & 1.16 & 1200R & 130$\times$4$\times$2 & 1.21 \\
9     &  60678.246 & 267.1      & 300V  & 480$\times$4$\times$2  & 1.26 & --    & --  & --   \\
\hline
\end{tabular}
{$^a$}{Relative to the estimated time of \textcolor{black}{the shock breakout} at MJD 60411.03.} \\
{$^b$}{Observations carried out with two exposures each at four different half-wave-plate angles.} \\
{$^c$}{Dataset discarded owing to very bad seeing ($\sim 4.8''$).} 
\caption{\textbf{Log of VLT Spectropolarimetry of SN\,2024ggi.~\label{Table:log_specpol}}}
\end{table}

%\begin{table} % Do NOT use \begin{table*}
%	\centering
%	% Captions go above tables
%	\caption{{All captions must start with a short bold sentence, acting as a title.}
%		Then explain what is being listed in the table, the meaning of each column etc.		Captions are placed above tables.}
%	\label{tab:example} % give each table a logical label name
%	
%	\begin{tabular}{lccc} % four columns, alignment for each
%		\\
%		\hline
%		Sample & $A$ & $B$ & $C$\\
%		 & (unit) & (unit) & (unit)\\
%		\hline
%		First & 1 & 2 & 3\\
%		Second & 4 & 6 & 8\\
%		Third & 5 & 7 & 9\\
%		\hline
%	\end{tabular}
%\end{table}

\clearpage
\section*{Materials and Methods}

\subsection*{Interstellar Polarization}~\label{sec:isp}
To investigate the polarization intrinsic to SN\,2024ggi, before proceeding to detailed discussions of the observed polarization, we derive the interstellar polarization (ISP) induced by the dust grains along the SN-Earth line of sight. The ISP estimation was carried out using the earliest (day 1.1) and the \textcolor{black}{day 80.8 polarization}, where a global axially symmetric photosphere can be inferred from the presence of a prominent dominant axis. 
\textcolor{black}{
%Without 
After correction for the ISP, the dominant axis is seen as a straight line passing through the origin, as expected for an axisymmetric configuration~\cite{2010A&A...510A.108P, 2021A&A...651A..19D},} and the intersection of the dominant axes on days 1.1 and 80.8 at $Q_{\rm ISP}=-0.40\pm0.05$\%, $U_{\rm ISP}=0.54\pm0.04$\% yields the ISP (fig.~\ref{fig:isp_data}). 

The ISP is weakly dependent on wavelength, in particular within the low-ISP regime~\cite{2010A&A...510A.108P}. \textcolor{black}{In} the coordinate system defined by the Stokes $Q$ and $U$ parameters (i.e., the $Q-U$ plane~\cite{2001ApJ...550.1030W}), the effect of the ISP is largely the introduction of
%the 
uncertainties of the zero point. It is also expected to be time-independent~\cite{1996ApJ...459..307H} and manifests as an offset in the Stokes $Q-U$ diagram without affecting the morphological patterns of the data points. High-resolution spectroscopic observations of the Na\,I\,D absorption doublet have led to the conclusion that the line-of-sight extinction toward SN\,2024ggi can be decomposed into a Galactic ($E(B-V)_{\rm MW}=0.120\pm0.028$\,mag) and a host-galaxy ($E(B-V)_{\rm host}=0.034$\,mag~\cite{2024ApJ...972L..15S}) component. Since interstellar extinction and polarization are both induced by dust grains~\cite{1975ApJ...196..261S}, the stronger Galactic extinction suggests that the Galactic polarization is the dominant component of the ISP. In the case of SN\,2024ggi, the exact ISP level is difficult to estimate with the widely used standard methods (e.g., Refs.~\cite{2017MNRAS.469.1897S, 2023MNRAS.519.1618Y}). In particular, the absence of resolved cores of emission profiles dominated by unpolarized photons released by recombination is a handicap.

%A prominent axial symmetry of an electron-scattering structure leads to a wavelength-independent PA of the continuum in the $Q-U$ plane. For data points with different wavelengths, their distance from the origin (the polarization degree $p$) varies owing to the different optical depths and, hence, scattering efficiency. Together they form a straight line known as the dominant axis. 
%Additional 
Polarization by dust grains in the interstellar matter shifts the dominant axis away from the origin in the $Q-U$ plane. For SN ejecta with a high degree of axial symmetry, the ISP would be located at one of the ends (or beyond them) of the dominant axis~\cite{2001ApJ...556..302H, 2003ApJ...591.1110W, 2008ARA&A..46..433W}. If the variability of an object causes the dominant axis of the intrinsic polarization to rotate, the rotation angle is independent of the chosen value of the ISP because the latter only introduces a displacement of the data points from the origin. However, careful subtraction of the ISP is of paramount importance when determining the shape of an object from its intrinsic polarization. 
%Since a dominant axis can be clearly identified at both days 1.1 and 80.8, we assign the intersection of the two dominant axes to the ISP, namely \textcolor{black}{$Q_{\rm ISP}= -0.40\pm0.05$\%, $U_{\rm ISP} = 0.54\pm0.04$\%}
%$Q_{\rm ISP}= -0.396\pm0.045$\%, $U_{\rm ISP} = 0.536\pm0.041$\% 
%(Figure~\ref{fig:isp_data}). This level is consistent with the fact that the ISP should be located close to one of the two ends of the data distribution in the $Q-U$ plane. 

Another approach to estimate the total line-of-sight ISP assumes that the emission \textcolor{black}{peak} of the strong P~Cygni profiles of the Balmer lines are unpolarized during the photospheric phase of Type II/IIP SNe~\cite{2021A&A...651A..19D}. We estimate the error-weighted mean polarization within a wavelength range of 6550--6750\,\AA\ to be 
\textcolor{black}{$Q_{\rm ISP}^{\rm +33\,d}= -0.32\pm0.04$\%, $U_{\rm ISP}^{\rm +33\,d} = 0.55\pm0.08$\%,}
%$Q_{\rm ISP}^{\rm +33\,d}= -0.318\pm0.040$\%, $U_{\rm ISP}^{\rm +33\,d} = 0.551\pm0.081$\%, 
consistent with the estimate presented above.

\textcolor{black}{
We also estimate the ISP from the spectropolarimetry of SN\,2024ggi at day 267.1. Since the ejecta expand and the electron-scattering cross section decreases as $\propto t^{2}$, the SN has entered the nebular phase at such a late epoch. 
\textcolor{black}{Except for several polarized blue-shifted absorption components of the P Cygni profile (see, Methods:~\nameref{sec:vlt}),} the continuum spectrum during the nebular phase can be treated as an unpolarized source dominated by significantly blended emission lines from various iron-group elements, which are free from electron scattering and intrinsically unpolarized. Therefore, the continuum polarization on day 267.1 also measures the ISP toward the SN. We measure the error-weighted mean continuum polarization over 4000--6300\,\AA\ as: 
$Q_{\rm ISP}^{\rm +267\,d} = -0.25\pm0.24$\,\%, 
$U_{\rm ISP}^{\rm + 267\,d} = 0.62\pm0.24$\,\%, consistent with the other methods. 
%We also estimate the ISP based on the spectropolarimetry of SN\,2024ggi at day 267.1. Since the ejecta expand and the electron-scattering cross section decreases $\propto t^{2}$, the SN has entered the nebular phase at this late epoch. The continuum spectrum during the nebular phase can be treated as an unpolarized source dominated by the significantly blended emission lines from various iron-group elements, which are free from electron scattering and, therefore, intrinsically depolarized. Accordingly, the continuum polarization at day 267.1 also measures the ISP toward the SN. We measure the error-weighted mean continuum polarization over 4000--6300\,\AA\ as $Q_{\rm ISP}^{\rm +267\,d} = -0.25\pm0.24$\,\%, $U_{\rm ISP}^{\rm + 267\,d} = 0.62\pm0.24$\,\%, 
%$Q_{\rm ISP}^{\rm +267\,d} = -0.245\pm0.237$\,\%, 
%$U_{\rm ISP}^{\rm + 267\,d} = 0.623\pm0.236$\,\%, consistent with the other methods.
}
Throughout this paper, 
\textcolor{black}{$Q_{\rm ISP}=-0.41\% \pm 0.05$\%, $U_{\rm ISP}=0.55\% \pm 0.04$\%}
%$Q_{\rm ISP}=-0.409\% \pm 0.048$\%, $U_{\rm ISP}=0.550\% \pm 0.041$\% 
is adopted for the intrinsic polarization of SN\,2024ggi. 
\textcolor{black}{These approaches provide different values compared to the Galactic ISP sampled by a bright star $\sim1^{\circ}$ away from SN\,2024ggi. The result of this sanity check is discussed in the Supplementary Text and fig.~\ref{fig:isp}.
}

\textcolor{black}{The wavelength-dependent polarization of SN\,2024ggi at day 1.1 shows a remarkable resemblance to the characteristic wavelength-dependent Serkowski law. In the low ISP regime, the observed wavelength dependence can be well fitted by a single ISP component, consistent with the single-cloud interpretation based on a comprehensive investigation on the effects of ISP induced by various interstellar dust contents~\cite{2025A&A...698A.168M}. However, high-resolution spectroscopy of SN\,2024ggi obtained at $\approx$3 days after its explosion reveals at least three major discrete absorbing components~\cite{2024ApJ...972L..15S}. Therefore, the ISP towards SN\,2024ggi may not follow a single cloud model that accounts for the day 1.1 observation.
}

\textcolor{black}{We also conducted a sanity test to verify that the wavelength dependence of the polarization across the observed wavelength range on day 1.1 is intrinsic to the SN. 
The wavelength ($\lambda$) dependence of the ISP can be approximated by the empirical Serkowski law~\cite{1975ApJ...196..261S},
\begin{equation}
p(\lambda)/p(\lambda_{\rm max}) = {\rm exp}[-K\ \rm {ln}^2 (\lambda_{\rm max} / \lambda)],
\label{Eqn_Ser}
\end{equation}
where $\lambda_{\rm max}$ and $p(\lambda_{\rm max})$ represent the wavelength and the level of the maximum polarization, respectively. The parameter $K$ characterizes the width of the peak of the ISP. By attributing the wavelength-dependent polarization of SN\,2024ggi on day 1.1 to the ISP, we ﬁtted a Serkowski law to the polarization spectra and present the results \textcolor{black}{in the left panel of fig.~\ref{fig:serk}.} However, as illustrated in fig.~\ref{fig:serk_results}, the removal of the wavelength dependence derived based on the day 1.1 observation would introduce significant wavelength-dependent polarization at all other epochs. 
As neither the endpoints nor the line segment passes through the origin on the $Q-U$ plane from days 5.8 to 80.8, we conclude that the ISP cannot be naturally accounted for by the wavelength-dependent polarization on day 1.1. The latter, which persisted only briefly after the SN explosion, is therefore intrinsic to the SN and traces the geometry of the shock breakout.}

\textcolor{black}{In the right panel of fig.~\ref{fig:serk}, we overlay the best-fit Serkowski law to the day 1.1 observation onto the polarization of SN\,2024ggi on day 267.1, when the SN has entered the nebula phase. We investigated the wavelength dependence of the day 267.1 polarization by resampling the data with large (150\,\AA) wavelength bins. The result does not reproduce Serkowski’s fit to the day 1.1 observations, further strengthening the claim that the day 1.1 polarization is intrinsic to the SN, rather than the ISP. 
}

\subsection*{Spectropolarimetry of SN\,2024ggi}~\label{sec:vlt}
Spectropolarimetry of SN\,2024ggi was carried out with the FOcal Reducer and low-dispersion Spectrograph 2 (FORS2~\cite{1998Msngr..94....1A}) on Unit Telescope~1 (UT1, Antu) of the Very Large Telescope (VLT) at the European Southern Observatory's Paranal site in Chile. 
%The observations were carried out with the FORS2 instrument~\cite{1998Msngr..94....1A} on the Very Large Telescope (VLT) of ESO's La Silla Paranal Observatory in Chile (Method~\ref{sec:vlt}). 
Each observation in the Polarimetric Multi-Object Spectroscopy (PMOS) mode consists of \textcolor{black}{eight} exposures at retarder-plate angles of 0, 22.5, 45, and 67.5 degrees. All observations were carried out using the 300V grism and a 1$''$-wide slit. 
%, giving a spectral resolving power of $R \approx 440$.
Therefore, \textcolor{black}{the resolving power and the intrinsic width of each resolution element near its central wavelength at 5849\,\AA\ are $R\sim440$ and $\sim 13.3$\,\AA, respectively, corresponding to $\sim 680$\,km\,s$^{-1}$ \cite{Anderson_etal_2018}.} The observing log is available in Table~\ref{Table:log_specpol}. 

Preprocessing of the two-dimensional (2D) images obtained at each retarder plate angle and the extraction of the ordinary (o) and extraordinary (e) beams were carried out with standard procedures within IRAF~\cite{1986SPIE..627..733T, 1993ASPC...52..173T}. Wavelength calibration of each individual spectrum was performed separately, with a typical root-mean-square accuracy of $\sim 0.20$\,\AA. Following the prescriptions in Ref~\cite{2006PASP..118..146P}, we then derived the Stokes parameters and calculated the observed polarization degree ($p_{\rm obs}$) and position angle (PA$_{\rm obs}$),
\begin{equation}
\begin{aligned}
p_{\rm obs} = \sqrt{Q^2 + U^2}, \ 
{\rm PA}_{\rm obs} = \frac{1}{2} {\rm arctan} \bigg{(} \frac{U}{Q} \bigg{)}, 
\ 
\end{aligned}
\label{Eqn_stokes0}
\end{equation}
where $Q$ and $U$ denote the intensity ($I$)-normalized Stokes parameters. An additional debiasing procedure was applied because the true value of the polarization degree is nonnegative \cite{1985A&A...142..100S}: 

\begin{equation}
\begin{aligned}
p = \bigg{(} p_{\rm obs} - \frac{\sigma_{p}^2}{p_{\rm obs}}\bigg{)} \times h(p_{\rm obs} - \sigma_p); \\
{\rm PA} = {\rm PA}_{\rm obs}, 
\end{aligned}
\label{Eqn_stokes1}
\end{equation}
where $\sigma_{p}$ and $h$ denote the $1\sigma$ uncertainty in $p_{\rm obs}$ and the Heaviside step function, respectively. 
%r3Finally, we corrected for the $\lesssim 0.1$\% instrumental polarization \cite{2017MNRAS.464.4146C}. 
Following the prescription described in previous work (e.g., Refs~\cite{2020ApJ...902...46Y, 2023MNRAS.519.1618Y}), \textcolor{black}{where} 
the wavelength-dependent instrumental polarization in the PMOS mode of FORS2 ($\lesssim$0.1\%) was characterized to remain stable over time, this effect was corrected according to the characterization by Ref~\cite{2017MNRAS.464.4146C}. The low instrumental polarization during the campaign of SN\,2024ggi polarimetry is consistent with that inferred from the observations of polarized and unpolarized standard stars carried out in each night 
\textcolor{black}{with observations for our program.}
%of our ToO triggering under ESO Program ID 60.A-9203(E). 

\textcolor{black}{Throughout the paper, all spectra and spectropolarimetry data of SN\,2024ggi were corrected to the rest frame adopting the heliocentric recession velocity of NGC\,3621 of 730\,km\,s$^{-1}$ ($z\approx$0.002435, \cite{2004AJ....128...16K}}. 
\textcolor{black}{Spectropolarimetry of SN\,2024ggi obtained from days 1.1 to 267.1 is displayed in figs.~\ref{fig:iqu_ep1}--\ref{fig:iqu_ep10}.} 
All data are presented in the rest frame and before correcting for the ISP. Principal-component decomposition of the SN\,2024ggi spectropolarimetry is \textcolor{black}{shown} in fig.~\ref{fig:pca_24ggi} to better visualize the temporal evolution of the total-flux spectra and the polarization spectra projected onto the dominant axis and the orthogonal axis. 

%\subsection{The Stokes $Q-U$ Plane}~\label{sec:quplane}

\subsection*{Polarization Across the Photoionized Features}~\label{sec:linepol}
The exceptionally early-epoch polarimetry includes the short-lived \textcolor{black}{photoionization-powered} emission lines during the first days of SN\,2024ggi. In the first $\sim 2$ days, the total-flux emission profiles consist of a prominent emission peak and a weak, broad underlying component with \textcolor{black}{full width at half-maximum intensity (FWHM)} $\approx 1000$--2000\,km\,s$^{-1}$ (fig.~\ref{fig:linepol_early_iqu}). 
\textcolor{black}{We also computed the polarized flux density $p\times f_{\lambda}$ across the flash features and found no significant deviation from the adjacent continuum.
}
The broad wings are due to scattering by free electrons in the unshocked, \textcolor{black}{photoionized} CSM~\cite{2001MNRAS.326.1448C, 2009MNRAS.394...21D, 2018MNRAS.475.1261H}. The polarization of the electron-scattering wings traces the spatial distribution of the associated ionic species. The spectral-line-specific geometric diagnostics are best derived in the Stokes $Q-U$ plane by comparing, epoch by epoch, the location of a given spectral line and that of the continuum~\cite{2001ApJ...550.1030W}. The slope of the distribution of the data points represents the orientation 
%in the sky plane 
\textcolor{black}{of} the symmetry axis of the feature in question (line or continuum), 
\textcolor{black}{projected onto the plane of the sky.}
%r3A perfectly axially symmetric structure will produce a single straight ``dominant axis'' in the Stokes $Q-U$ plane~\cite{2003ApJ...591.1110W}. 

\textcolor{black}{High-ionization lines (e.g., O\,V\,$\lambda$5597) appear only at the earliest phases and are generally thought to form in the relatively inner part of the CSM and close to the shock front,} where the highest temperatures produce the highest ionization states. 
%Accordingly, the polarization of the continuum and the most highly ionised lines at days +1.1 trace the geometry of the shock. 
In the case of a spherically symmetric shock breakout and the resulting concentric ionization front, \textcolor{black}{their} identical shapes would manifest as a single dominant axis in the continuum and for all early-time emission lines. In SN\,2024ggi (fig.~\ref{fig:linepol_early_method}), 
 \textcolor{black}{the polarization PAs of the spectral lines with the highest ionization potentials (e.g., O\,V, $\chi=113.9$\,eV) follow that of the underlying continuum, while other lines such as C\,IV $\lambda$5807 ($\chi=64.5$\,eV) and H$\alpha$ ($\chi=13.6$\,eV) exhibit distinctly different dominant axes than the continuum.} 
\textcolor{black}{Due to a saturation issue within the rest-frame wavelength range of 4630--4710\,\AA\ that covers the He\,II\,$\lambda$4686 ($\chi=$54.5\,eV) emission line at day 1.1, this region was excluded from the analysis of the line polarization.}

Although both H$\alpha$ and H$\beta$ arise from the recombination to the \textcolor{black}{second excitation level of hydrogen,} the transition probability expressed as the form of weighted oscillator strength (log($gf$)) of H$\alpha$ is a factor of $\sim5$ higher than that of the H$\beta$. Therefore, higher polarization can be expected for H$\alpha$ wherever an energy level of 13.6\,eV is reached. Compared to H$\alpha$, H$\beta$ would mainly form at a much narrower region. The polarization is also weaker and only becomes dominant close to the photosphere, thus effectively tracing the geometry of the ionization front as early as day 1.1. 
\textcolor{black}{
The agreement between the dominant axes of the continuum and the distribution of the high-excitation lines on the Q$-$U diagram as presented in Fig.~\ref{fig:linepol_early} further strengthen the interpretation of the axially symmetric configuration of the shock breakout. 
Portraits of the early-phase photoionized spectral features are offered in fig.~\ref{fig:linepol_early_method}. The O\,V line itself, whose electron-scattering wings are likely to arise from the CSM confined to the most energetic shock-ionization front, exhibits a relatively clear dominant axis that is similar to that of the continuum.
}

The line polarization behaviour is also compatible with the picture inferred from the continuum polarization. As the shock-ionised emission preferentially emerges promptly from the less dense regions perpendicular to the plane of the CSM disc,
%~\cite{2023ApJ...956...46S}, 
\textcolor{black}{the shock would propagate faster toward the perpendicular directions when the ejecta have not yet overrun the CSM.} Consequentially, the faster shock heats the post-shock gas to a higher temperature, thus producing the earliest prolate geometry that is aligned with the less-dense regions perpendicular to the CSM plane. In contrast, the denser CSM plane will decelerate the shock more strongly, resulting in a lower post-shock temperature. The lower-ionization lines would preferentially be developed in this lower-temperature region and occupy a broad range in CSM-plane azimuth angle.

\subsection*{Modeling the Polarization for an Expanding Envelope}~\label{sec:pcygni}
Following the general assumptions of the Sobolev approximation (e.g.,~\cite{2002ApJ...565..380K}), we treat the SN atmosphere with a low velocity gradient in its inner region, below some velocity cutoff $v_{\rm cut}$ of a few thousand km\,s$^{-1}$, that radiates as a blackbody and is surrounded by an expanding atmosphere with a significantly larger velocity power-law exponent $n$. 
%%YY\textcolor{teal}{DBA2YY:  Would it be better to speak of a velocity power-law exponent?}\textcolor{violet}{YY2DBA: Revised.}
The density of the atmosphere at a given time ($t$) after the explosion and different radial velocities ($v_{r}$) 
\textcolor{black}{below and above the layer with $v_{\rm cut}$ are}
given by 
\begin{align}
    \rho_{\rm in} (t) \propto \bigg{(} \frac{t}{t_{0}} \bigg{)}^{-2}, \ 
    \rho_{\rm out} (v_{r}, t) \propto \bigg{(} \frac{v_{\rm cut}} {v_{r}} \bigg{)}^{n} \times \bigg{(} \frac{t}{t_{0}} \bigg{)}^{-2}, 
\end{align}
\textcolor{black}{respectively.} 
\noindent We denote as $u$ and $v$ the two components of $v_{r}$ that are projected onto and perpendicular to the plane of the sky, respectively. Therefore, 
\begin{equation}
    \theta = {\rm tan}^{-1} \bigg{(} \bigg{|} \frac{u}{v} \bigg{|} \bigg{)}, \\
    v_{r} = \sqrt{u^2+v^2}, \\
    \mu = \sqrt{1- \bigg{(} \frac{u}{v_{ph}} \bigg{)}^2}\, ,
\end{equation}

\noindent where $v_{\rm ph}$ represents the \textcolor{black}{rest-frame} velocity of the photosphere at a given $t$. For an atmosphere of free electrons governed by Thomson scattering, the intensities of the electric vectors parallel ($I_{l}$) and perpendicular ($I_{r}$) to the plane of the sky were adopted from Equations 122 and 123 of Ref~\cite{1946ApJ...103..351C}. With this, the polarization degree $p$ and the intensity-normalised Stokes $Q$ and $U$ parameters can be derived as 
\begin{align}
I_{\rm 0}(\mu) = I_{r}(\mu) + I_{l}(\mu), \ 
p = \frac{I_{r}(\mu) - I_{l}(\mu)}{I_{r}(\mu) + I_{l}(\mu)}, \ 
Q = p\cos(2\phi), \ 
U = p\sin(2\phi)\, .
\end{align}
%I_{l} = (\mu + 0.705927 - (1 - \mu^2)(0.1402646 / (1.0 + 2.718381\mu) + 0.06791696 / (1.0 + 1.118216\mu)) + 0.00718392(1 - 3.458589\mu) + 0.01861255(1 - 1.327570\mu) - 0.0328664(1.0 - 1.046766\mu)) * 3/8
Here, $\phi$ is the longitude measured toward the line of sight. 
%%YY\textcolor{teal}{DBA2YY:  From where is $\phi$ measured?  In what plane?}\textcolor{violet}{YY2DBA: There's only one longitude, I see no need to mention what plane. From where: doesn't matter as it can be easily offset by the viewing angle $\phi_{0}$, $\theta_{0}$.}

Following Refs.~\cite{2000PASP..112..217B, 2002ApJ...565..380K}, we \textcolor{black}{calculated} the shape of the P~Cygni profile of the H$\alpha$ line under the Sobolev approximation. The flux density profile $f_{\rm env}$ was computed separately for the blue side ($v < -v_{\rm ph}$), the middle region ($-v_{\rm ph} \leq v < 0$\,km\,s$^{-1}$), and the red side ($v\geq 0$\,km\,s$^{-1}$). This prescription assumes a 
%homologous and 
spherical atmosphere established soon after the SN explosion. In order to account for the effect of asphericity, we introduce a geometric factor $A(\theta, \phi)$. By multiplying by the optical depth calculated for specific line velocities \textcolor{black}{in the rest frame}, this function characterises the directional dependence of the emission.

To investigate the overall geometric properties of the line-emitting region, we adopted for the sake of simplicity an oblate spatial distribution of the optical depth, namely  
\begin{align}
    \frac{x^2 + y^2}{a^2} + \frac{z^2}{c^2} = 1\, ,
    \label{con:Eq11}
\end{align}
\begin{align}
A(\theta, \phi) = \bigg{[}\frac{\cos(\theta)^2 \sin(\phi)^2}{a^2} + 
                  \frac{\sin(\theta)^2 \sin(\phi)^2}{a^2} + 
                  \frac{\cos(\phi)^2}{c^2} \bigg{]}^{-\frac{1}{2}}\, .
\end{align}~\label{eqn:A}
%\noindent
Therefore, 
\begin{align}
    I = f_{\rm env} A(\theta, \phi)\, .
\end{align}

\noindent
The polarization is then calculated as
\begin{align}
    q = I p \cos2\phi \sin2\theta\, , \\
    u = I p \sin2\phi \sin2\theta\, .
\end{align}

%The polarized flux, namely Stokes $Q$ and $U$ 
%\textcolor{teal}{DBA2YY:  Here we are back to capital Q and U.}
%across a P Cygni profile integrated over $\phi=[0,\ 2\pi]$ can, therefore, be estimated by replacing $I_{0}$ with $p \cos(2\phi)$ and $p \sin (2\phi)$. Figure~\ref{fig:polflux} illustrates the modelling of the P Cygni profile of the H$\alpha$ line at day $+$33.0. 

\subsection*{The Misaligned Symmetry Axes of the Shock Breakout and the Ejecta-CSM Interaction}~\label{sec:rotate}
With the most plausible scenario suggested by the 
\textcolor{black}{temporal evolution of the polarization of} SN\,2024ggi (Figs.~\ref{fig:contpol} and ~\ref{fig:schem_early}, Methods:~\nameref{sec:isp}), we expect a $180^{\circ}$ difference between the PA estimated at Stages I and III since the transition from a prolate to oblate geometry must go through a point with zero polarization and flip the signs of the Stokes $Q-U$ parameters. 
A basic flip in the orientation of the $Q$ and $U$ polarization distribution is illustrated in the bottom row of fig.~\ref{fig:schem} for the case when the prolate and oblate components have a common symmetry axis. For this model, the polarization of the electron-scattering emitting region is calculated for an expanding envelope (Methods:~\nameref{sec:pcygni}), which can be linearly decomposed into a ``prolate'' and an ``oblate'' component. The former and the latter represent the prompt and the later emission that mainly originate from the directions perpendicular to and within the CSM plane, respectively. 

On day 2.0, the continuum polarization jumps to its peak, i.e., from $[Q, U]^{\rm day\,1.1} = [-0.043\% \pm0.074\%, 0.046\%\pm0.077\%]$ to $[Q, U]^{\rm day\,2.0} = [+0.110\%\pm0.075\%, 0.381\%\pm0.069\%]$ (Fig.~\ref{fig:contpol}), computed as the error-weighted mean values over 3800--7800\,\AA. The continuum polarization subsequently decreases monotonically. 
%Of the three straight lines identified in Figure~\ref{fig:contpol_qu}, the one that corresponding to the phase with the photosphere residing in the CSM (the blue dashed line) significantly deviates from the other two. 
%For a simple representation of the differences, the schematic in Figure~\ref{fig:schem_early} adopts a moderate density enhancement towards an arbitrarily defined plane, namely the CSM plane. To illustrate the implications for the geometric structure of SN\,2024ggi, we proceed with some schematic geometric models. 
We hereby break down the possible configurations of the ejecta and the CSM displayed in Fig.~\ref{fig:schem_early}. In Fig.~\ref{fig:schem_early}A both the ejecta and the CSM are spherical, so that there will be no net polarization. In Fig.~\ref{fig:schem_early}B the shock emerges from the star spherically symmetrically and the \textcolor{black}{asphericity}
%asymmetry 
is entirely due to the CSM. 
%This geometry might result from bipolar winds or common-envelope evolution. 
In Fig.~\ref{fig:schem_early}C prolate ejecta will lead to a prompt diffusion of photons from a spherical CSM along certain directions. Configurations illustrated by Figs.~\ref{fig:schem_early}B and~\ref{fig:schem_early}C exhibit only one symmetry axis, producing a single dominant axis in the $Q-U$ diagram \cite{2008ARA&A..46..433W}. The breakout emission would thus emerge promptly toward the direction where a shorter diffusion time is achieved (Methods:~\nameref{sec:diffusion}), producing a prolate photosphere as represented by the equal-arrival-time contour. 
As a consequence, the dominant axis would shrink monotonically and \textcolor{black}{its} orientation remains constant over time until a flip of the signs of $Q$ and $U$ takes place \textcolor{black}{(see the bottom row of fig.~\ref{fig:schem}).} The blue and green dashed lines in Fig.~\ref{fig:contpol_qu} would also coincide. 
The fact that we observed two \textcolor{black}{distinctly} different axes in Fig.~\ref{fig:contpol} disfavor the schematic scenario presented in Fig.~\ref{fig:schem_early}B. A similar argument applies to the alternative where the ejecta are %asymmetric 
\textcolor{black}{aspherical}
and the CSM is spherically symmetric (Fig.~\ref{fig:schem_early}C). 

%However, these features were not observed in SN\,2024ggi, making Figure~\ref{fig:schem_early}b--c implausible for SN\,2024ggi. 
%A flattened CSM would result in a change of the geometry from prolate to oblate as the photosphere evolves to deeper layers. Stronger ejecta-CSM interaction would be seen close to the CSM plane since higher density CSM would provide higher optical depth and emission emerging from the oblate component would become progressively dominant. 

%The breakout emission would thus emerge promptly towards the direction perpendicular to the CSM plane due to a shorter diffusion time (Method~\ref{sec:diffusion}), producing a prolate photosphere as represented by the iso-travel-time contour at this early phase (Figure~\ref{fig:schem_early}b). 

If an aspherical shock breaks through the surface of the progenitor into a nonspherical CSM (Fig.~\ref{fig:schem_early}D), %the explosion is intrinsically nonspherical and 
\textcolor{black}{the behavior is more complex.} The early polarization would also tend to be \textcolor{black}{that of a prolate structure} but aligned with the axes of neither the ejecta nor the CSM. \textcolor{black}{There would be a complex interaction between the ejecta and the CSM, and the polarization \textcolor{black}{tends} to show an oblate geometry as the photosphere recedes toward the H-rich envelope of the SN ejecta.} At later times when the CSM becomes transparent, the polarization should probe the geometry of deeper layers of the ejecta. This qualitative behavior is reflected in our observations. The gradually rotating distribution of the polarization of SN\,2024ggi in the Stokes $Q-U$ plane from days 1.1 to 6.9, which can only be disclosed by the time-resolved data, suggests a misalignment between the aforementioned symmetry axes (see the top row of fig.~\ref{fig:schem}). Consequently, the continuum polarization paints a ``loop-like'' trajectory over time (Fig.~\ref{fig:contpol_qu}). 

Models illustrating the gradual rotation of the direction of the 
\textcolor{black}{centers of the data cloud}
%data clouds center 
on the Stokes $Q-U$ plane over time, assuming one prolate (blue) and one oblate (red) emission component each, are presented in fig.~\ref{fig:schem}. The top and the bottom rows display the two symmetry axes misaligned by $\sim20^{\circ}$ and aligned scenarios, respectively. The presented models adopt $v_{\rm cut}=1000$\,km\,s$^{-1}$, a density index of $n=1.5$, and a viewing angle of $\theta_{0}=90^{\circ}$, $\phi_{0}=70^{\circ}$. Coefficients that are arbitrarily assigned to describe the relative strengths of the prolate and the oblate components (i.e., [$c_{p}$, $c_{o}$]) for the four epochs \textcolor{black}{are} 
[0.5, 0.0], [0.4, 1.2], [0.2, 1.6], [0.1, 2.0]. 
\textcolor{black}{We note that the aim of fig.~\ref{fig:schem} is to provide a schematic illustration of the polarization time evolution for the cases of a time-variant and a fixed axisymmetry, as presented in the top and the bottom rows, respectively. The continuum polarization of SN\,2024ggi draws a `loop-like' trajectory (fig.~\ref{fig:contpol_qu}), suggesting that the ejecta-CSM interaction exhibits a different geometry compared to that measured at the shock breakout and the H-rich ejecta.}

\subsection*{Schematic Evolution of the Geometry of the Ionization Front}~\label{sec:diffusion}
As a simple representation of the aspherical CSM profile that would produce the proposed prolate-to-oblate geometric transformation, 
\textcolor{black}{outside the expanding early ejecta, we adopt a spherical CSM envelope that exhibits density variation throughout the azimuth, where the highest density is achieved near the denoted CSM plane. We assume that the CSM is centered on the SN and quickly swept up by the matter ejected by the SN explosion. The time between 
\textcolor{black}{the emission of a photon from the surface of the SN ejecta and its diffusion out of}
%a photon emitted from the surface of the ejecta and diffuses out
the surrounding CSM can be estimated for any point at the outer CSM boundary as}
\begin{equation}
    t_{d} = \frac{\kappa \rho(r, \theta) \Delta R^2}{c}\, ,
\end{equation}
where $\kappa=0.34$\,cm$^{2}$\,g$^{-1}$ gives the opacity, $\rho(r, \theta)$ represents the number density of the medium at distance $r$ and viewing angle $\theta$, $\Delta R$ denotes the diffusion length from the ejecta to the location ($r, \theta$) in the CSM, and $c$ is the speed of light. 

The density profile of the CSM can be described as 
%r1The density profile of a disc-concentrated CSM can be described as 
\begin{equation}
    \rho(r, \theta) = \rho_{0} r^{-n} (|\cos(\theta)| + \rho_{\rm min})\, ,
\end{equation}
where $\rho_{\rm min}$ indicates the minimum density of the CSM at a latitude angle of $\phi$, and $n$ denotes the power-law index of the radial density distribution. We estimate a characteristic %diffusion time by adding up the inverse of the 
%diffusion time
\textcolor{black}{isodiffusion-time contour by adding up the distances between a given point of the CSM and all points on the ejecta surface, and dividing the lengths of each path by the associated photon travel speed. The former counts only the line segments that connect the given point on the outer boundary of the CSM to the ejecta surface, while the latter is dependent on the number density of the CSM where the photon is traveling through (Fig.~\ref{fig:schem_early}). 
The schematic isodelay contour takes into account the asphericity of the ejecta as well as a disk-concentrated configuration of the CSM. The isodiffusion-time surface can then be sketched over the entire CSM surface.
}
%photon travel time across the whole ejecta to}
%for the paths between 
%a given point of the outer boundary of the CSM to all points at the surface of the ejecta (Figure~\ref{fig:schem_early}). 
The isodiffusion-time surface can then be sketched over the entire CSM surface.

When the shock propagates outward and progressively runs into the CSM, the shock breakout can be seen toward the less-dense regions as hinted by the dominant axes measured across the continuum, which aligned with the \textcolor{black}{photoionized} features (Fig.~\ref{fig:schem_early}). For a spherical CSM embedding a spherical shock breakout, photons will emerge from the CSM isotropically; thus, no polarization would be expected as a consequence of the persistent spherical symmetry (Fig.~\ref{fig:schem_early}A). Any deviation from spherical symmetry in the CSM \textcolor{black}{or} the ejecta would produce an aspherical isophoton-travel-time surface. Examples for the former case with a less-dense CSM toward the perpendicular directions and the latter case with a stretched ejecta are given in Figs.~\ref{fig:schem_early}B and~\ref{fig:schem_early}C, respectively. When both configurations are aspherical and misaligned by a certain angle, the prolate geometry is manifested as an interplay between the internal shock breakout and the external CSM density distribution (Fig.~\ref{fig:schem_early}D). On day $1$, lower-excitation lines can be found over a wide range in azimuth. The emitting region traced by integrating the reciprocal of the diffusion time calculated over the SN ejecta exhibits a peanut shape (Fig.~\ref{fig:schem_early}D). For this configuration, the symmetry axis connects the perpendicular directions, which have the lowest CSM density.  
%As the ejecta continue to expand and in some directions reach the outer boundary of the ambient CSM, the iso-diffusion surface becomes stretched along the direction of the dominant axis of the ejecta. 
%However, unlike the prevailing axial symmetry seen in the photospheric phase of other events, e.g., SN\,2021yja, \cite{2024MNRAS.527.3106V}, no compelling evidence of a persistent symmetry axis can be found in SN\,2024ggi after day 10.9 (Figures~\ref{fig:contpol} and ~\ref{fig:contpol_qu}). 

%For the phase after the ejecta have expanded beyond the CSM, estimation of the diffusion time at the corresponding boundary point was carried out by considering the locations of the continuum emission by the inner ejecta and of the line emission from the outer ejecta, 
%\textcolor{teal}{DBA2YY:  This is totally confusing to me.  Is my wording just approximately right?}
%between which the photons travel at the speed of light, i.e., are not delayed by scattering or similar. Therefore, as the outer boundary of the CSM becomes progressively engulfed by the ejecta, the photon-diffusion effect will become eventually negligible, and the geometry of the emitting zone will gradually settle to that determined by the shape of the ejecta. 
%\textcolor{teal}{DBA2YY:  Would it be possible to replace 'photon diffusion time' with 'photon travel time'?  This would be more intuitive.  At least for me.}

\subsection*{Polarization of the Prolate and Oblate Geometric Configurations}~\label{sec:contpolmodel}
We use the 3D Monte Carlo Polarization Simulation Code (MCPSC~\cite{2023ApJ...955....9W}) for electron-scattering-dominated photospheres to estimate the deviation from spherical symmetry of SN\,2024ggi at various phases. Following the prescription provided by Ref.~\cite{1995ApJ...441..400C}, this technique has been implemented in many SN polarization calculations~\cite{1991A&A...246..481H,2003ApJ...593..788K,2005A&A...429...19L,2015MNRAS.450..967B}. We discretise the space by a $100 \times 100 \times 100$ 3D grid with uniform density ($\rho$) and electron-scattering opacity ($\kappa_{\rm{es}}$). Unpolarized Monte Carlo photon packets are emitted from an electron-scattering-dominated photosphere with an even surface brightness, where the Stokes parameters of each photon packet are initialized as
\begin{equation}
I = 
\left( \begin{array}{c}
I\\ Q\\ U
\end{array} \right)
= 
\left( \begin{array}{c}
1\\ 0\\ 0
\end{array} \right).
\end{equation}
%The propagation of the emitted photon packets through the ellipsoidal envelope of SN\,2024ggi will be tracked and recorded. 
For different ellipsoidal envelope configurations, Equation~\ref{con:Eq11} can be rewritten in cylindrical coordinates as
\begin{equation}
    \frac{r^{2}}{A^{2}} + z^{2} = c^{2}\, ,
\end{equation}
where we introduce the axis ratio ($A=a/c$), with $A<1$ and $A>1$ representing the prolate and the oblate configurations, respectively. The ellipsoidal envelope along radial isodensity surfaces can be expressed as 
\begin{equation}
    \rho(\xi) = \rho_{0}\xi^{-n}\, ,
\end{equation}
where
\begin{equation}
    \xi = \sqrt{\frac{r^{2}}{A^{2}} + z^{2}}\, .
\end{equation}
In all calculations we adopt a power-law index $n = 12$~\cite{2003MNRAS.345..111C} considering the rather dense and steep density profile of the outer layers of the ejecta within the first few days after the SN explosion. The maximum photosphere radius ($R_{\rm{ph}}$) is determined by the position where the electron-scattering optical depth ($\tau$) along the semimajor axis of the ellipsoidal envelope equals unity, where
\begin{equation}
\tau=\int_{R_{\rm{ph}}}^{\xi_{\rm{out}}}\kappa_{\rm{es}}\rho(\xi)dz=1\, . 
\end{equation}
Here $\xi_{\rm{out}}$ denotes the outer boundary of the ellipsoidal envelope. Each photon packet is assigned a random optical depth $\tau = -\ln(z)$ ($0 < z \leq 1$) during its propagation, indicating that scattering will occur whenever an electron packet reaches this optical depth while traversing the medium. Each scattering would alter the Stokes parameters of the photon packet through multiplying the rotation (${L}(\phi)$) and the phase (${R}(\Theta)$) matrices: 
\begin{equation}
{I}_{\rm{out}} = {L}(\pi - i_{\rm{2}}) R(\Theta) {L}(-i_{\rm{1}}) {I}_{\rm {in}}\, ,
\end{equation}
where ${I}_{\rm{in}}$ and ${I}_{\rm{out}}$ denote the set of Stokes parameters in the rest frame prior to and 
%post 
\textcolor{black}{after} a certain scattering event, respectively. Terms $i_{\rm 1}$ and $i_{\rm 2}$ denote the angles on the spherical triangle as defined in Ref.~\cite{1960ratr.book.....C}. 
The rotation matrix yields 
\begin{equation}
{L}(\phi) =
\left( \begin{array}{ccc}
   1    &   0        &   0        \\
   0    & \cos 2\phi & \sin 2 \phi \\
   0    & -\sin 2\phi & \cos 2 \phi \\
\end{array} \right), 
\end{equation}
and the phase matrix in the scattering frame can be written as
\begin{equation}
{R}(\Theta) = \frac{3}{4}
\left( \begin{array}{ccc}
\cos^2 \Theta + 1 & \cos^2 \Theta - 1 &  0 \\
\cos^2 \Theta - 1 & \cos^2 \Theta + 1 &  0 \\
0                 &        0          &  2 \cos \Theta
\end{array} \right),
\end{equation}
where $\Theta$ is the scattering angle in the scattering plane, which has been chosen by sampling its probability distribution function ($f_{\rm{pdf}}$),
\begin{equation}
f_{\rm{pdf}} (\Theta, i_{\rm 1}) = \frac{1}{2} (\cos^2 \Theta + 1) + \frac{1}{2} (\cos^2 \Theta -1)
(\cos 2i_{\rm 1} Q_{\rm in}/I_{\rm in} - \sin2i_{\rm 1} U_{\rm in}/I_{\rm in}).
\end{equation}
Therefore, $i_{\rm 1}$ and $\cos \Theta$ can be sampled from a uniform distribution,
\begin{equation}
\begin{aligned}
     i_{\rm 1} &=  2 \pi \xi_{\rm{1}}\, ,\\
    \cos \Theta &= 1-2\xi_{\rm{2}}\, .
\end{aligned}
\end{equation}
The random seeds $\xi_{\rm{1}}$ and $\xi_{\rm{2}}$ are chosen from a uniform distribution on the interval [0, 1]. After each scattering, the photon packet will travel along the new direction determined by $i_{\rm 1}$ and $\cos \Theta$. This process continues until the photon packet escapes the computational boundary and will be collected in different viewing angle ($\theta$) bins depending on its final direction $\vec{n}$. The continuum polarization of prolate and oblate photospheres seen from different viewing angles $\theta$ are presented in figs.~\ref{fig:model_prolate} and~\ref{fig:model_oblate}, respectively. 

We remark that the purpose of these calculations is to provide a rough quantitative justification of the inferred bipolar shock breakout and the subsequent prolate-to-oblate geometric transformation. The latter is also naturally reproduced by the calculations for a prolate (days 1.1 and 2.0, fig.~\ref{fig:model_prolate}) and an oblate (days 5.8 to 10.9, fig.~\ref{fig:model_oblate}) configuration. A schematic of the temporal evolution of the emission component as approximated by the combination of these two is also illustrated in fig.~\ref{fig:schem}. Within the optically thick regime ($\tau > 1$), the peak of the polarization degree decreases as the optical depth increases until it reaches its asymptotic value at $\tau \gtrsim 4$ (Fig.~1 of~\cite{1991A&A...246..481H}). Therefore, the estimated ellipticities yield a lower bound of the departure of the photosphere from spherical symmetry.

\subsection*{Polarization Across the Balmer and He\,II Lines}~\label{sec:balmer}
%The systematically blueshifted H$\alpha$ profile with emission peak velocities of $\sim -3000$ to $-2000$\,km\,s$^{-1}$ from days 10.9 to 33.0 (Figure~\ref{fig:linepol_late_iqu}) indicates the presence of an extended, aspherical scattering atmosphere during this phase. 
\textcolor{black}{The systematically blueshifted H$\alpha$ profile with emission peak velocities of $\sim-$3000 to $-$2000\,km\,s$^{-1}$ from days 10.9 to 33.0 (fig.~\ref{fig:linepol_late_iqu}, see also Refs~\cite{2024A&A...688L..28P, 2024ApJ...972L..15S}) suggests a rather steep radial density structure of the H-rich envelope of SN\,2024ggi. This can be understood as an enhanced occultation of the receding side of the ejecta, \textcolor{black}{namely extincted by gas on the approaching side}, leading to a suppressed emission towards the red end of the emission profile~\cite{2005A&A...437..667D}.}
Furthermore, a rather steep density gradient during the early recombination phase is expected~\cite{2003MNRAS.345..111C}. Under such a high-density regime, the Balmer emission is driven by a combination of electron scattering and collisional bound-bound excitation. The line polarization profile may thus be formed due to a combined effect of the aspherical limb of the photosphere and the line excitation~\cite{1984MNRAS.210..829M}. The latter may lead to an \textcolor{black}{uneven}
blocking of the underlying photosphere that induces polarization. The universal symmetry axis shared by the prolate shock breakout and the oblate H-rich envelope further strengthens the proposal of a \textcolor{black}{aspherical explosion of SN\,2024ggi, with a well-defined symmetry axis
}

Additionally, our monitoring of the polarization of SN\,2024ggi \textcolor{black}{until day 80.8} shows several distinct temporal trends (fig.~\ref{fig:linepol_late_method}). Portraits of the spectral evolution of the H$\alpha$ and H$\beta$ lines are provided in fig.~\ref{fig:linepol_early_method}. First, a dominant axis with 
\textcolor{black}{2\,PA$_{\rm +80.8\,d}=37^{\circ}.0_{-5^{\circ}.1}^{+5^{\circ}.9}$ can be identified (Fig.~\ref{fig:contpol}).} 
A similar PA is measured after excluding the He\,I and Balmer lines, 
\textcolor{black}{2\,PA$=34^{\circ}.5_{-5^{\circ}.2}^{+4^{\circ}.6}$ (fig.~\ref{fig:isp_data}).} 
Second, the He\,I\,$\lambda$5876 line has emerged, the polarization of which tightly follows a well-defined dominant axis of 2\,PA$^{\rm HeI}=+19^{\circ}.0_{-4^{\circ}.7}^{+4^{\circ}.9}$ that is $\sim33^{\circ}$ off from the symmetry axis shared by the earliest prolate and the later oblate configurations (Fig.~\ref{fig:linepol_late}). A rather small misalignment between the well-defined dominant axis of the ejecta and that of He\,I\,$\lambda$5876 is inferred, 2\,PA$_{\rm +80.8\,d}-$2\,PA$^{\rm HeI} \approx16^{\circ}$. 
\textcolor{black}{The presence of both strong Balmer and He\,II lines and their different dominant axes, therefore, suggests that the mixing of helium into the still optically thick H-rich envelope exhibits a different symmetry axis.}
%r1The presence of both strong Balmer and He~II lines and their different dominant axes, therefore, suggests that the photosphere at day 80.8 is located in a transition zone between the outermost H-rich and the inner He-rich layers, both inherently having large-scale asymmetry. 
%%YY\textcolor{black}{JCW: I'm not clear on the justification for this statement. YY2JCW: Rewrite the whole stentence. Besides, the light curve is still on plateau. I didn't quote it because I'd prefer to leave the photometric analysis to others.} 

%Despite 
Determining the He-rich layer geometry requires spectropolarimetry after the photosphere recedes through the H-rich envelope, the inner ejecta exhibit a symmetry axis as indicated by the dominant axis fitted to the continuum at day 80.8, which is misaligned with the outermost H-rich envelope as defined by a prolate-to-oblate geometry transformation since the shock breakout. A more complicated inner geometry can be inferred from the deviations from axial symmetry in moderate scales as indicated by the departure from the dominant axis at day 80.8 (Fig.~\ref{fig:contpol}, lower-right panel, and fig.~\ref{fig:isp}, right panel). This is also compatible with the main features of the neutrino-driven explosion that manifests on a large scale: bubbles and fractured structures~\cite{2021Natur.589...29B}. Fallback-induced accretion may be involved in reshaping the inner geometry of the ejecta~\cite{1999ApJ...524L.107K, 2001AIPC..586..459H}. 

%%%%%%%%%%%%%%%% REFERENCES %%%%%%%%%%%%%%%

\clearpage % Clear all remaining figures and tables then start a new page
%\printbibliography

% The list of references goes after the main text and before the acknowledgements
% When preparing an initial submission, we recommend you use BibTeX, like this:
%
%\bibliography{science_template} % for a file named science_template.bib
%\bibliographystyle{sciencemag}
\printbibliography
% After the paper has completed peer review and been revised ready for acceptance,
% you should comment out the lines above and copy-paste the contents of your .bbl
% file here instead. This will help ensure that our conversion software works correctly.
% Remember to re-run BibTeX first - check the timestamp!
%
% Example of the first three entries copy-pasted from science_template.bbl:
%
%\begin{thebibliography}{1}
%
%\bibitem{example}
%A.~N. {Author}, An example reference. \emph{Journal of Improbable Research}
%  {1}, 67 (2020).
%
%\bibitem{example2}
%F.~M. {Surname}, S.~{Author}, A second example. \emph{Interesting Research
%  Letters} {32}, 897 (2019).
%
%\bibitem{example_preprint}
%P.~{One}, P.~{Two}, P.~{Three}, {An unpublished preprint}. \emph{preprint}
%  (2021), arXiv:2101.12345.
%
%\end{thebibliography}

%%%%%%%%%%%%%%%% ACKNOWLEDGEMENTS %%%%%%%%%%%%%%%

\section*{Acknowledgments}
We are grateful to the European Organisation for Astronomical Research in the Southern Hemisphere (ESO) for the generous allocation of Director's Discretionary Time (DDT; program ID 113.27R1, PI Y.\,Yang) that enabled this study at ESO's La Silla Paranal Observatory. \textcolor{black}{We express our sincere appreciation to the staff of the Paranal Observatory for their proficient and motivated support of this project in service mode --- in particular, the prompt evaluation of our submission of the DDT proposal, followed by the scheduling of the requested observations. These efforts resulted in the earliest polarimetric observation of any transient.} 
%The polarimetric studies in this work are based on observations collected at ESO's La Silla Paranal Observatory of ESO under programme ID 113.27R1 (PI Y.~Yang). 
%\textcolor{black}{Y.~Yang was a Bengier-Winslow-Robertson Fellow in Astronomy.}
%\textcolor{black}{K.~C.~Patra was a Nagaraj-Noll-Otellini Graduate Fellow in Astronomy. 
%S.~S.~Vasylyev is a Steven Nelson Graduate Fellow in Astronomy.}
%J.\,C.\,Wheeler is supported by NSF grant AST1813825.
%A.\,V.\,Filippenko acknowledges financial assistance from the Christopher R. Redlich Fund and many other donors. 
%S.\,Schulze is partially supported by LBNL Subcontract 7707915. 
\textcolor{black}{A.\,V.\,Filippenko is grateful for the hospitality of the Hagler Institute for Advanced Study as well as the Department of Physics and Astronomy at Texas A\&M University during part of this investigation.} 

\paragraph*{Funding:}
\textcolor{black}{
Y.\,Yang's research is partially supported by the Tsinghua University Dushi Program. 
L.\,Wang acknowledges the U.S. National Science Foundation (NSF) for support through award AST-1817099.}
X.\,Wang is supported by the National Natural Science Foundation of China (NSFC grants 12288102 and 12033003), the Newcorner Stone Foundation and the Ma Huateng Foundation. 
J.\,C.\,Wheeler is supported by NSF grant AST1813825. 
A.\,Gal-Yam’s research is supported by the ISF GW excellence centre, an IMOS space infrastructure grant and BSF/Transformative and GIF grants, as well as the Andr\'{e} Deloro Institute for Space and Optics Research, the Center for Experimental Physics, a WIS-MIT Sagol grant, the Norman E Alexander Family M Foundation ULTRASAT Data Center Fund, and Yeda-Sela; A.\,Gal-Yam is the incumbent of the Arlyn Imberman Professorial Chair. 
%X.~Wang is supported by the National Natural Science Foundation of China (NSFC grants 11178003 and 11325313). 
P.\,Hoeflich acknowledges support from NSF grant AST-1715133 (``Signatures of Type Ia Supernovae, New Physics, and Cosmology''). 
S.\,Schulze is partially supported by LBNL Subcontract 7707915. Generous financial support was provided to A.\,V.\,Filippenko’s supernova
group at U.C. Berkeley by the Christopher R.
Redlich Fund, Steven Nelson, Sunil Nagaraj, Landon Noll,
Sandra Otellini, Gary and Cynthia Bengier, Clark and
Sharon Winslow, Sanford
Robertson (S.\,S.\,V. is a Steven Nelson Graduate Fellow
in Astronomy, K.\,C.\,P. was a Nagaraj-Noll-Otellini Graduate Fellow in Astronomy, Y.\,Y. was a Bengier-Winslow-Robertson Fellow in Astronomy), and many other donors.
\textcolor{black}{The work of A.\,Cikota is supported by NOIRLab, which is managed by the Association of Universities for Research in Astronomy (AURA) under a cooperative agreement with the NSF.}\\
%U.S. National Science Foundation.}\\

%List the grants, fellowships etc. that funded the research; use initials to specify which author(s) were supported by each source. Include grant numbers when appropriate or required by the funding agency. For example: F.~A. was funded by the Generous Science Agency grant~2372.
\paragraph*{Author contributions:}
%List each author’s contributions to the paper. Use initials to abbreviate author names.
Conceptualization: Y.Y., X.Wen., L.W., D.B., J.C.W., A.V.F., J.M., S.S., P.H., X.Wang., F.P., and S.S.V. Methodology: Y.Y., X.Wen., L.W., X.Wang., M.B., P.H., G.L., J.M., and F.P. Investigation: Y.Y., X.Wen., L.W., A.V.F., P.H., J.M., and F.P. Visualization: Y.Y., X.Wen., and L.W. Supervision: Y.Y., L.W., J.C.W., A.V.F., X.Wang., and F.P. Writing-original draft: Y.Y., X.Wen., L.W., and X.Wang. Writing-review \& editing: Y.Y., X.Wen., L.W., D.B., J.C.W., A.V.F., A.G., J.M., X.Wang., C.A., M.B., A.C., H.G., P.H., D.M., K.C.P., and S.S.V. Validation: Y.Y., X.Wen., L.W., J.C.W., X.Wang., J.M., and F.P. Formal analysis: Y.Y., X.Wen., L.W., X.Wang., and S.Y. Funding acquisition: Y.Y., L.W., A.V.F., and X.Wang. Data curation: Y.Y., X.Wen., L.W., and J.M. Software: Y.Y., X.Wen., and L.W. Project administration: Y.Y. and X.Wang. Resources: Y.Y., X.Wen., L.W., and S.S.V. 
%Y.\,Y. initiated the project, wrote the DDT proposal, planned the observations, reduced and analyzed the data, carried out all analyses, modeling, and physical interpretations (except Methods:~\nameref{sec:contpolmodel}, Supplementary figs~\ref{fig:model_prolate} and~\ref{fig:model_oblate}), and wrote the manuscript. 
%X.\,Wen calculated the continuum polarization of the prolate and oblate configurations (Methods:~\nameref{sec:contpolmodel}) and contributed to the manuscript. 
%L.\,W. provided key thoughts to the data interpretation and contributed to the manuscript. 
%D.\,B. and J.\,C.\,W. contributed to the manuscript. 
%J.\,M. contributed to the observation planning and data reduction. 
%C.\,A., M.\,B., A.\,V.\,F., A.\,G., P.\,H., S.\,S., S.\,V., and X.\,Wang contributed to the data interpretation and provided comments on the manuscript. 
%Other coauthors provided comments on the manuscript. 

\paragraph*{Competing interests:}
There are no competing interests to declare.

\paragraph*{Data and materials availability:}
\textcolor{black}{
All data needed to evaluate the conclusions in the paper are present in the paper and/or the Supplementary Materials. 
All data presented in this study are based in part on observations collected at the European Organisation for Astronomical Research in the Southern Hemisphere under ESO program 113.27R1 (PI Y.\,Yang) \textcolor{black}{and can be accessed via: \url{https://archive.eso.org/cms.html}}. 
{IRAF} is distributed by the National Optical Astronomy Observatories, which are operated by the Association of Universities for Research in Astronomy, Inc., under cooperative agreement with the NSF.
PyRAF, PyFITS, and STSCI$\_$PYTHON are products of the Space Telescope Science Institute (STScI), which is operated by the Association of Universities for Research in Astronomy, Inc., under NASA contract NAS5-26555. This research has made use of NASA's Astrophysics Data System Bibliographic Services, the SIMBAD database, operated at CDS, Strasbourg, France, and the NASA/IPAC Extragalactic Database (NED) which is operated by the Jet Propulsion Laboratory, California Institute of Technology, under contract with NASA.}
\subsection*{Supplementary materials}
Supplementary Text\\
Figs.~\ref{fig:isp_data} to~\ref{fig:linepol_late_method}\\
References \textit{(121--123)}\\ % automatically fills out the last reference number
% (filling out the other numbers automatically is possible but fiddly and liable to break)
%Movie S1\\
%Data S1
%%%%%%%%%%%%%%%% END OF MAIN TEXT %%%%%%%%%%%%%%%
\newpage

\clearpage

%%%%%%%%%%%%%%%% START OF SUPPLEMENT %%%%%%%%%%%%%%%

% Figures, tables, equations and pages in the supplement are numbered S1, S2 etc.
\renewcommand{\thefigure}{S\arabic{figure}}
\renewcommand{\thetable}{S\arabic{table}}
\renewcommand{\theequation}{S\arabic{equation}}
\renewcommand{\thepage}{S\arabic{page}}
\setcounter{figure}{0}
\setcounter{table}{0}
\setcounter{equation}{0}
\setcounter{page}{1} % not 0 as \newpage already started a supplementary page
% References continue the numbering from the main text.

%%%%%%%%%%%%%%%% SUPPLEMENT TITLE PAGE %%%%%%%%%%%%%%%

\begin{center}
\section*{Supplementary Materials for\\ \scititle}
\end{center}

% Fill out the numbers for each type of supplementary material,
% and delete any lines that aren't applicable.
% These are just example numbers that don't match the rest of this template.
\subsection*{This PDF file includes:}
\begin{figure}
    \centering
    \includegraphics[trim={0.0cm 0.0cm 0.0cm 0.0cm},clip,width=1.0\textwidth]{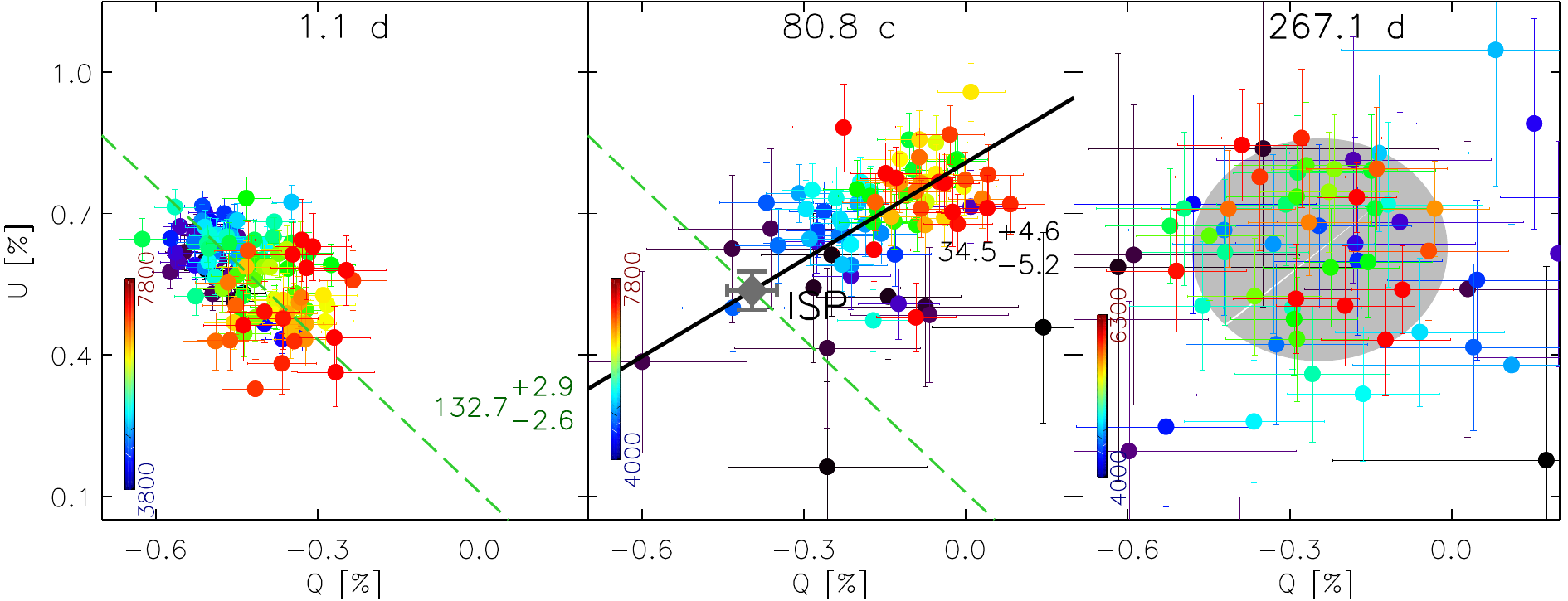}
%{figures/sn2024ggi_contpol_early_late_30A.eps}
    \caption{\textcolor{black}{\textbf{Estimation of the interstellar polarization.} 
    Small filled circles show the measurements in 30\,\AA\,bins, and their colors identify the wavelengths according to the color bar. The green dashed and the black solid lines in the left and the middle panels ﬁt the dominant axes to the observed polarization in the wavelength range 3800--7800\,\AA\ on days 1.1 and 80.8, respectively. The filled gray circle (with 1$\sigma$ error bars) in the middle panel marks the estimated ISP: $Q_{\rm ISP} = -0.40\pm0.05$\,\%, $U_{\rm ISP} = 0.54\pm0.04$\,\%. 
The right panel shows the polarization at day 267.1, when SN\,2024ggi has entered the nebular phase and the emission is dominated by a blend of \textcolor{black}{intrinsically} unpolarized Fe-group features. The ISP estimated by the error-weighted mean polarization within 4000--6300\,\AA\ gives $Q_{\rm ISP}^{\rm +267\,d} = -0.25\pm0.24$\,\%, $U_{\rm ISP}^{\rm +267\,d} = 0.62\pm0.24$\,\% as indicated by the gray-shaded ellipse.
}
    }~\label{fig:isp_data}
\end{figure}

Supplementary Text\\
Figs.~\ref{fig:isp_data} to \ref{fig:linepol_late_method}\\
References \textit{(121 to 123)}\\ 
%References \textit{(94-\arabic{enumiv})}\\ 

%\subsubsection*{Other Supplementary Materials for this manuscript:}
%Movies S1 to S2\\
%Data S1 to S2
\newpage

%%%%%%%%%%%%%%%% MATERIALS AND METHODS %%%%%%%%%%%%%%%
\section*{Supplementary Text}
\textcolor{black}{As a sanity check, we}
also attempted to estimate the Galactic ISP caused by Galactic extinction with observations of the ``probe star'' CD328000 ($\alpha =$11:18:10.676, $\delta = -$32:51:10.729; J2000), which is $143''.85$\,W and $55''.43$\,S of SN\,2024ggi 
%($\alpha = 11:18:22.09$, $\delta = -32:50:15.3$; J2000) 
and has a distance of $1081 \pm 21$\,pc from the Sun as measured by Gaia~\cite{2020yCat.1350....0G}. The 3D extinction map estimates the Galactic reddening toward CD328000, $E(B-V)^{\rm Map} = 0.09_{-0.02}^{+0.01}$\,mag~\cite{1989ApJ...345..245C, 2011ApJ...737..103S}, which is smaller than the Galactic extinction $E(B-V)_{\rm Gal} = 0.120\pm0.028$ mag~\cite{2024ApJ...972L..15S} estimated by adding the reddening derived from the two Na~I~D1 and D2 components from the rest velocity ($E(B-V)_{\rm Gal} = 0.054\pm0.020$ mag) and a Galactic intervening cloud at redshift $z=0.00039$ ($E(B-V)_{\rm Cloud} = 0.066\pm0.020$). 

\begin{figure}
    \centering
    \includegraphics[trim={0.0cm 0.0cm 0.0cm 0.0cm},clip,width=0.6\textwidth]{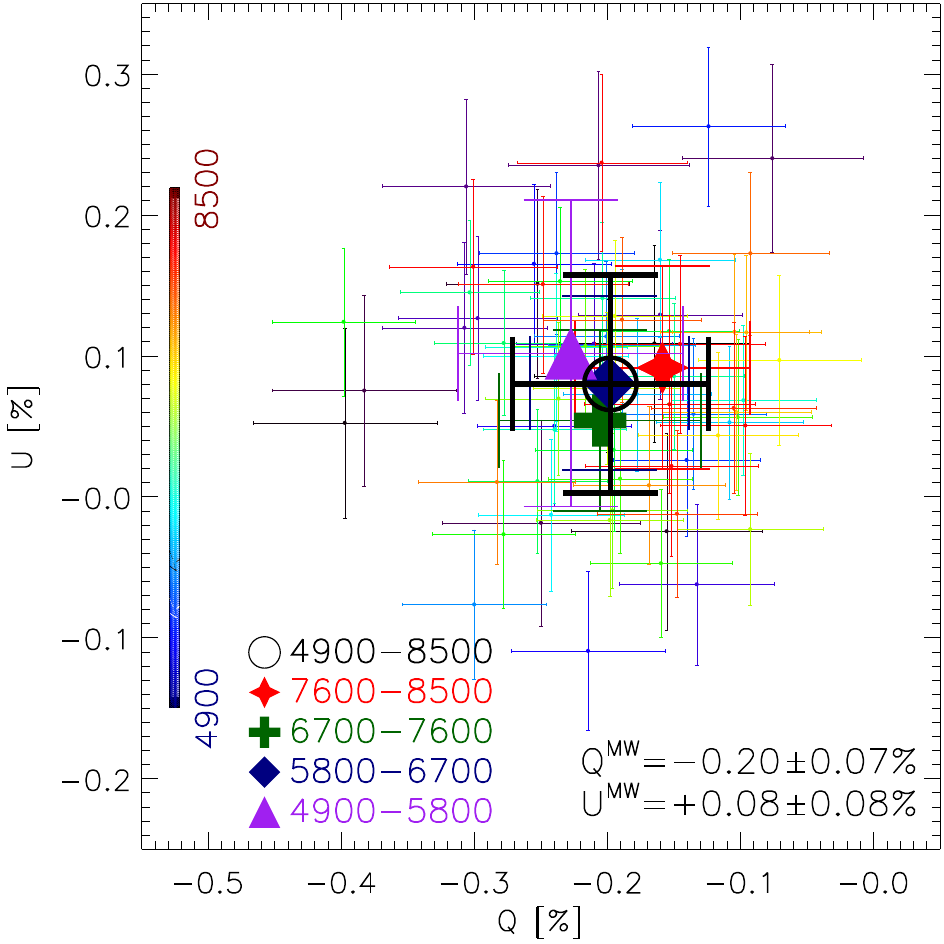}
    %{figures/isp_contpol_regions_50A.eps}
    \caption{\textbf{Polarization of the Galactic ISP probe star.} 
    Four filled symbols represent the weighted-mean polarization in four different wavelength ranges (as labeled).  The small dots show the measurements in 50\,\AA\ bins, and their colors identify the wavelengths according to the color bar.  The black circle (with 1$\sigma$ error bars) marks the overall ISP, the $Q$ and $U$ values of which appear in the legend. 
}
    \label{fig:isp}
\end{figure}

\begin{figure}
    \centering
    \includegraphics[trim={0.0cm 0.0cm 0.0cm 0.0cm},clip,width=1.0\textwidth]{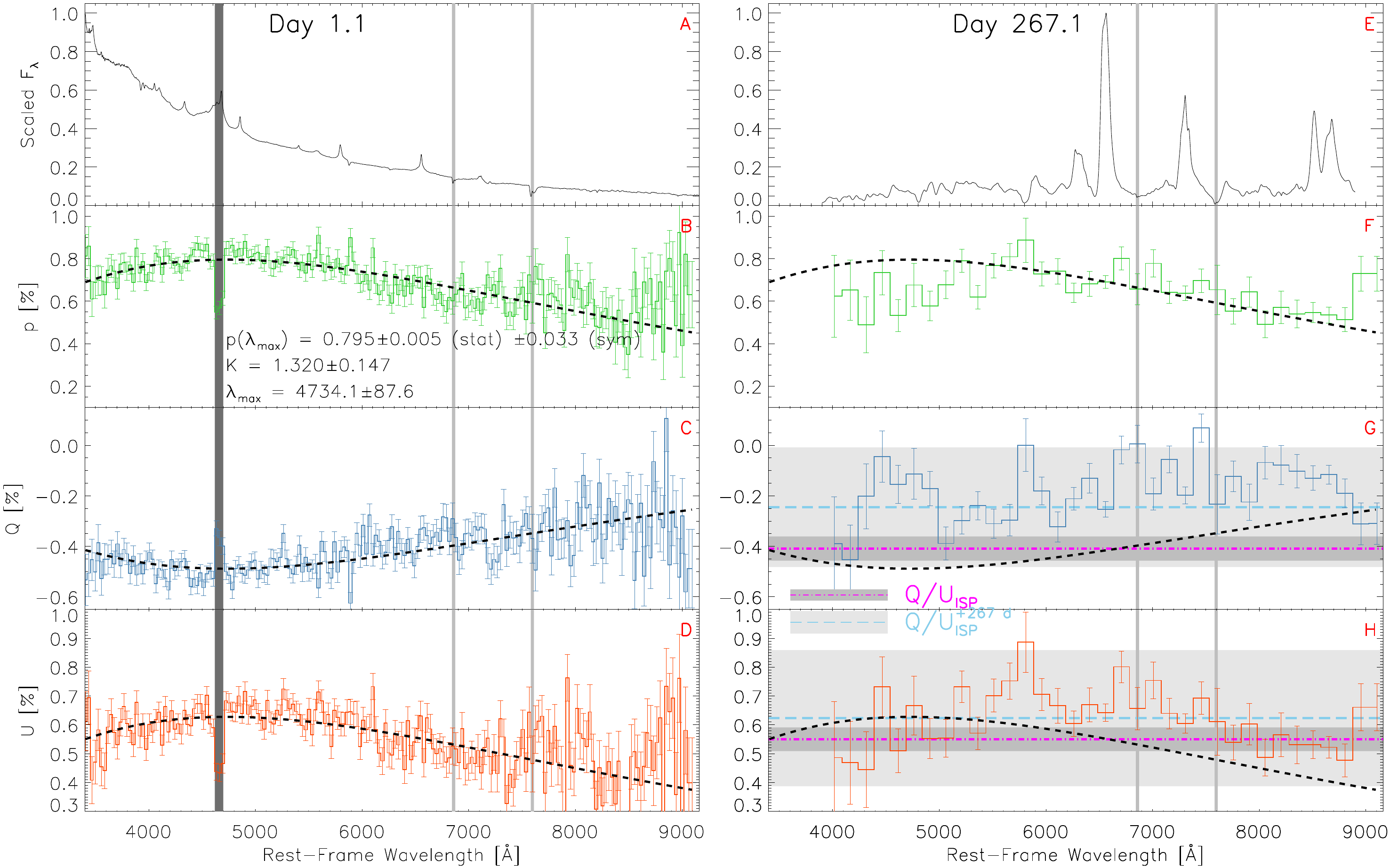}
        %{figures/sn2024ggi_serk_30A.eps}
    \caption{
    \textbf{Fitting the day 1.1 spectropolarimetry of SN\,2024ggi with a Serkowski law.} Panel {\bf A} shows the ﬂux spectrum at day 1.1 normalized to the maximum value within the observed spectral range. Panels {\bf B}--{\bf D} present the degree of linear polarization, the Stokes $Q$ and $U$ for the same epoch, respectively, with the black dashed curve indicating the best ﬁt using a Serkowski law. \textcolor{black}{The result parameters are also labeled. The first and the second error terms in $p(\lambda_{\rm max})$ denote the statistical uncertainty due to the Serkowski-law fitting and the systematic uncertainty as represented by the median error in the $p$ spectrum at day 1.1, respectively. } The data have been rebinned to 30\,\AA. 
    The region of the dark-gray-shaded band suffers from detector saturation and has been excluded from the ﬁtting. 
    \textcolor{black}{The right panel shows the observation on day 267.1 and compares the polarization with the best-fit Serkowski law on day 1.1. The former exhibits significant departures from the latter, indicating the wavelength-dependent polarization at day 1.1 is not caused by the ISP. The data have been rebinned to 150\,\AA\ for clarity. The horizontal gray and light gray-shaded bands overlay the ISP estimated from the intersection of the dominant axes on days 1.1 and 80.8 and the nebular phase data at day 267.1, respectively.}
    }~\label{fig:serk}
\end{figure}

\begin{figure}
    \centering
    \includegraphics[trim={0.0cm 0.0cm 0.0cm 0.0cm},clip,width=1.0\textwidth]    {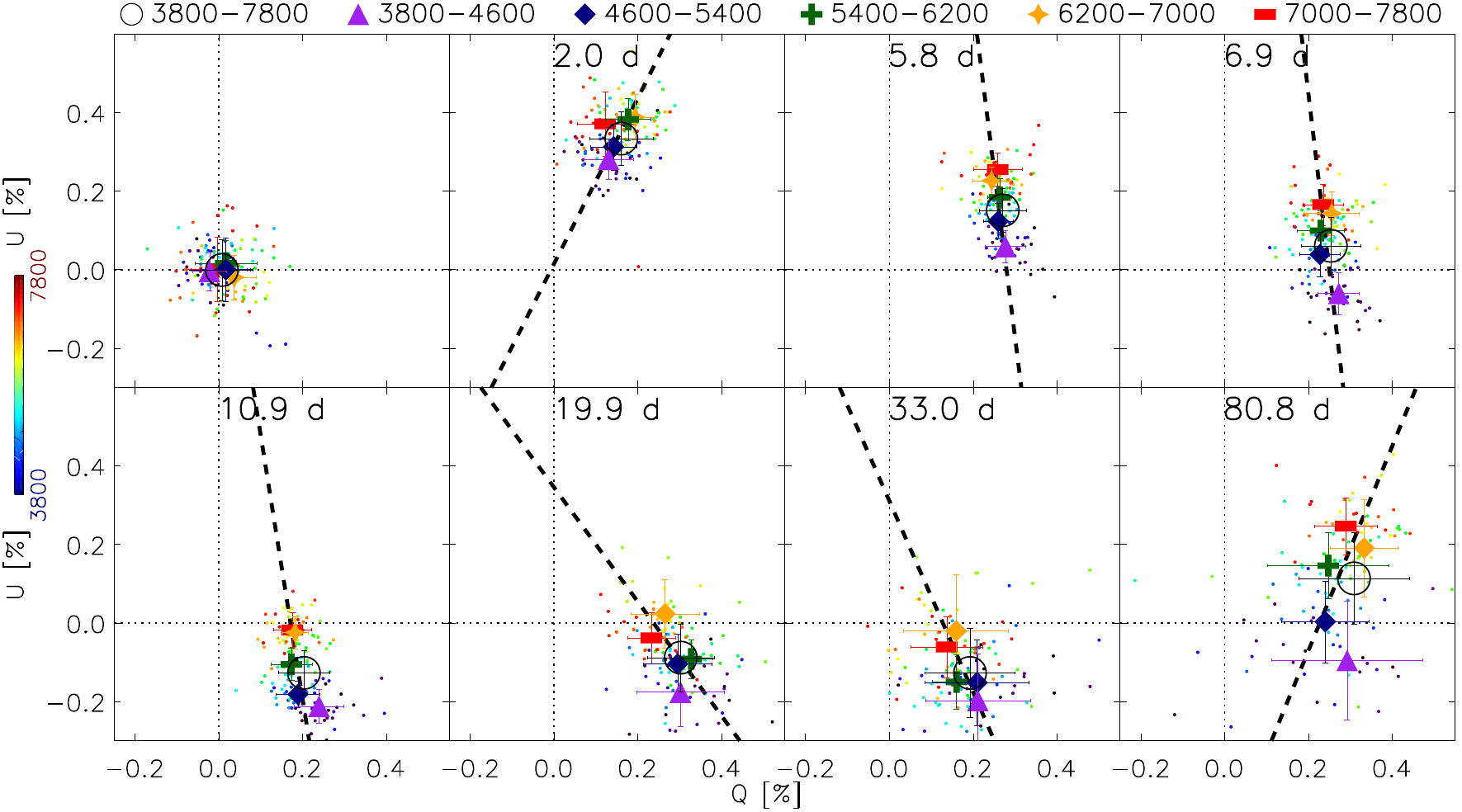}
    \caption{
    \textcolor{black}{
    \textbf{Reproduction of Fig.~\ref{fig:contpol} after removing the wavelength dependence of the day 1.1 spectropolarimetry.} By attributing the latter to the ISP that follows a Serkowski law described in fig.~\ref{fig:serk}, the subtraction would introduce wavelength-dependent polarization at all other epochs as indicated by the best fit to the data shown by the black dashed lines.}
    }~\label{fig:serk_results}
\end{figure}

\begin{figure}
    \centering
    \includegraphics[trim={0.0cm 0.0cm 0.0cm 0.0cm},clip,width=1.0\textwidth]{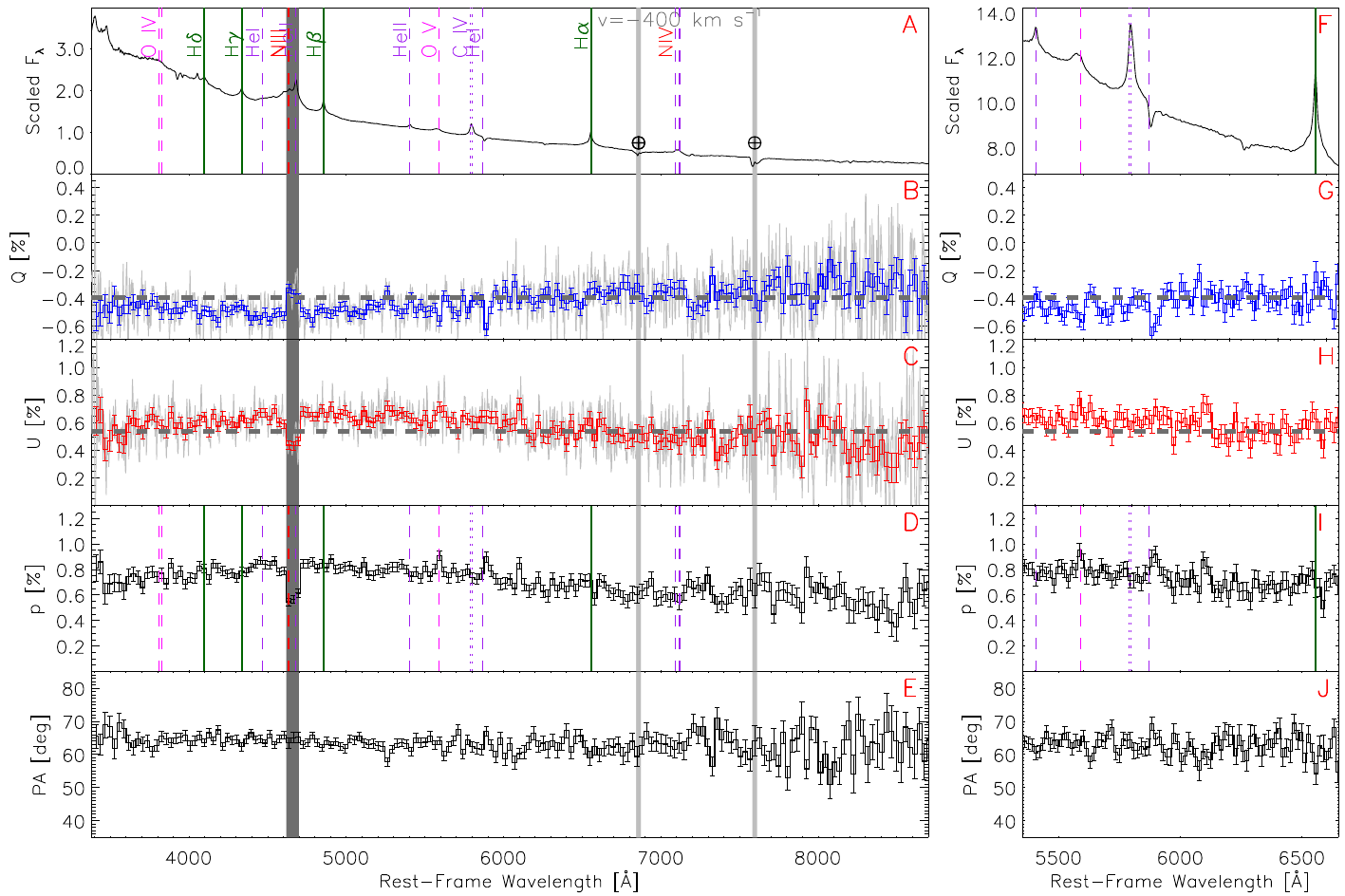}
    %{figures/sn2024ggi_iqu_vltall_ep1_30A.eps}
    \caption{\textbf{Spectropolarimetry of SN\,2024ggi on day 1.1 (epoch 1).} The five panels on the left (from top to bottom) display ({\bf A}) the arbitrarily scaled total-flux spectrum with major spectral features identified; ({\bf B}, {\bf C}) the intensity-normalized Stokes parameters $Q$ and $U$, respectively, \textcolor{black}{with the level of ISP indicated by the horizontal gray dashed lines}; ({\bf D}) the polarization spectrum ($p$); and ({\bf E}) the polarization position angle. 
    %(PA). 
    \textcolor{black}{Panels ({\bf B})--({\bf E}) represent the polarimetry before ISP correction, using 30\,\AA\ bins for clarity.} The light-gray-shaded vertical bands identify regions of telluric contamination while the region of the dark-gray-shaded band suffers from detector saturation. 
    \textcolor{black}{The panels on the right ({\bf F})--({\bf J}) repeat some of the data in the left panels at higher resolution using 15\,\AA\ bins.} 
     }~\label{fig:iqu_ep1}
\end{figure}

\begin{figure}
    \centering
    \includegraphics[trim={0.0cm 0.0cm 0.0cm 0.0cm},clip,width=0.78\textwidth]{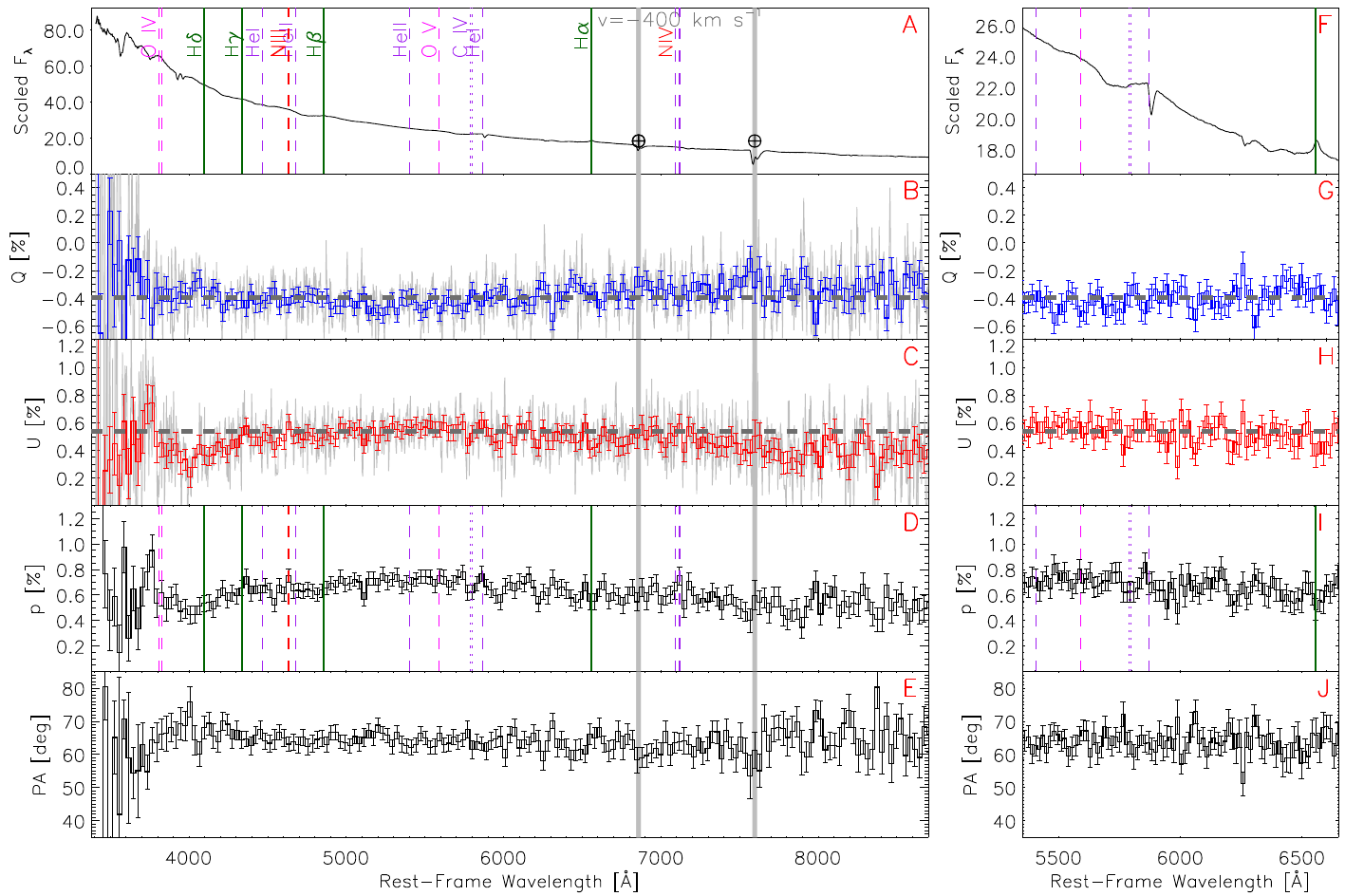}
    %{figures/sn2024ggi_iqu_vltall_ep3_30A.eps}
    \caption{
    \textbf{Spectropolarimetry of SN\,2024ggi on day 2.0 (epoch 2).} The layout is the same as that of fig.~\ref{fig:iqu_ep1}, namely the Stokes $I$, $Q$, $U$, $p$, and PA, from top to bottom rows, respectively. 
    }~\label{fig:iqu_ep2}
\end{figure}

\begin{figure}
    \centering
    \includegraphics[trim={0.0cm 0.0cm 0.0cm 0.0cm},clip,width=0.78\textwidth]{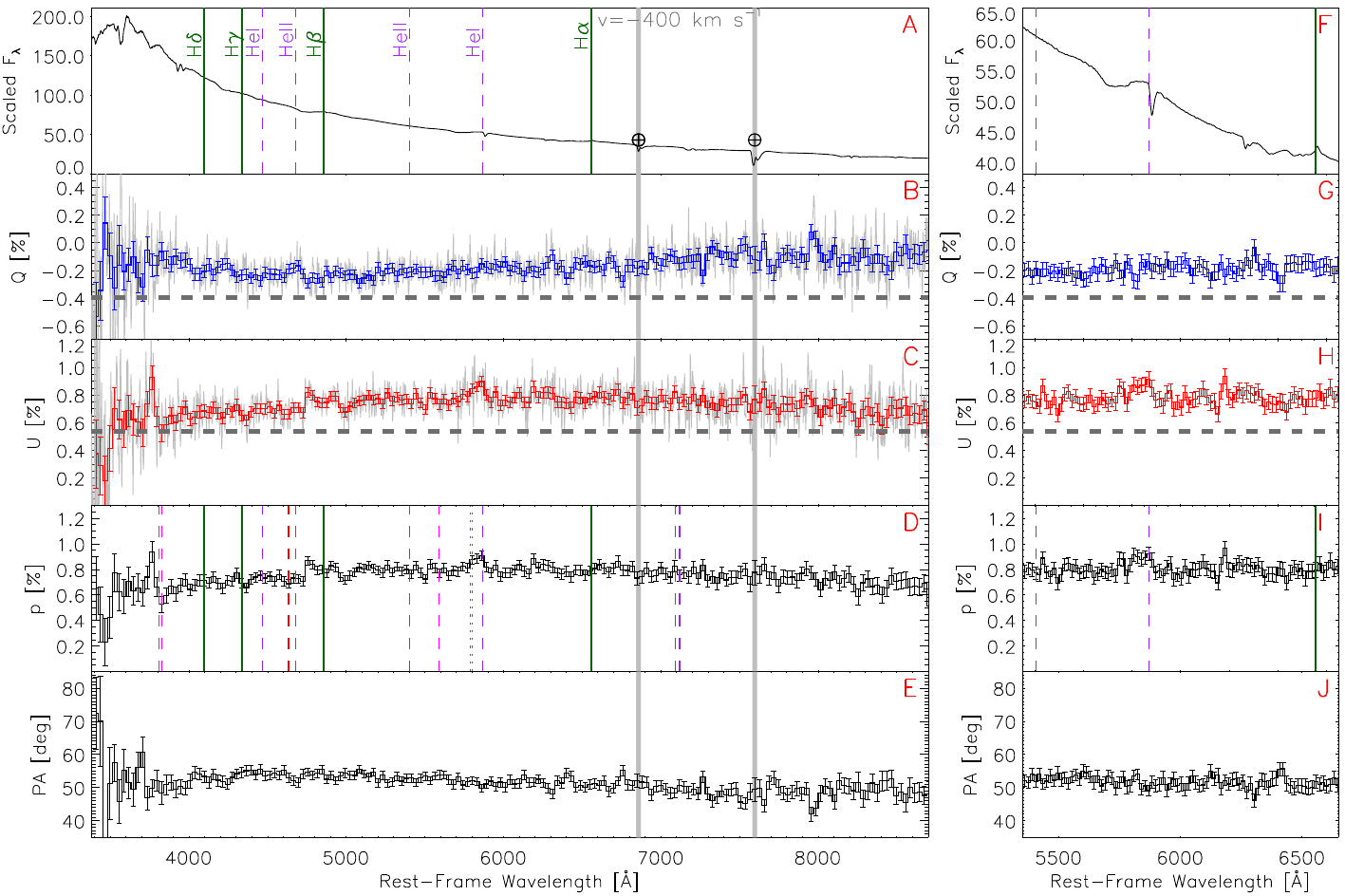}
    %{figures/sn2024ggi_iqu_vltall_ep4_30A.eps}
    \caption{\textbf{
    Spectropolarimetry of SN\,2024ggi on day 5.8 (epoch 3).} The layout is the same as that of fig.~\ref{fig:iqu_ep1}, namely the Stokes $I$, $Q$, $U$, $p$, and PA, from top to bottom rows, respectively.
%%YY    \textcolor{teal}{DBA2YY:  Starting with this figure, the epoch counts seems to be wrong.}\textcolor{violet}{YY2DBA: I intended to omit epoch 3 (4.8$''$ seeing).}
    }~\label{fig:iqu_ep4}
\end{figure}

\begin{figure}
    \centering
    \includegraphics[trim={0.0cm 0.0cm 0.0cm 0.0cm},clip,width=0.78\textwidth]{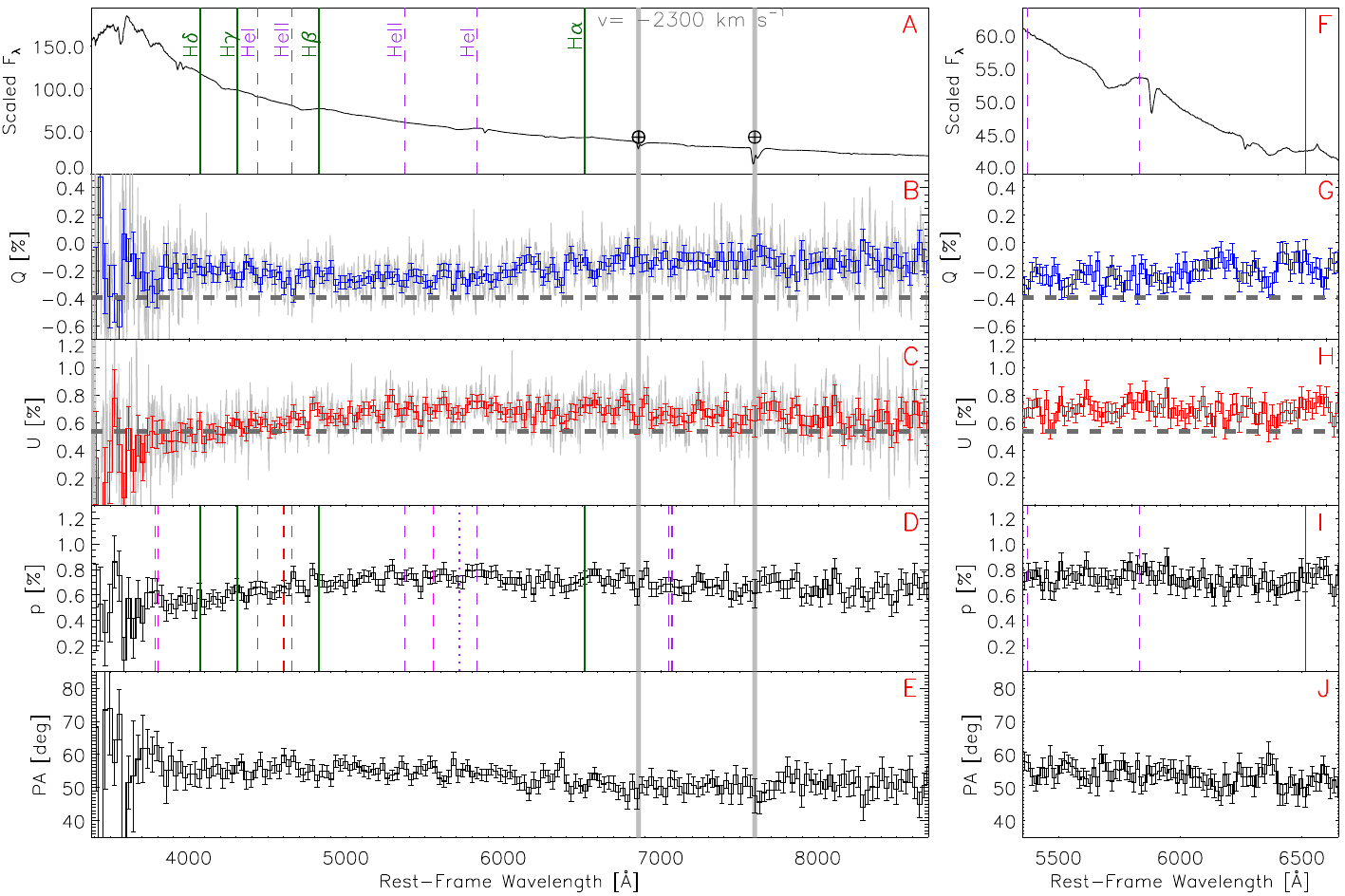}
    %{figures/sn2024ggi_iqu_vltall_ep5_30A.eps}
     \caption{\textbf{
     Spectropolarimetry of SN\,2024ggi on day 6.9 (epoch 4).} The layout is the same as that of fig.~\ref{fig:iqu_ep1}, namely the Stokes $I$, $Q$, $U$, $p$, and PA, from top to bottom rows, respectively.
     }~\label{fig:iqu_ep5}
\end{figure}

\begin{figure}
    \centering
    \includegraphics[trim={0.0cm 0.0cm 0.0cm 0.0cm},clip,width=0.78\textwidth]{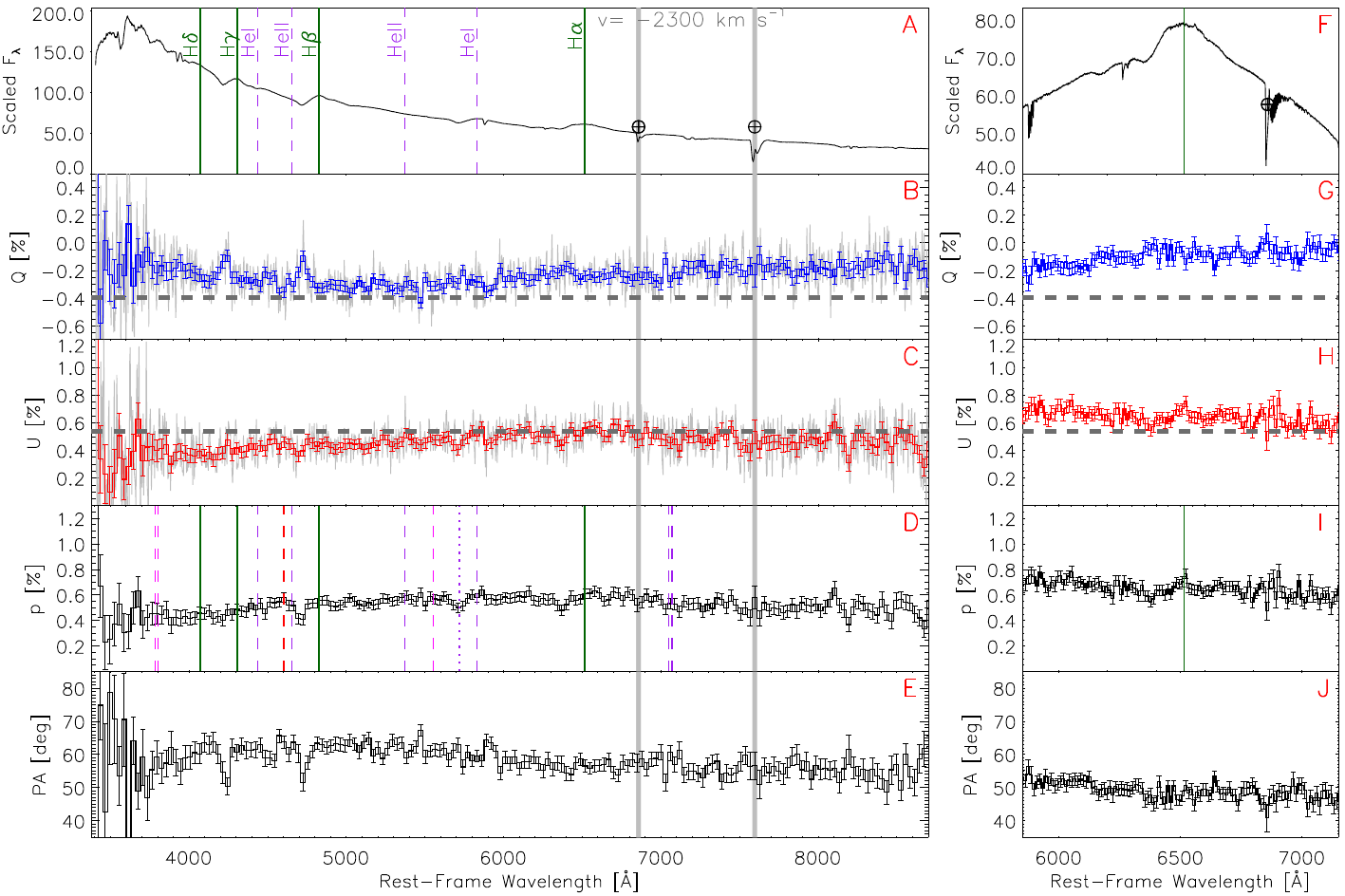}
    %{figures/sn2024ggi_iqu_vltall_ep6_30A.eps}
    \caption{\textbf{
     Spectropolarimetry of SN\,2024ggi on day 10.9 (epoch 5).} The layout is the same as that of fig.~\ref{fig:iqu_ep1}, namely the Stokes $I$, $Q$, $U$, $p$, and PA, from top to bottom rows, respectively.
    }~\label{fig:iqu_ep6}
\end{figure}

\begin{figure}
    \centering
    \includegraphics[trim={0.0cm 0.0cm 0.0cm 0.0cm},clip,width=0.78\textwidth]{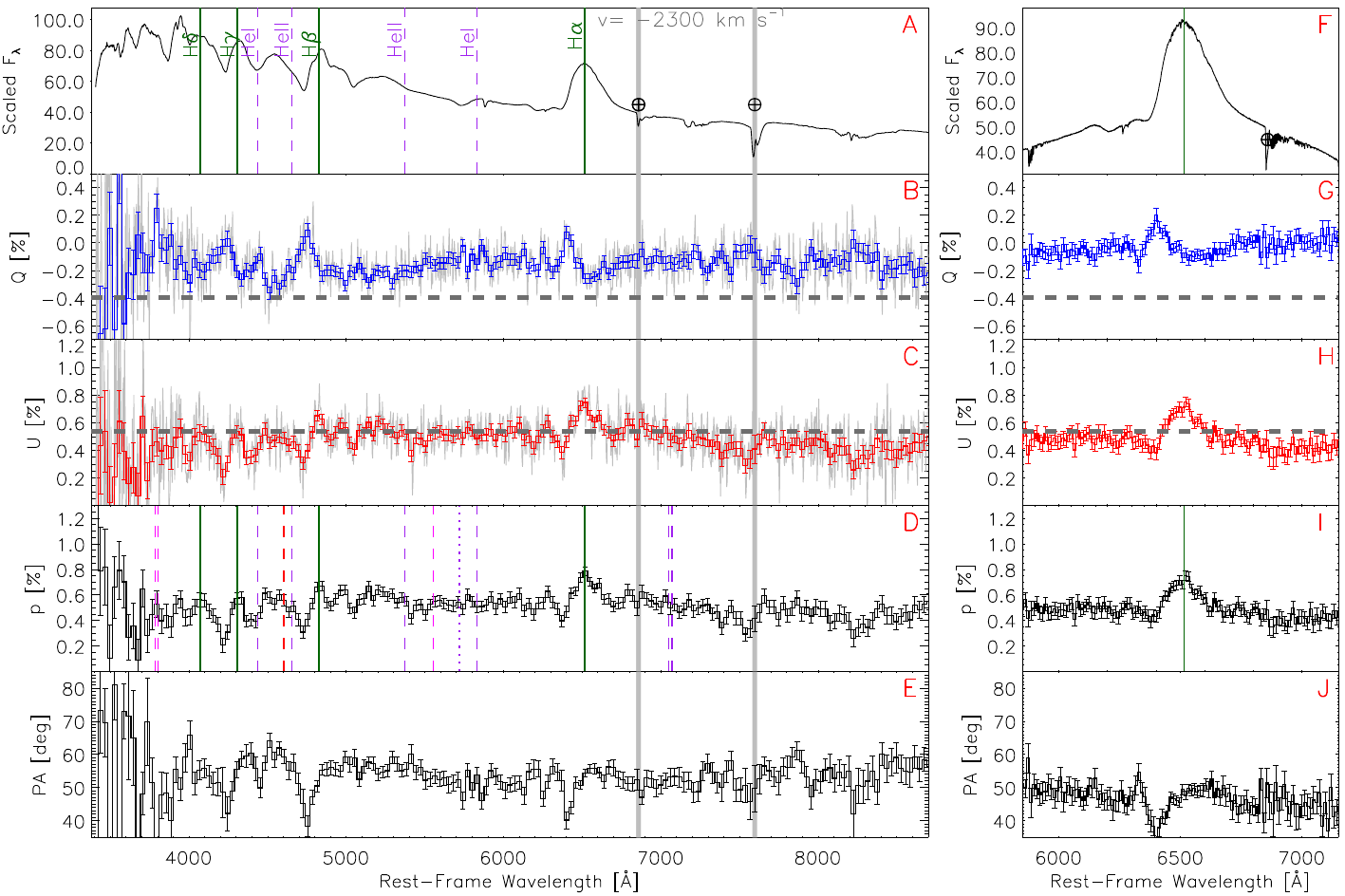}
    %{figures/sn2024ggi_iqu_vltall_ep7_30A.eps}
     \caption{\textbf{
     Spectropolarimetry of SN\,2024ggi on day 19.9 (epoch 6).} The layout is the same as that of fig.~\ref{fig:iqu_ep1}, namely the Stokes $I$, $Q$, $U$, $p$, and PA, from top to bottom rows, respectively.
     }~\label{fig:iqu_ep7}
\end{figure}

\begin{figure}
    \centering
    \includegraphics[trim={0.0cm 0.0cm 0.0cm 0.0cm},clip,width=0.78\textwidth]{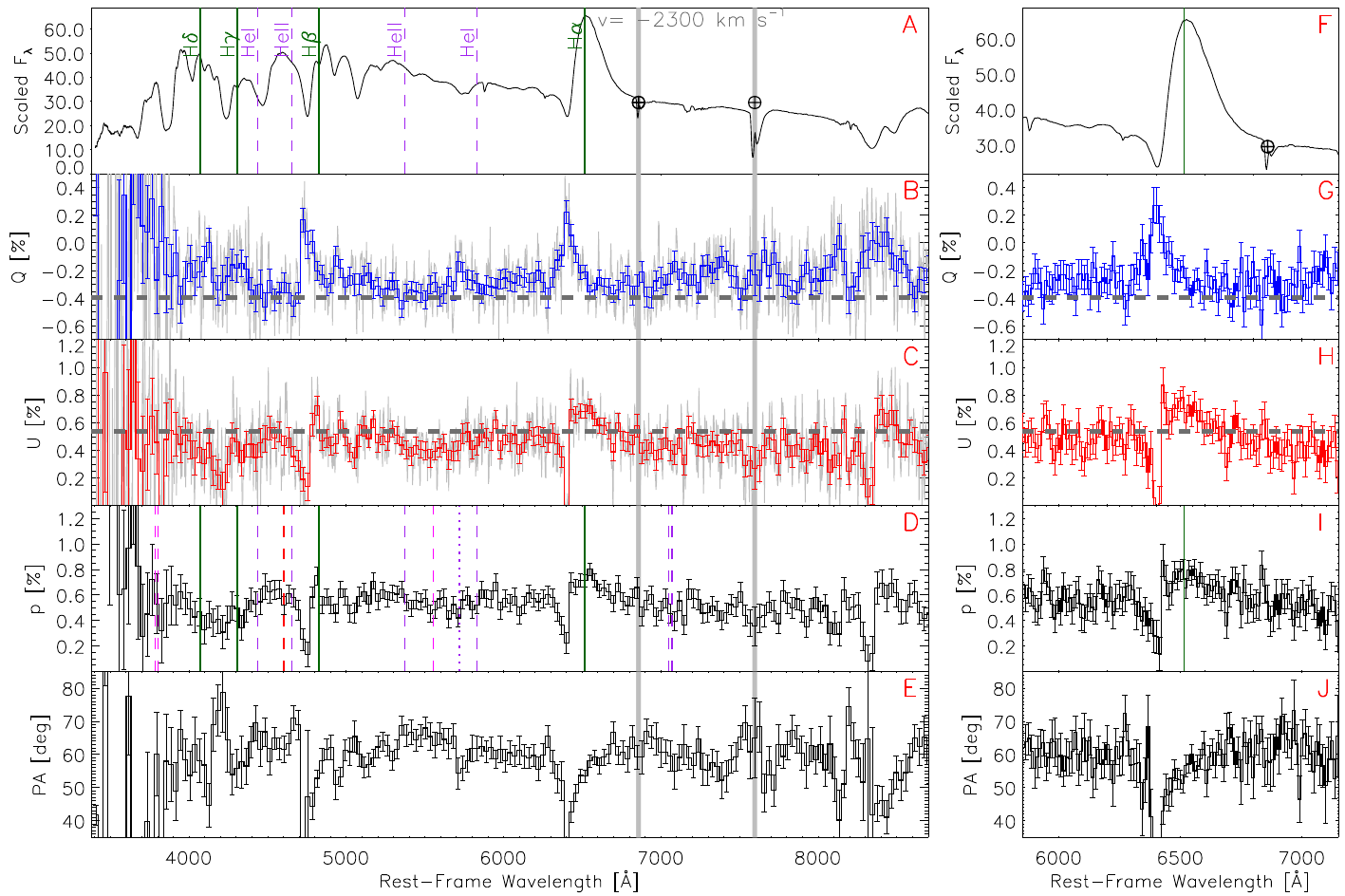}
    %{figures/sn2024ggi_iqu_vltall_ep8_30A.eps}
     \caption{\textbf{
     Spectropolarimetry of SN\,2024ggi on day 33.0 (epoch 7).} The layout is the same as that of fig.~\ref{fig:iqu_ep1}, namely the Stokes $I$, $Q$, $U$, $p$, and PA, from top to bottom rows, respectively.
     }~\label{fig:iqu_ep8}
\end{figure}

\begin{figure}
    \centering
    \includegraphics[trim={0.0cm 0.0cm 0.0cm 0.0cm},clip,width=0.78\textwidth]{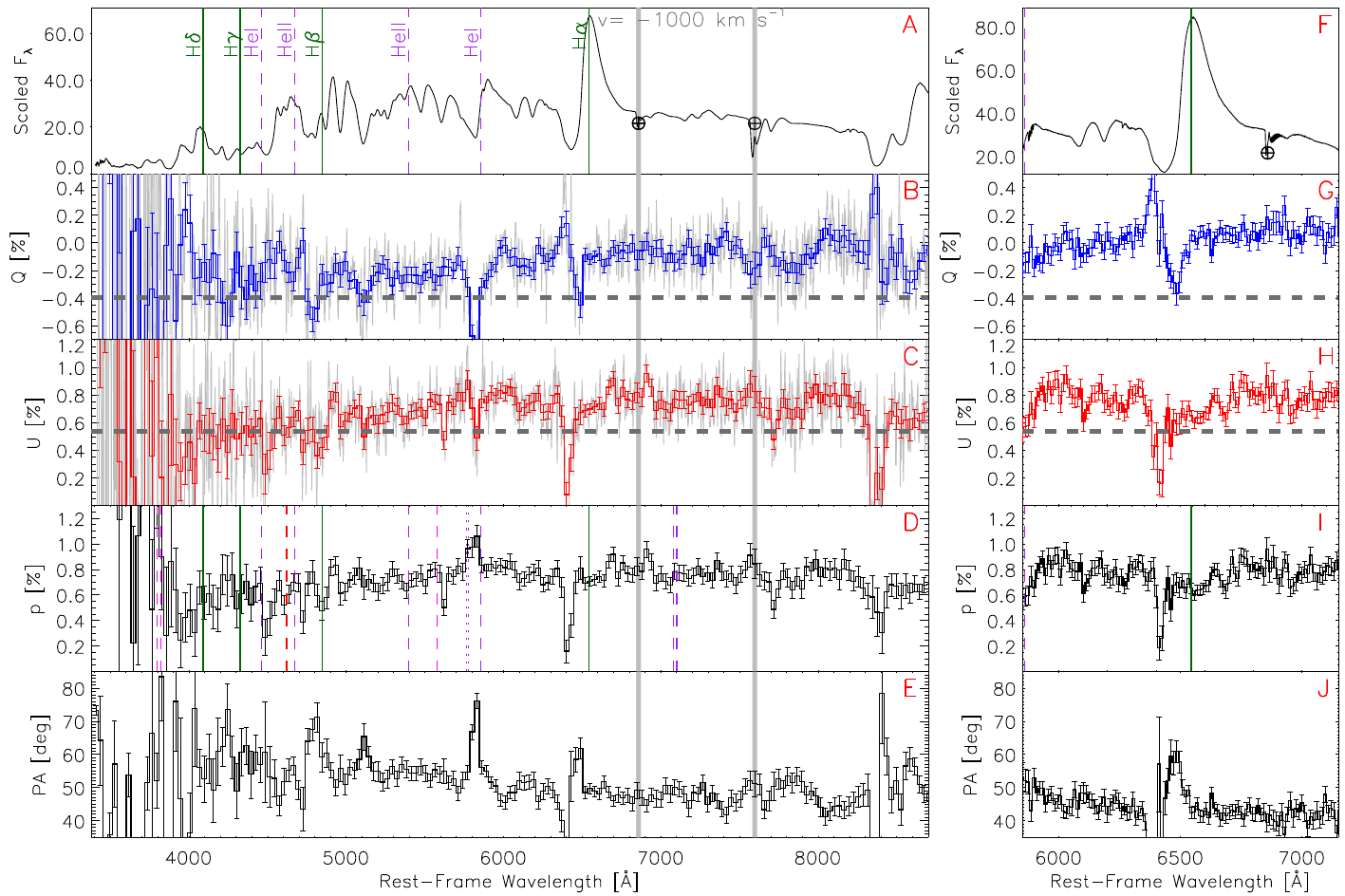}
    %{figures/sn2024ggi_iqu_vltall_ep9_30A.eps}
     \caption{\textbf{
     Spectropolarimetry of SN\,2024ggi on day 80.8 (epoch 8).} The layout is the same as that of fig.~\ref{fig:iqu_ep1}, namely the Stokes $I$, $Q$, $U$, $p$, and PA, from top to bottom rows, respectively.
     }~\label{fig:iqu_ep9}
\end{figure}

\begin{figure}
    \centering
    \includegraphics[trim={0.0cm 0.0cm 0.0cm 0.0cm},clip,width=0.78\textwidth]{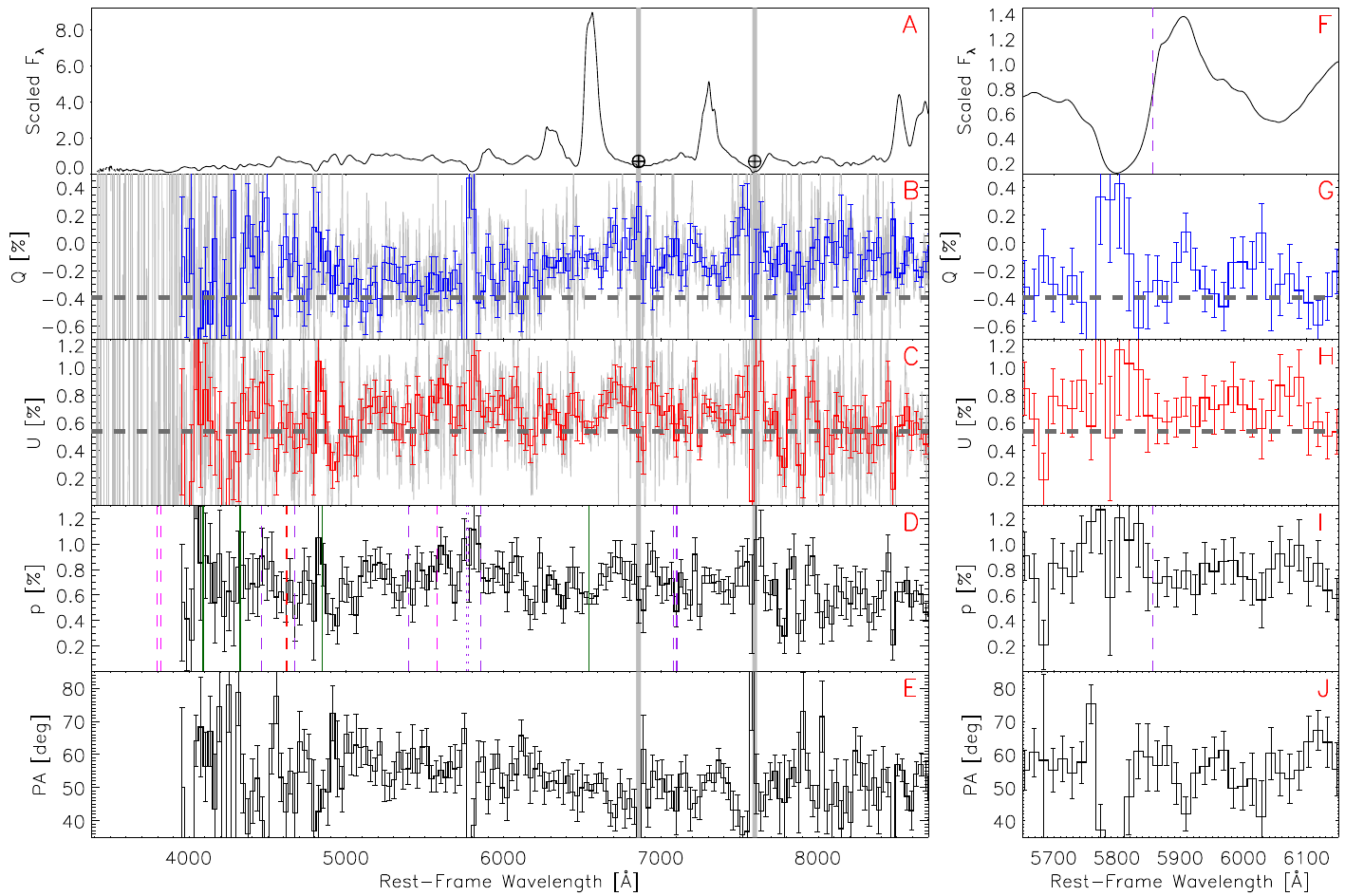}
    %{figures/sn2024ggi_iqu_vltall_ep10_30A.eps}
     \caption{\textbf{\textcolor{black}{
     Spectropolarimetry of SN\,2024ggi on day 267.1 (epoch 9).}} The layout is the same as that of fig.~\ref{fig:iqu_ep1}, namely the Stokes $I$, $Q$, $U$, $p$, and PA, from top to bottom rows, respectively.
     }~\label{fig:iqu_ep10}
\end{figure}

\begin{figure}
    \centering
    \includegraphics[trim={0.0cm 0.0cm 0.0cm 0.0cm},clip,width=1.0\textwidth]{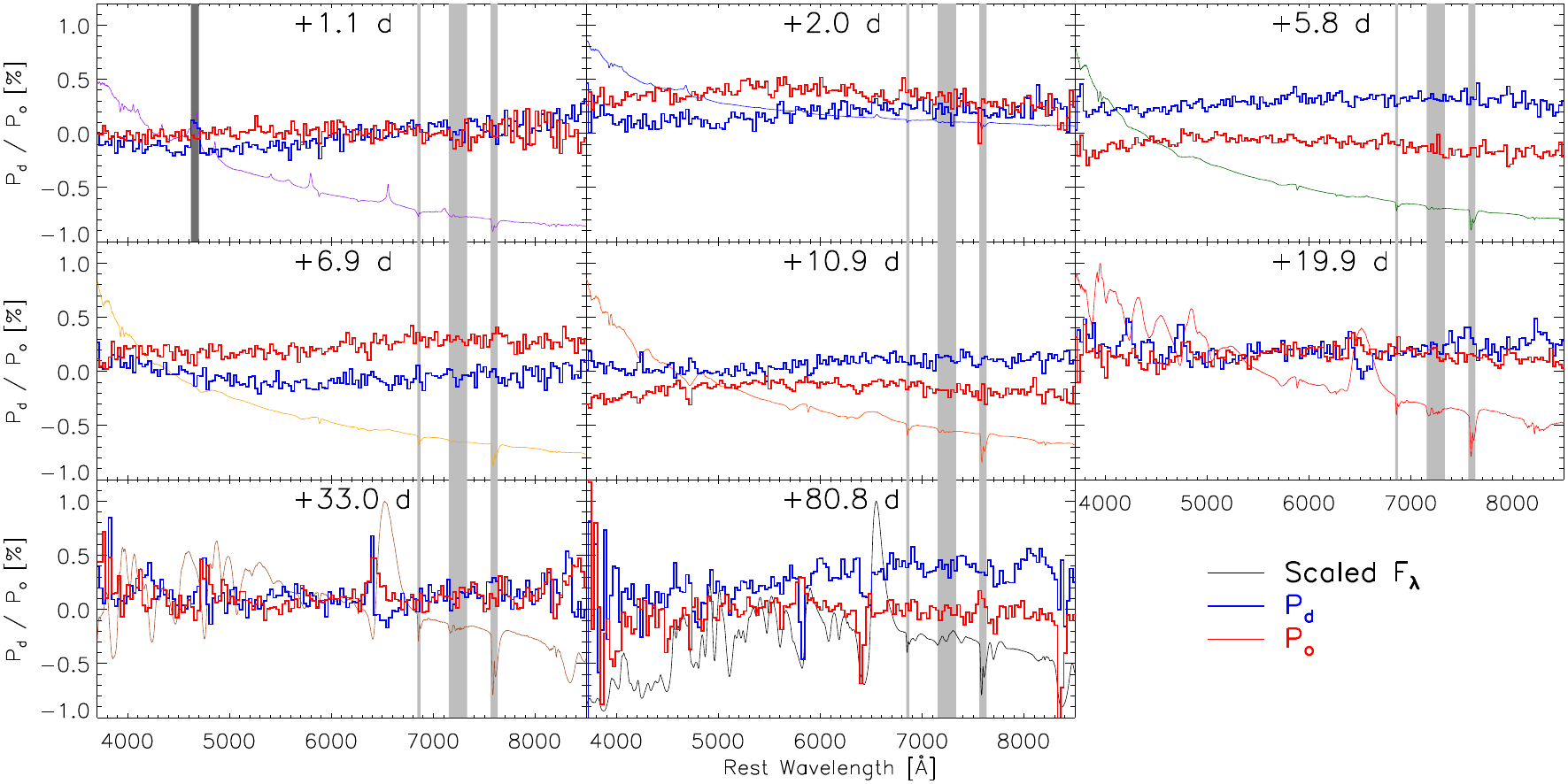}
    %{figures/sn2024ggi_pca_30A.eps}
    \caption{\textbf{Principal-components decomposition of the SN\,2024ggi spectropolarimetry obtained between days 1.1 and 80.8.} 
\textcolor{black}{In each panel, the color-coded line} represents the arbitrarily scaled total-flux spectrum. The blue and the red histograms present the polarization spectrum projected onto the dominant ($P_{d}$) and the orthogonal axes ($P_{o}$), respectively, fitted across the entire observed wavelength range. 
}
    \label{fig:pca_24ggi}
\end{figure}

\begin{figure}
    \centering
    \includegraphics[trim={0.0cm 0.0cm 0.0cm 0.0cm},clip,width=1.0\textwidth]{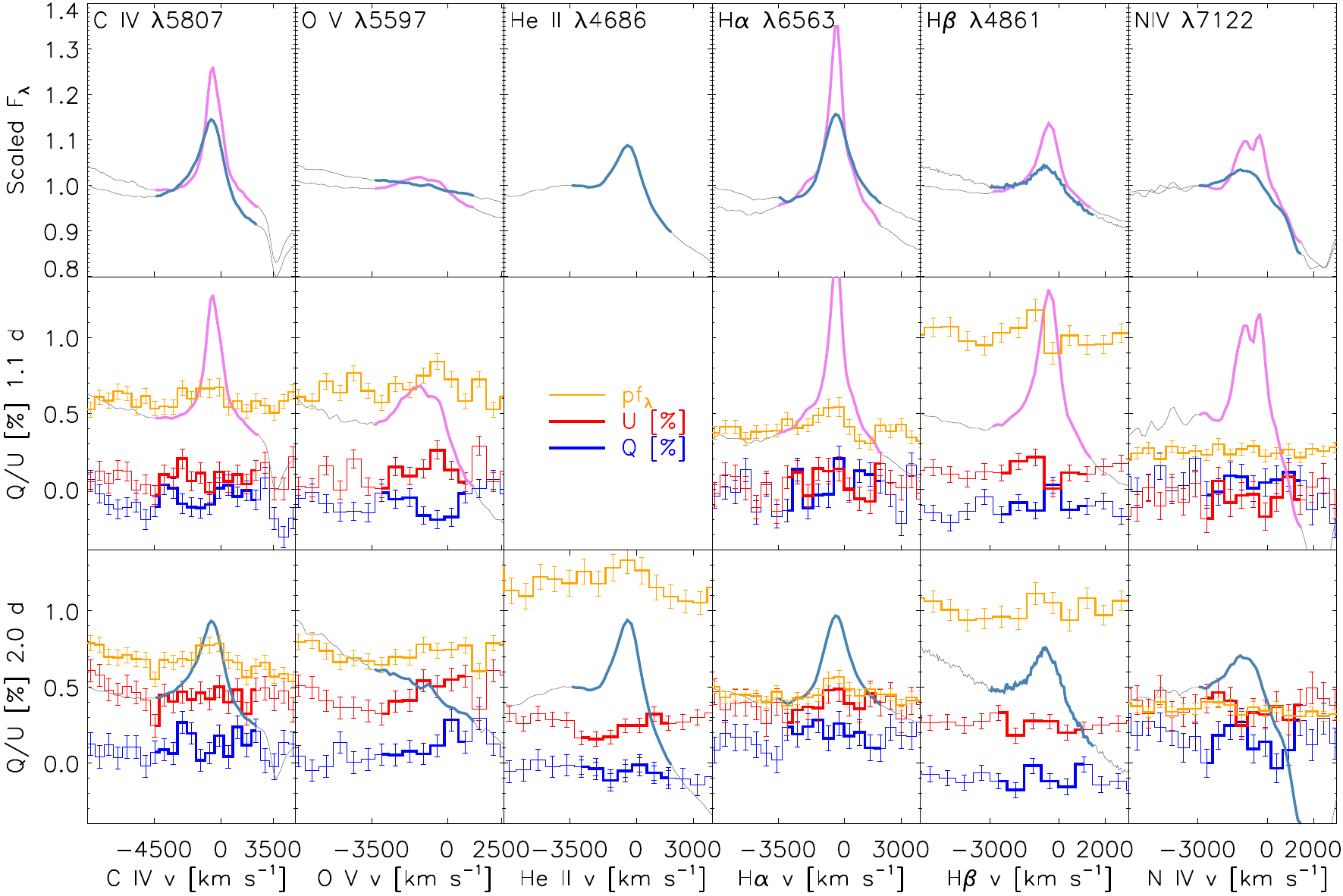}
    %{figures/sn2024ggi_linepol_iqu_12A.eps}
    \caption{\textbf{Portrait gallery of the early-phase \textcolor{black}{photoionized} spectral features in Fig.~\ref{fig:linepol_early}.} The top row compares the scaled flux profiles (Stokes $I$) on days 1.1 (violet lines) and 2.0 (steel-blue lines). The spectral lines are identified above the top panels. The middle and bottom rows present the measurements of Stokes $Q$ (blue histograms) and $U$ (red histograms), respectively, for day 1.1 (middle row) and day 2.0 (bottom row) with arbitrarily scaled Stokes $I$ overlaid. 
    \textcolor{black}{The orange histogram traces the polarized flux density $p\times f_{\lambda}$, which displays no significant deviation from the adjacent continuum.}
    \textcolor{black}{The color-coded} wavelength segments identify the ranges over which the dominant axes of the continuum polarization shown in Fig.~\ref{fig:linepol_early} have been fitted in the Stokes $Q-U$ plane. 
}
    \label{fig:linepol_early_iqu}
\end{figure}

\begin{figure}
    \centering
    \includegraphics[trim={0.0cm 0.0cm 0.0cm 0.0cm},clip,width=1.0\textwidth]{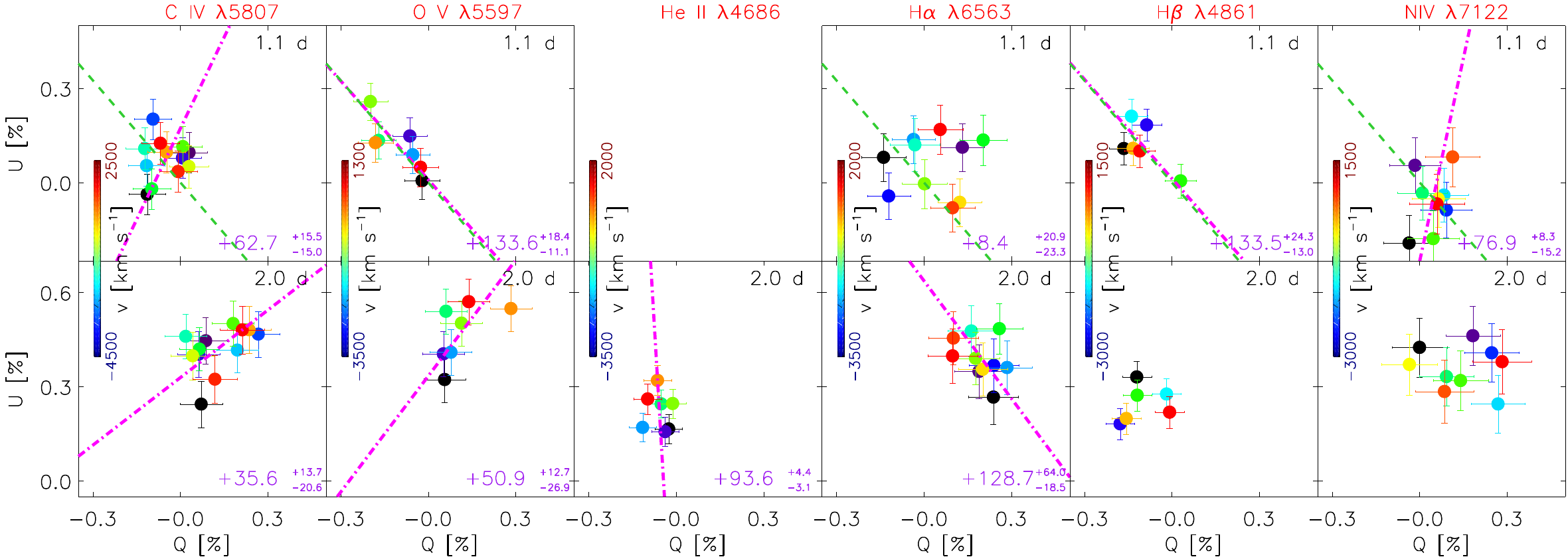}
    %{figures/sn2024ggi_linepol_regions_12A.eps}
    \caption{
    \textbf{Evolution of the line polarization of SN\,2024ggi within two days of the discovery.} 
    $Q-U$ plots \textcolor{black}{with a 15\,\AA\ binning adopted} are shown for five spectral lines as labeled and days 1.1 (top) and 2.0 (bottom) after the discovery 
    (on day 1.1, the He\,II\,$\lambda$4686 emission was saturated and thus excluded from the analysis of the line polarization). In each panel, the magenta dash-dotted line fits the polarization measured at different velocity intervals \textcolor{black}{in the rest frame} identified by color; note that the color bars have different velocity ranges. The green dashed lines are the dominant axes of the continuum polarization (copied from Fig.~\ref{fig:contpol}).     In the top row (day 1.1), the dominant axes of the spectral features with high excitation potential (e.g., O\,V, H$\beta$) closely follow that of the continuum, while other lines, which are formed farther out in the ionization front, have different orientations. 
}
    \label{fig:linepol_early_method}
\end{figure}

\begin{figure}
    \centering
    \includegraphics[trim={0.0cm 0.0cm 0.0cm 0.0cm},clip,width=1.0\textwidth]{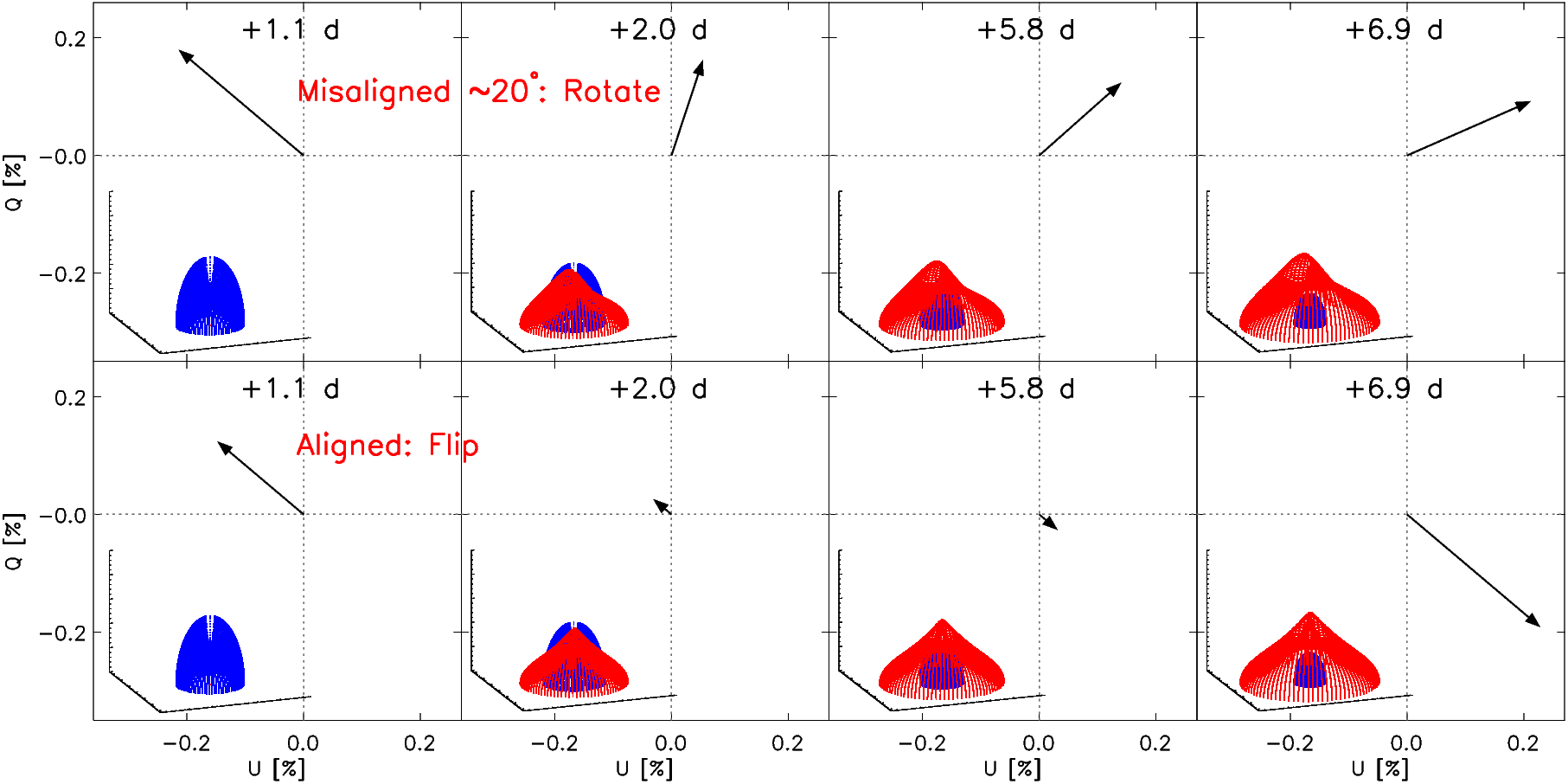}
%    {figures/sn2024ggi_prolate_oblate_arrow.png}
    \caption{
    \textbf{Schematic illustration of the geometric prolate-to-oblate transition in one hemisphere that accounts for the clockwise rotation in the $Q-U$ plane of the dominant axis from days 1.1 to 6.9}. In each panel of the top row, the arrow points towards the direction of the continuum polarization computed from a linear combination of a prolate (blue) and an oblate (red) scattering atmosphere shown in the inset in the lower-left corner. The latter is tilted at an angle of 20$^{\circ}$ relative to the prolate structure and grows monotonically over time \textcolor{black}{representing the way, in which} the emission from the CSM concentrated in a plane becomes dominant. The bottom row illustrates the temporal evolution of the same quantity with the prolate and the oblate components aligned with each other, in which case the direction of the dominant axis exhibits a flip instead of a rotation. 
    %The dark-to-light-blue colour code represents the short-to-long wavelengths. 
    The \textcolor{black}{relative strengths of the prolate and oblate} emission components and the wavelengths were all arbitrarily assigned for illustration purposes. 
}
    \label{fig:schem}
\end{figure}

\begin{figure}
   \begin{minipage}[t]{0.48\textwidth}
     \centering
     \includegraphics[width=1.02\linewidth]{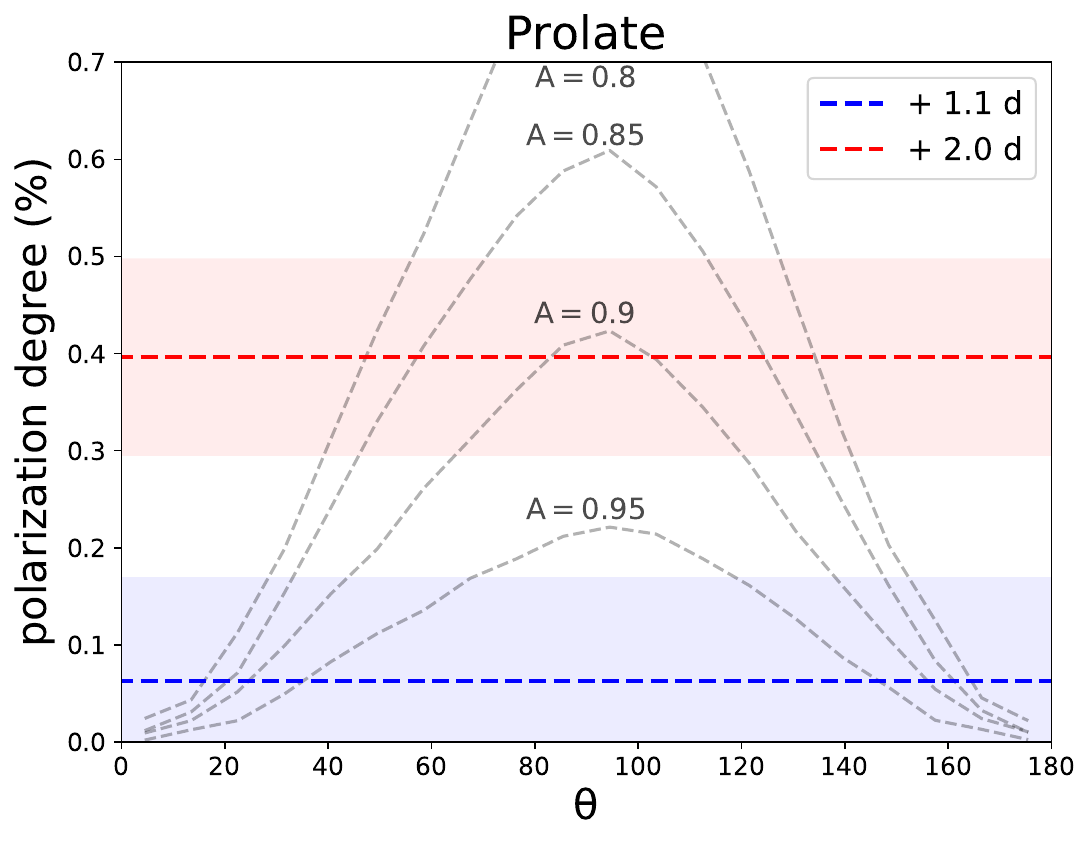}
     %{figures/pol_prolate.pdf} 
     \caption{\textbf{Continuum polarization of a prolate geometry.} Results with $A=a/c<1$ seen from different viewing angles $\theta$ are presented. The ejecta have a radial density structure of $\rho(r) \propto r^{-12}$. Photons are emitted from the photosphere with an optical depth $\tau=1$.
     }\label{fig:model_prolate}
   \end{minipage}\hfill
   \begin{minipage}[t]{0.48\textwidth}
     \centering
     \includegraphics[width=1.02\linewidth]{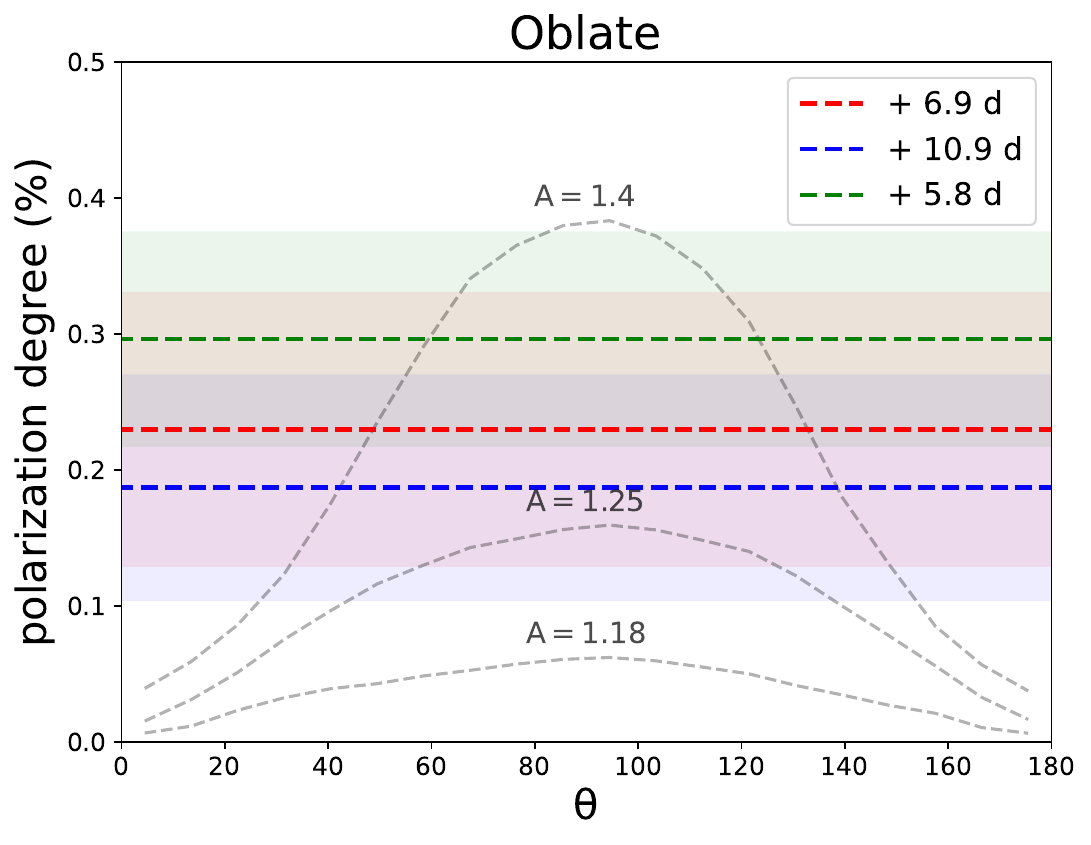}
     %{figures/pol_oblate.pdf}
     \caption{
     \textbf{Continuum polarization of a oblate geometry.} Results with $A=a/c>1$ seen from different viewing angles $\theta$ are presented. 
     Calculations and layout are similar to that of fig.~\ref{fig:model_prolate}.
     }\label{fig:model_oblate}
   \end{minipage}
\end{figure}

\begin{figure}
    \centering
    \includegraphics[trim={0.0cm 0.0cm 0.0cm 0.0cm},clip,width=1.0\textwidth]{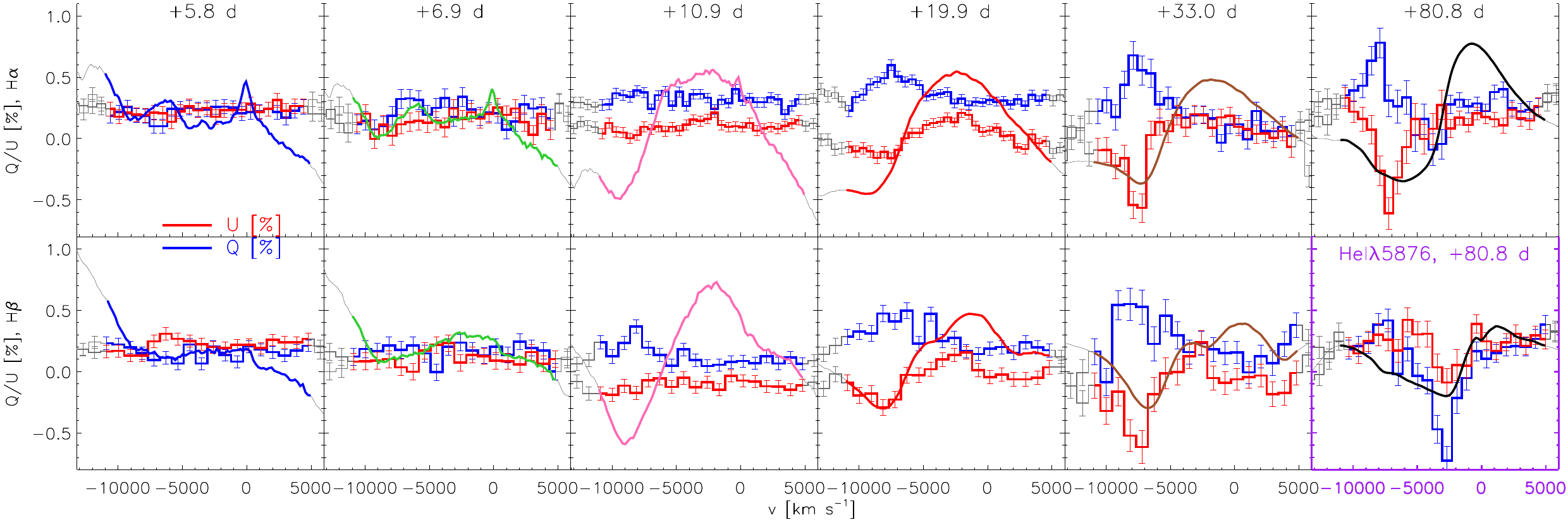}
    %{figures/sn2024ggi_linepol_late_hahb_iqu_15A.eps}
    \caption{\textbf{Stokes $Q-U$ diagrams showing the H$\alpha$ (top row) and H$\beta$ features (bottom row) of SN\,2024ggi.} 
    The layout of the figure is similar to that of fig.~\ref{fig:linepol_early_iqu} except for the epochs being days 5.8 (left column) to 33.0 (right column).
}
    \label{fig:linepol_late_iqu}
\end{figure}

\begin{figure}
    \centering
    \includegraphics[trim={0.0cm 0.0cm 0.0cm 0.0cm},clip,width=1.0\textwidth]{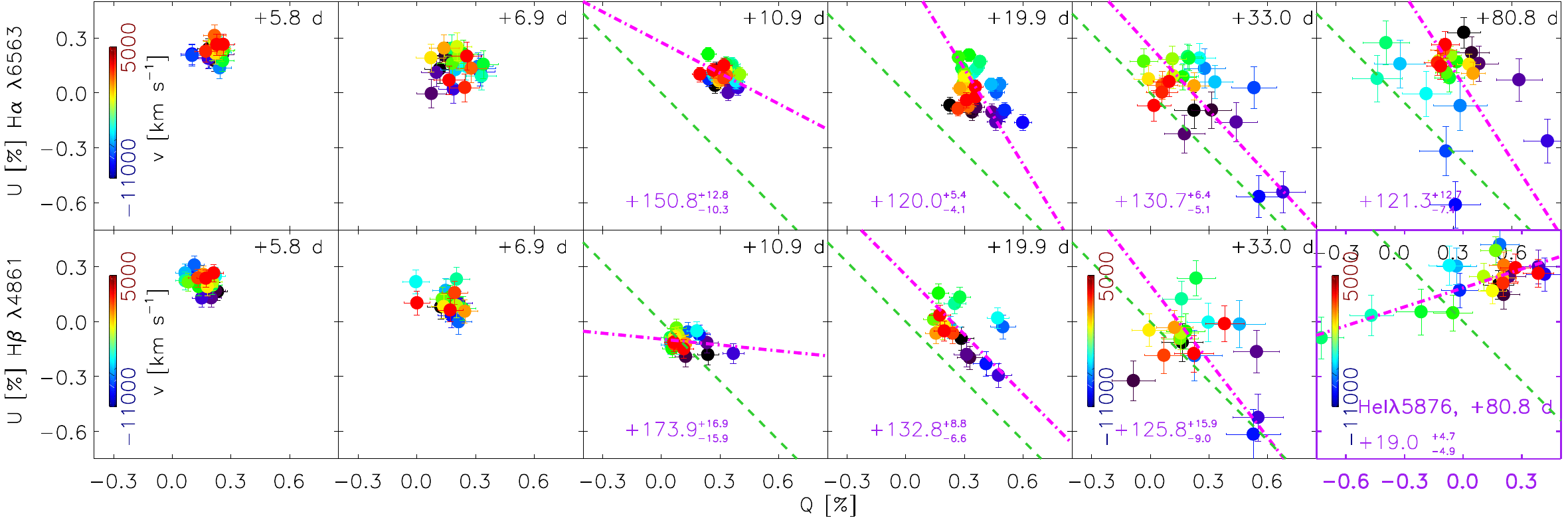}
    %{figures/sn2024ggi_linepol_late_regions_hahb_15A.eps}
    \caption{\textbf{Evolution of the H$\alpha$ (top row) and H$\beta$ (bottom row) polarization of SN\,2024ggi from days 5.8 (left) to 80.8 (right column).} The colors encode velocities according to the color bars. 
    %In each panel, the thick black line represents the shared axial symmetry between the prolate and the oblate configurations as fitted in Figure~\ref{fig:contpol_qu}. 
    In each panel, the magenta dot-dashed line fits the polarization distribution measured at different velocities that cover the corresponding spectral feature. The green dashed lines in the third to sixth columns overplot the dominant axis at day 1.1, which appear to be aligned with that of the H envelope that progressively emerged after day 6.9 (magenta dot-dashed line). The bottom-right panel presents the polarization across the He\,I\,$\lambda$5876 profile \textcolor{black}{on day 80.8. 
    The dominant axis fitted to this feature
    %whose a dominant axis
    } 
    is misaligned with that shared by the prolate and the oblate configurations.}
    \label{fig:linepol_late_method}
\end{figure}

In fig.~\ref{fig:isp} we present VLT spectropolarimetry of 
\textcolor{black}{CD328000, as a sanity check for the Galactic component of the ISP.}
%the ISP probe star. 
The \textcolor{black}{error-}weighted mean values of the continuum polarization calculated for different wavelength ranges are all consistent.  Their combined value is 
\textcolor{black}{($Q^{\rm ISP}_{\rm probe}$, $U^{\rm ISP}_{\rm probe}$) = ($-$0.20\% $\pm$ 0.07\%, $+$0.08\% $\pm$ 0.08\%).}
%($Q^{\rm ISP}_{\rm probe}$, $U^{\rm ISP}_{\rm probe}$) = ($-$0.198\% $\pm$ 0.074\%, $+$0.080\% $\pm$ 0.077\%). 
\textcolor{black}{If the total ISP is proportional to the ratio between the total Galactic extinction along the SN-Earth line of sight and the extinction measured from the} probe star (assuming it to be intrinsically unpolarized which is the case for most single stars), 
\textcolor{black}{the total Galactic ISP of SN\,2024ggi will then amount to} 
($Q^{\rm ISP}_{\rm MW}$, $U^{\rm ISP}_{\rm MW}$)$ = (Q^{\rm ISP}_{\rm probe}$, $U^{\rm ISP}_{\rm probe}$) $\times E(B-V)^{\rm MW} / E(B-V)^{\rm Map} = (Q^{\rm ISP}_{\rm probe}$, $U^{\rm ISP}_{\rm probe}$) $\times 1.508\pm0.389 = [-0.30\%\pm0.14\%, 0.12\%\pm0.12\%]$.

The estimated Galactic ISP component is consistent with an empirical relation between extinction and dichroic extinction-induced polarization, $p_{\rm ISP} [\%] \textless 9 \% \times E(B-V)$, which has been found for dust in the Milky Way \cite{1975ApJ...196..261S}.  However, while different reddening components can be added because they are scalar quantities, this is not true for the ISP values because polarization is a pseudovector. \textcolor{black}{We also assume that both the Galactic and the SN\,2024ggi host dust follow a similar $R_{V}=3.1$ reddening law \cite{1989ApJ...345..245C}.} Depending on the orientation difference between the ISP in the host galaxy and the MW, we estimate a range of the total ISP toward SN\,2024ggi as $-0.38\%\pm0.17\% < Q^{\rm ISP} < -0.21\%\pm0.10\%$, $0.09\%\pm0.09\% < U^{\rm ISP} < 0.16\%\pm0.16\%$. 
%This roughly estimated ISP does not agree with that derived based on the spectropolarimetry observations at days 1.1 and 80.8. In addition to the invalidity of any assumptions based on the Galactic probe star, the discrepancy can be attributed to large variations of the extragalactic dust properties at small scales \cite{2017ApJ...834...60Y}. 

%\begin{figure}
%    \centering
%    \includegraphics[trim={0.0cm 0.0cm 0.0cm 0.0cm},clip,width=0.7\textwidth]{figures/sn2024ggi_polflux.eps}
%    \vspace{-0.0 cm}
%    \caption{{Modeling the P Cygni profile of the H$\alpha$ line of SN\,2024ggi at day $+$33.0.} 
%    The top panel compares the observed flux spectrum with that of the model described in Section~\ref{sec:pcygni}. The bottom panel compares the observed and the modeled Stokes $Q$ and $U$ spectra. From the observed polarization spectra, the ISP has been subtracted, and the underlying continuum polarization has been arbitrarily added to the model spectra to match the observations. 
%}
%    \label{fig:polflux}
%\end{figure}
\clearpage

\end{document}